\documentclass[aps,pra,amssymb,superscriptaddress,twocolumn]{revtex4-1}
\usepackage[pdftex]{graphicx}
\usepackage[dvipsnames]{xcolor}
\usepackage{bm}
\usepackage{amsmath}
\usepackage{ascmac}
\usepackage{color}
\usepackage{setspace}
\usepackage{amssymb}
\usepackage{latexsym}
\usepackage{mathtools}
\usepackage{lipsum}
\usepackage{braket}
\usepackage[colorlinks,linkcolor=RoyalBlue,citecolor=RoyalBlue,urlcolor=RoyalBlue]{hyperref}
\usepackage{MnSymbol}

\usepackage{amsthm}
\theoremstyle{plain}

\allowdisplaybreaks

\begin{document}

\title{Exact Anomalous Current Fluctuations in Quantum Many-Body Dynamics} 

\author{Kazuya Fujimoto}
\affiliation{Department of Physics, Institute of Science Tokyo, 2-12-1 Ookayama, Meguro-ku, Tokyo 152-8551, Japan}

\author{Taiki Ishiyama}
\affiliation{Department of Physics, Institute of Science Tokyo, 2-12-1 Ookayama, Meguro-ku, Tokyo 152-8551, Japan}

\author{Taiga Kurose}
\affiliation{Department of Physics, Institute of Science Tokyo, 2-12-1 Ookayama, Meguro-ku, Tokyo 152-8551, Japan}

\author{Takato Yoshimura}
\affiliation{Department of Mathematics, King's College London}
\affiliation{Rudolf Peierls Centre for Theoretical Physics, University of Oxford, 1 Keble Road, Oxford OX1 3NP, U.K.}
\affiliation{All Souls College, Oxford OX1 4AL, U.K.}

\author{Tomohiro Sasamoto}
\affiliation{Department of Physics, Institute of Science Tokyo, 2-12-1 Ookayama, Meguro-ku, Tokyo 152-8551, Japan}

\begin{abstract}
Fluctuations of integrated currents have attracted considerable interest over the past decades in the context of statistical mechanics. Recently, anomalous current fluctuations, characterized by the M-Wright function, were obtained exactly in a classical automaton [Ž. Krajnik et al., Phys. Rev. Lett. {\bf 128}, 160601 (2022)], and previous studies have shown that the anomalous behavior can arise in a variety of classical systems. Despite the rapidly growing interest in such anomalous behaviors, which capture a universal aspect of one-dimensional many-body transport, the exact derivation of the M-Wright function in quantum many-body systems has remained elusive. In this Letter, we present the first exact microscopic derivation of the M-Wright function in quantum many-body dynamics by analyzing the integrated spin current in a one-dimensional Fermi-Hubbard model with infinitely strong repulsive interactions. Our results lay the groundwork for exploring anomalous integrated currents in a broad class of quantum many-body systems.
\end{abstract}

\maketitle

\let\oldaddcontentsline\addcontentsline
\renewcommand{\addcontentsline}[3]{}

{\it Introduction.--}
Transport constitutes a central class of nonequilibrium phenomena that emerge in a wide variety of classical and quantum many-body systems~\cite{Altland_Simons_2010,Gaspard_2022}. Among the key quantities characterizing transport is the integrated current, which measures the total number of particles that cross a given point—typically the origin—during time evolution. This quantity has played a fundamental role in exploring universal aspects of nonequilibrium phenomena, as exemplified by the establishment of macroscopic fluctuation theory~\cite{Derrida2007,Bertini2015,Mallick2015} and the Kardar-Parisi-Zhang (KPZ) universality~\cite{Sasamoto2007,Krug2010,CORWIN2012,Quastel2015,takeuchi2018}. Such integrated currents have recently attracted considerable attention in one-dimensional quantum and classical systems, particularly due to their nontrivial connection to the KPZ universality~\cite{Ljubotina2019,Dipankar2019,Krajnik2020_CS1,Michele2020,Bingtian2022,Dipankar2023,Moca2023,Moca2025,Takeuchi2025}: various numerical and theoretical studies suggest that dynamical two-point correlators are well described by the stationary KPZ universality class, whereas the probability for the current fluctuations is not. This trend has been accelerated by recent advances in both theoretical and experimental studies of integrated currents in quantum and classical many-body systems~\cite{Moriya2019,Doyon2020,Myers2020,Fujimoto2020,Perfetto2021,Fujimoto2021,Krajnik2022-2,Mallick2022,Fujimoto2022, Nardis2023, McCulloch2023, McCarthy2024,Bhakuni2024,Sreemayee2024,Tater2025,Krajnik2025,Angelo2025,Muzzi2025,Ishiyama2025_1,Ishiyama2025_2,Wei2022,Rosenberg2024}.

\begin{figure}[t]
\begin{center}
\includegraphics[keepaspectratio, width=\linewidth]{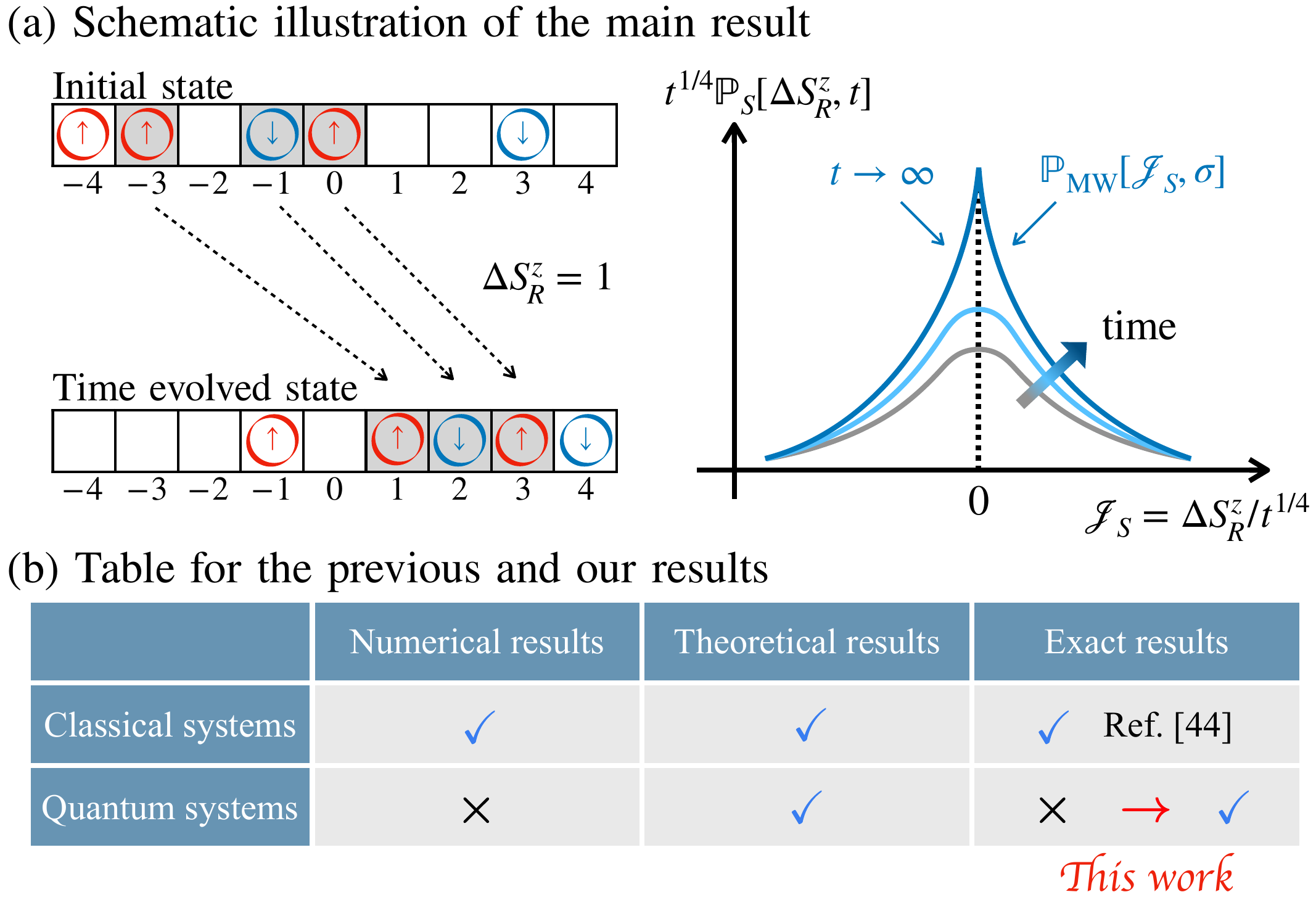}
\caption{(a) Schematic illustration for the main result. 
The left figure depicts the initial and time evolved configurations of fermions on a one-dimensional lattice. The numbers below the lattice denote labels of sites, and a circle with an up (down) arrow represents a fermion with an up (down) spin. In this figure, an integrated spin current $\Delta S^z_R$ (see Eq.~\eqref{eq:def_PS} for the precise definition) is unity. The right figure shows the probability $\mathbb{P}_S[\Delta S_R^z,t]$ of $\Delta S_R^z$ at time $t$ as a function of a scaled integrated spin current $\mathcal{J}_S = \Delta S_R^z/t^{1/4}$. In this work, we exactly demonstrate that it obeys the M-Wright function $\mathbb{P}_{\rm MW}[\mathcal{J}_S,\sigma]$ in the long-time limit. 
(b) Table for previous results and ours. The symbol of $\checkmark~(\times)$ means presence (absence) of previous works reporting $\mathbb{P}_{\rm MW}[\mathcal{J}_S,\sigma]$ for integrated currents. The theoretical results mean that $\mathbb{P}_{\rm MW}[\mathcal{J}_S,\sigma]$ is derived under some approximations or assumptions.
} 
\label{fig1} 
\end{center}
\end{figure}

One of the most intriguing recent findings on integrated currents is the emergence of anomalous current fluctuations; specifically, the probability of a one-dimensional integrated current $\mathcal{J}$ obeys a strongly non-Gaussian distribution whose probability density function is the M-Wright function $\mathbb{P}_{\rm MW}[\mathcal{J},\sigma] \coloneqq \int^{\infty}_{0}  d s  e^{ - s^2/(2\sigma^2) - \mathcal{J}^2 /(2s) }/(\pi \sigma \sqrt{s})$ with a parameter $\sigma$~\cite{Mainardi2010,comment1}. This anomalous behavior was first derived exactly by Krajnik et al. in a particular classical automaton model~\cite{Krajnik2022}. Interestingly, subsequent theoretical works have uncovered that the anomalous current fluctuations associated with the M-Wright function can emerge in a variety of classical many-body systems, including anisotropic spin chains~\cite{Kormos2022,Gopalakrishnan2024-1,Gopalakrishnan2024-2,Krajnik2024-1,Krajnik2024-2,Krajnik2025_2,Yoshimura2025-1,Yoshimura2025-2}.
Thus, the anomalous current fluctuations capture a universal feature of one-dimensional many-body transport physics. Despite this intriguing universal behavior and the recently growing interest in integrated currents of quantum many-body dynamics, no exact microscopic derivation of the M-Wright function has been achieved in any quantum many-body system to date, although a few theoretical works~\cite{Kormos2022,Gopalakrishnan2024-1} based on some approximations exist. Therefore, it is important, challenging, and timely to demonstrate exactly the appearance of the M-Wright function in quantum many-body dynamics by microscopic calculations. 

In this Letter, we present the first exact microscopic derivation of the M-Wright function in quantum many-body dynamics, thereby establishing a theoretical foundation for exploring the anomalous current fluctuations in quantum many-body systems. The theoretical model used in this work is a one-dimensional $t\textrm{-}0$ model, which is an effective model for a spin-1/2 Fermi-Hubbard model with strongly repulsive interactions~\cite{Essler2005}. We exactly calculate the probability for the scaled integrated spin current $\mathcal{J}_S$ of the $t\textrm{-}0$ model with a specific initial state, demonstrating that it converges to the M-Wright function $\mathbb{P}_{\rm MW}[\mathcal{J}_S,\sigma]$ in the long-time limit. Figure~\ref{fig1} summarizes our main result and its relation to previous works. Furthermore, we apply generalized hydrodynamics (GHD)~\cite{Olalla2016,Bertini2016,Doyon2020_Rev,Alba2021,Bouchoule2022,Essler2023} and ballistic macroscopic fluctuation theory (BMFT)~\cite{BMFT1,BMFT2,Friedrich2025,Kethepalli2025,Andrew2026} to the $t\textrm{-}0$ model, deriving the M-Wright function for dynamics from a general grand-canonical initial state. Finally, we perform a numerical study of a finite system similar to the cold-atom experimental setup of Ref.~\cite{Scherg2021}, discussing the possibility of experimentally observing our theoretical prediction.

{\it Setup.--}
Suppose that we have a one-dimensional lattice $\Lambda \coloneqq \{-N+1,-N+2, ..., N\} $ with a positive integer $N$ and denote an annihilation operator for a fermion with spin $\sigma \in S \coloneqq \{  \uparrow, \downarrow\}$ at site $j \in \Lambda$ by $\hat{f}_{\sigma,j}$. Then, the Hamiltonian of the $t\textrm{-}0$ model~\cite{Essler2005} is given by
\begin{align} 
\hat{H} \coloneqq - \hat{P}\sum_{\sigma \in S} \sum_{j=-N+1}^{N-1} \left( \hat{f}_{\sigma,j+1}^{\dagger} \hat{f}_{\sigma,j} + {\rm h.c.}  \right) \hat{P}, 
\label{t0}
\end{align}
where we introduce $\hat{P} \coloneqq \prod_{j \in \Lambda} \left( 1 - \hat{n}_{\uparrow,j} \hat{n}_{\downarrow,j}\right)$ with $\hat{n}_{\sigma,j} \coloneqq \hat{f}_{\sigma,j}^{\dagger} \hat{f}_{\sigma,j}$, which is a projector to Hilbert space without any double occupancy. This model is the effective description of the Fermi-Hubbard model with large repulsive interactions~\cite{Essler2005}, and its name comes from the fact that Eq.~\eqref{t0} is the spin interaction strength $J=0$ case of the famous $t\textrm{-}J$ model. Let us denote a density matrix at time $t \in \mathbb{R}_{ \geq 0}$ by $\hat{\rho}(t)$, which obeys the conventional von-Neumann equation. Then, the time-evolved density matrix is given by $\hat{\rho}(t) = \hat{U}(t) \hat{\rho}(0) \hat{U}(t)^{\dagger}$ with $\hat{U}(t) \coloneqq e^{- {\rm i} \hat{H}t}$, where the Dirac's constant is set to unity. The initial density matrix $\hat{\rho}(0)$ for the exact calculation is assumed to be 
\begin{align} 
\hat{\rho}(0) = \frac{1}{4^N} \prod_{j \in \Lambda} \left( 1 - \dfrac{1}{2} \hat{n}_j \right) \label{eq:initial}
\end{align}
with $\hat{n}_{j} \coloneqq \hat{n}_{\uparrow, j} + \hat{n}_{\downarrow, j}$. The term on a site $j$ for $\hat{\rho}(0)$ has the simple interpretation that a fermion can occupy the site $j$ with a probability of $1/2$ and its spin is up (down) with a probability of $1/2$.

The physical quantity of interest is the probability $\mathbb{P}_{S}[\Delta S_{R}^z ,t]$ of an integrated spin current $\Delta S_{R}^z$ at the origin and time $t$. 
To define this probability, we introduce an operator $\hat{S}^z_R \coloneqq \sum_{j=1}^N \hat{s}^z_j$ with $\hat{s}^z_j \coloneq  \hat{n}_{\uparrow,j} -  \hat{n}_{\downarrow,j}$, where we omit a factor $1/2$ for the sake of simplicity.
Then, $\mathbb{P}_{S}[\Delta S_{R}^z ,t]$ is defined by two-time measurements~\cite{Moriya2019,McCulloch2023}:
\begin{align}
&\mathbb{P}_{S}[\Delta S_{R}^z ,t] \nonumber\\
&\coloneqq \sum_{S^z_R = -N}^{N} {\rm Tr} \left[  \hat{P}_{S}(S_R^z + \Delta S_R^z) \hat{U}(t) \hat{P}_{S}(S_R^z) \hat{\rho}(0) \hat{P}_{S}(S_R^z) \hat{U}(t)^{\dagger} \right], 
\label{eq:def_PS}
\end{align} 
where $\hat{P}_S(S^z_R)$ is a projector for eigenspace with an eigenvalue $S^z_R$ of $\hat{S}^z_R$. 
We denote the corresponding generating function by $G_{S}(\lambda,t) \coloneqq \sum_{\Delta S_{R}^z=-N}^{N} \mathbb{P}_{S}[\Delta S_{ R}^z ,t] e^{\lambda \Delta S_{R}^z}$.
By definition, we can express the probability via the generating function as $\mathbb{P}_{S}[\Delta S_{R}^z ,t] = \int_{C_r} dz~z^{-1-\Delta S_{R}^z} G_{S}(\lambda,t)/(2\pi {\rm i})$,  where we use a complex variable $z \coloneqq e^{\lambda}$ and $C_{r}$ is a contour with a radius $r$ enclosing the origin in the complex plane. In this work, we first calculate $G_{S}(\lambda,t)$ and subsequently derive $\mathbb{P}_{S}[\Delta S_{R}^z ,t]$ by performing the contour integral.

{\it Exact expression of $\mathbb{P}_{S}[\Delta S_{R}^z ,t]$.--}
We derive the exact expression of $\mathbb{P}_{S}[\Delta S_{R}^z ,t]$ that is convenient for implementing its asymptotic analysis. The fundamental point for our exact analysis is the spin-charge separation of the $t\textrm{-}0$ model~\cite{Gamayun2023}, for which the initial spin configuration of the fermions is independent of time and the spins can be transported only through the charge transport (see Sec.~\ref{sec:sup_Gs} of Supplemental Material (SM)~\cite{SM}). Thanks to this property, we exactly derive
\begin{align} 
G_{S}(\lambda,t) = \mathbb{P}_{C}[ 0, t] + 2 \sum^{N}_{\Delta N_{R} = 1} \left( \cosh(\lambda) \right)^{\Delta N_{R}}  \mathbb{P}_{C}[ \Delta N_{R},t]. \label{eq:Gs}
\end{align}
The derivation is given in Sec.~\ref{sec:sup_Gs} of SM~\cite{SM}. 
Here, we define $\mathbb{P}_{C}[ \Delta N_{R},t] \coloneqq  \int_{C_r} dz~ {\rm det}[\delta_{m,n} + \sinh(\lambda/2)^2 D_{m,n}(t)]_{m,n=1}^N  / (2\pi {\rm i} z^{\Delta N_R + 1} )$, where $D_{m,n}(t)~(m,n \in \Lambda)$ is the solution of ${\rm i} d D_{m,n}(t)/dt = D_{m+1,n}(t) + D_{m-1,n}(t) - D_{m,n+1}(t) - D_{m,n-1}(t)$ with the initial condition $D_{m,n}(0) = \delta_{m,n} \chi[ m \in \{-N+1,..., 0 \}]$. Here, $\chi[\bullet]$ is the indicator function and the boundary condition is open. The physical meaning of $\mathbb{P}_{C}[ \Delta N_{R},t] $ is the probability that the transferred particle number from the left region ($-N+1,-N+2,..., 0$) to the right region ($1,2,...,N$) at time $t$ is $\Delta N_{R}$ under the unitary time evolution for spinless free fermions. This interpretation is explained in Sec.~\ref{sec:sup_Pc} of SM~\cite{SM}.

\begin{figure}[t]
\begin{center}
\includegraphics[keepaspectratio, width=\linewidth]{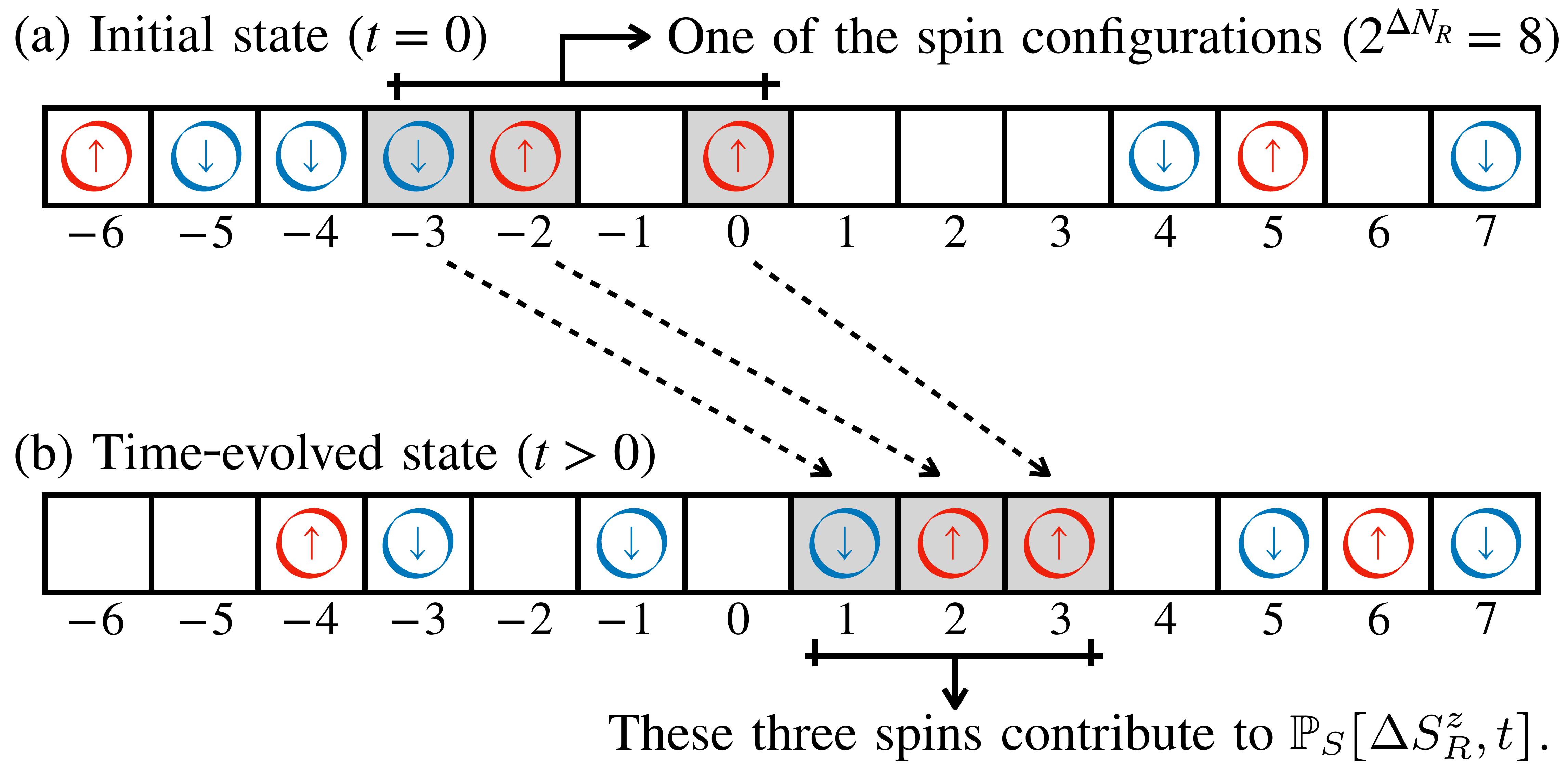}
\caption{
Schematic illustration for Eq.~\eqref{eq:exactPS} with $\Delta N_R = 3$ and $n=1$.
The three fermions occupying the shadowed cells only contribute to the integrated spin current at time $t$ due to $\mathbb{P}_{C}[ \Delta N_{R},t]$ in Eq.~\eqref{eq:exactPS}. The figure shows the specific spin configuration ($\downarrow, \uparrow, \uparrow$) for the three fermions, and our initial state of Eq.~\eqref{eq:initial} realizes $ _{\Delta N_R} C_{n}  = 3$ configurations having the same integrated spin current $\Delta S^z_R=1$ with equal probability [other spin configurations are ($ \uparrow, \downarrow,\uparrow$) and ($\uparrow, \uparrow,\downarrow$)]. The quantity $ _{\Delta N_R} C_{n} / 2^{\Delta N_R} = 3/8$ of Eq.~\eqref{eq:exactPS} reflects this probabilistic fact. 
} 
\label{fig2} 
\end{center}
\end{figure}

\begin{figure}[t]
\begin{center}
\includegraphics[keepaspectratio, width=\linewidth]{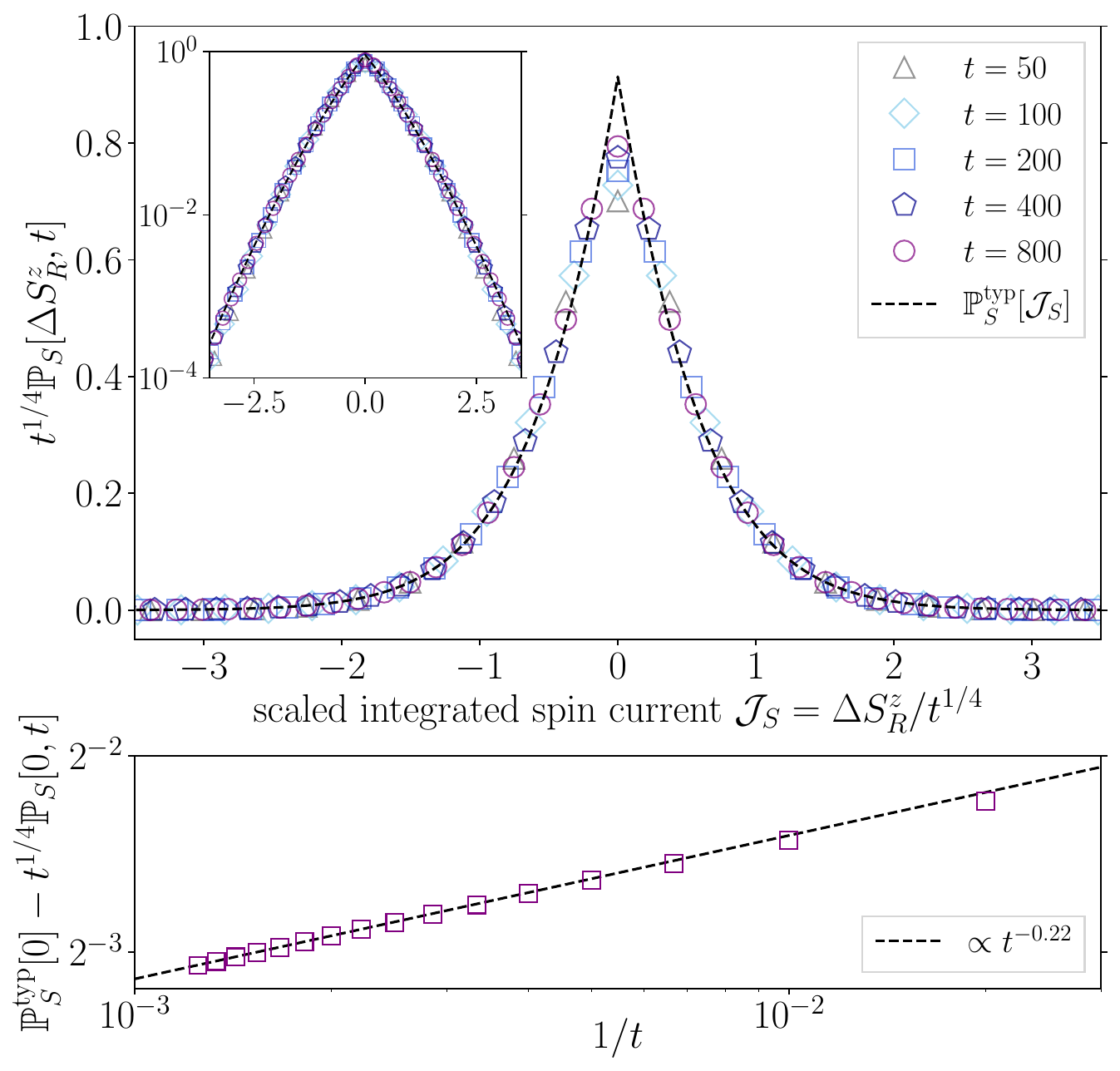}
\caption{
Numerical verification for the convergence of the scaled probability $t^{1/4}  \mathbb{P}_{S}[ t^{1/4}\mathcal{J}_S,t]$ to Eq.~\eqref{eq:Ps_limit}. 
The numerical method is explained in Sec.~\ref{sec:numerics} of SM~\cite{SM} and the system size is $2N = 2000$. 
(Upper panel) Scaled probability at $t=50, 100, 200, 400,$ and $800$. The ordinate and abscissa are the scaled probability $t^{1/4} \mathbb{P}_{S} [ \Delta S_R^z, t]$ and the scaled integrated spin current $\mathcal{J}_S = \Delta S_R^z/t^{1/4}$, respectively. The dashed line represents Eq.~\eqref{eq:Ps_limit}. The inset shows the probability of the main panel with the logarithmic ordinate. (Lower panel) Time evolution of the difference $ \mathbb{P}_{S}^{\rm typ} [0] - t^{1/4} \mathbb{P}_{S} [0, t] $. The dashed line represents the line proportional to $t^{-0.22}$.
} 
\label{fig3} 
\end{center}
\end{figure}

We substitute Eq.~\eqref{eq:Gs} into the contour integral for $\mathbb{P}_{S}[ \Delta S^z_{R},t]$, exactly deriving 
\begin{align} 
&\mathbb{P}_{S}[ \Delta S^z_{R},t] = \delta_{\Delta S^z_{R}, 0} \mathbb{P}_{C}[ 0, t] \nonumber\\
&+ 2 \sum^{N}_{\Delta N_{R} = 1}  \mathbb{P}_{C}[ \Delta N_{R},t] \sum_{n=0}^{\Delta N_R}   \dfrac{ ~_{\Delta N_R} C_{n}  }{2^{\Delta N_R }} \delta_{\Delta S^z_{R},\Delta N_{R}-2n}. \label{eq:exactPS}
\end{align}
The derivation is given in Sec.~\ref{sec:sup_Ps1} of SM~\cite{SM}, and this is the expression suitable for the asymptotic analysis. 

Before going to the asymptotic analysis, we note that the right hand side of Eq.~\eqref{eq:exactPS} has a simple probabilistic interpretation. From the probabilistic interpretation of $\mathbb{P}_{C}[ \Delta N_{R},t] $, the contribution to $\mathbb{P}_{S}[ \Delta S^z_{R},t]$ comes only from the $\Delta N_{R}$ fermions occupying the left region from the origin in the initial state. This fact is schematically displayed in Fig.~\ref{fig2}, where we consider the case with $\Delta N_{R}=3$. Considering this fact, we suppose that we have $\Delta N_{R} -n$ up-spin fermions and $n$ down-spin fermions in the left region of the initial state. At time $t$, all these fermions pass the origin because of $\mathbb{P}_{C}[ \Delta N_{R},t]$. Hence, the integrated spin current at time $t$ becomes $\Delta N_{R} - 2n$, which corresponds to the Kronecker delta of Eq.~\eqref{eq:exactPS}. The case with $\Delta N_{R}=3$ and $n=1$ is shown in Fig.~\ref{fig2}. Next, we consider possible initial spin configurations in the left region. In our initial density matrix, the positions of the particles and their spins are random via Eq.~\eqref{eq:initial}. Hence, the number of all possible configurations of up-spin and down-spin fermions in the left region is $2^{\Delta N_R}$ and that for $\Delta N_{R} -n$ up-spin fermions and $n$ down-spin fermions is $ ~_{\Delta N_R} C_n $. Hence, $_{\Delta N_R} C_n  / (2^{\Delta N_R )}$ is the probability of having $\Delta N_{R} - n$ up-spin fermions and $n$ down-spin fermions in the initial left region. Therefore, the right hand side of Eq.~\eqref{eq:exactPS} is the summation of the probability with the integrated spin current $\Delta N_{R} - 2n$ under the equal probability for such fermion configurations. 

{\it Asymptotic analysis of $\mathbb{P}_{S}[\Delta S_{R}^z ,t]$.--}
We shall study the long-time behavior of $\mathbb{P}_{S}[\Delta S_{R}^z ,t]$ after taking the thermodynamic limit ($N \rightarrow \infty$) of Eq.~\eqref{eq:exactPS}. 
The fundamental ingredient for this analysis is the probability $\mathbb{P}_{C}[ \Delta N_R,t]$ for the typical fluctuation of $\Delta N_R$, which is given by, for $t \gg 1$,   
\begin{align} 
\sqrt{t} \mathbb{P}_{C}[ \sqrt{t} \mathcal{J}_{C} ,t] \sim \dfrac{1}{\sqrt{2}} \exp\left(  -\dfrac{\pi \mathcal{J}_{C}^2}{2} \right). \label{eq:PN_asym}
\end{align}
This asymptotic form is derived in Sec.~\ref{sec:sup_Pc} of SM~\cite{SM}. Employing Eq.~\eqref{eq:PN_asym} and Stirling's formula, we obtain 
the limiting probability $\mathbb{P}_{S}^{\rm typ} [\mathcal{J}_{S}] \coloneqq \lim_{t \rightarrow \infty} t^{1/4}  \mathbb{P}_{S}[ t^{1/4}\mathcal{J}_S,t]$ for the typical fluctuation of $\Delta S^z_R$: 
\begin{align} 
\mathbb{P}_{S}^{\rm typ} [\mathcal{J}_{S}]  = \dfrac{1}{\sqrt{\pi}} \int^{\infty}_{0}  
\dfrac{ d\mathcal{J}_C }{  { \sqrt{\mathcal{J}_C }  } } 
\exp\left[  -\dfrac{\pi \mathcal{J}_{C}^2}{2} - \dfrac{\mathcal{J}_S^2}{2\mathcal{J}_C}\right],
\label{eq:Ps_limit}
\end{align}
which is derived in Sec.~\ref{sec:sup_Ps2} of SM~\cite{SM}. This expression is identical to the M-Wright function $\mathbb{P}_{\rm MW}[\mathcal{J}_{S},1/\sqrt{\pi}]$. Therefore, we derive the M-Wright function exactly in the quantum many-body dynamics.

We numerically check this asymptotic behavior. As detailed in Sec.~\ref{sec:numerics} of SM~\cite{SM}, in our numerical calculation, we solve the equation of motion for $D_{m,n}(t)$, then calculate the probability $\mathbb{P}_{C}[ \Delta N_{R},t]$ via the contour integral given just below Eq.~\eqref{eq:Gs}, and finally compute $\mathbb{P}_{S}[ \Delta S^z_{R},t]$ with Eq.~\eqref{eq:exactPS}. The upper panel and its inset of Fig.~\ref{fig3} display the numerical result for the time evolution of $\mathbb{P}_{S} [ \Delta S_R^z, t]$. We find that the edge of the probability well converges to the M-Wright function $\mathbb{P}_{\rm MW}[\mathcal{J}_S,1/\sqrt{\pi}]$, while the convergence around the origin is relatively slow~\cite{revise1}. In the lower panel of Fig.~\ref{fig3}, we plot the difference between $t^{1/4}\mathbb{P}_{S} [0, t]$ and $\mathbb{P}_{\rm MW}[0,1/\sqrt{\pi}]$, finding the tendency that it becomes small over time. From this numerical finding, we expect that the scaled probability $t^{1/4}  \mathbb{P}_{S}[ t^{1/4}\mathcal{J}_S,t]$ converges to $\mathbb{P}_{\rm MW}[\mathcal{J}_S,1/\sqrt{\pi}]$ in the long-time limit.

{\it Hydrodynamic derivation of the M-Wright function.--}
Just above, we derive the M-Wright function exactly in the dynamics with the specific initial state of Eq.~\eqref{eq:initial}, but its extension to general initial states remains challenging. Here, we explore another approach based on hydrodynamics to tackle this task and consider a grand-canonical density matrix as the initial state, which is given by 
\begin{align} 
\hat{\rho}(0) = \frac{1}{Z} e^{-\beta (\hat{H} - \mu \hat{N}  ) },  \label{eq:initial2}
\end{align}
where $\beta$, $\mu$, and $Z$ are the inverse nonzero temperature, the chemical potential, and the normalization constant, respectively, and we define $\hat{N}\coloneqq \hat{P}  \sum_{j \in \Lambda} \left( \hat{n}_{\uparrow,j} + \hat{n}_{\downarrow,j} \right) \hat{P}  $. As derived in End Matter, we apply GHD~\cite{Olalla2016,Bertini2016,Doyon2020_Rev,Alba2021,Bouchoule2022,Essler2023} and BMFT~\cite{BMFT1,BMFT2,Friedrich2025,Kethepalli2025,Andrew2026} to the $t\textrm{-}0$ model with Eq.~\eqref{eq:initial2}, deriving the M-Wright function:
\begin{align} 
\mathbb{P}_{S}^{\rm typ} [\mathcal{J}_{S}]  = \dfrac{1}{\pi d(\beta,\mu)} \int^{\infty}_{0}  
\dfrac{ d\mathcal{J}_C }{  { \sqrt{\mathcal{J}_C }  } } 
\exp\left[  -\dfrac{\mathcal{J}_{C}^2}{2 d(\beta,\mu)^2}  - \dfrac{\mathcal{J}_S^2}{2\mathcal{J}_C}\right]
\label{eq:Ps_limit_hydro}
\end{align}
with $d(\beta,\mu) \coloneqq \sqrt{ 2 [ 1/(2+e^{-\beta(2+\mu)}) - 1/(2+e^{\beta(2-\mu)}) ]/(\pi \beta) }$, which is related to the diffusion constant of the $t\textrm{-}0$ model~\cite{Gamayun2023}. This hydrodynamic result can reproduce the exact result of Eq.~\eqref{eq:Ps_limit} by taking the limit $\beta \rightarrow 0$ while fixing $\beta \mu = - \log 2$, for which Eq.~\eqref{eq:initial2} converges to the initial state of Eq.~\eqref{eq:initial} for the exact analysis. This result implies the robustness of the M-Wright function in the $t\textrm{-}0$ model against the choice of the initial states. Note that one can study the probability for a general filling factor $0 < \nu \leq 1$ using Eq.~\eqref{eq:Ps_limit_hydro}. For $0 < \nu < 1$, the probability is the M-Wright function, while it becomes the $\delta$ function at $\nu = 1$.

{\it Discussion.--}
We shall discuss three topics: (i) the relation between the classical automaton and the $t\textrm{-}0$ model, (ii) the possibility of experimental observation for our theoretical prediction, and (iii) the finite-interaction effect of the Fermi-Hubbard model.

\begin{figure}[t]
\begin{center}
\includegraphics[keepaspectratio, width=\linewidth]{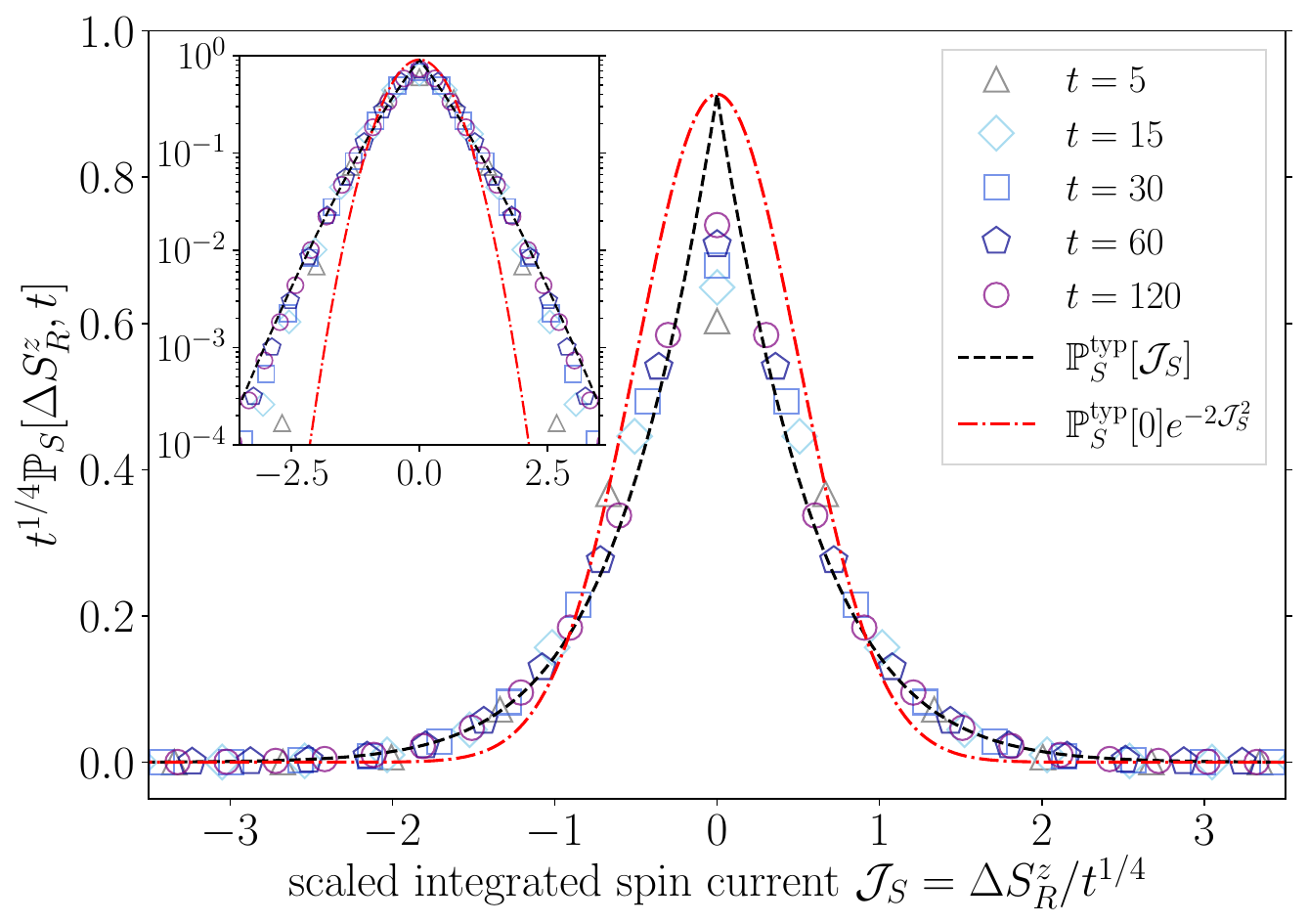}
\caption{
Numerical study for the integrated spin current of the $t\textrm{-}0$ model in a realistic experimental setup. 
The numerical method is the same as that of Fig.~\ref{fig3} and the system size to be $2N = 280$. The ordinate and abscissa are the scaled probability $t^{1/4} \mathbb{P}_{S} [ \Delta S_R^z, t]$ and the scaled integrated spin current $\mathcal{J}_S = \Delta S_R^z/t^{1/4}$, respectively. The inset shows the main panel with the logarithmic ordinate. The dashed and dash-dotted lines represent Eq.~\eqref{eq:Ps_limit} and the Gaussian function $\mathbb{P}_{S}^{\rm typ} [0] e^{-2\mathcal{J}_S^2}$, respectively.
} 
\label{fig4} 
\end{center}
\end{figure}

Regarding (i), we compare the two exact results for the $t\textrm{-}0$ model and the classical automaton~\cite{Krajnik2022}, where the anomalous integrated current characterized by the M-Wright function was studied exactly. The key point of this comparison is that the exact calculation in the $t\textrm{-}0$ model is based on Eq.~\eqref{eq:exactPS}. 
Although we derived Eq.~\eqref{eq:exactPS} for the $t\textrm{-}0$ model in this work, a similar expression also holds for the classical automaton of Ref.~\cite{Krajnik2022}, as shown in Sec.~\ref{sec:sup_CA} of SM~\cite{SM}:
\begin{align} 
&\mathbb{P}_{S}^{\rm CA}[\Delta S_R^z, t] = \delta_{\Delta S_R^z,0} \mathbb{P}_{C}^{\rm CA}[0, t] \nonumber \\
+&2 \sum_{\Delta N_R=1}^t \mathbb{P}_{C}^{\rm CA}[\Delta N_R, t] \sum_{n=0}^{\Delta N_R} \dfrac{ ~_{\Delta N_R}C_{n} }{2^{\Delta N_R}} \delta_{\Delta S_R^z, \Delta N_R - 2n}, 
\label{eq:exactPS_CA}
\end{align}
where $\mathbb{P}_{S}^{\rm CA}[\Delta S_R^z, t]$ and $\mathbb{P}_{C}^{\rm CA}[\Delta N_R, t]$ denote the probabilities of the integrated spin and charge currents in the classical automaton, respectively.
As shown in Sec.~\ref{sec:sup_CA} of SM~\cite{SM}, $\mathbb{P}_{C}^{\rm CA}[ \Delta N_{R},t]$ becomes Gaussian in the long-time dynamics, thereby leading to $\mathbb{P}_{\rm MW}[\mathcal{J}_S,\sigma]$ in the classical automaton, in the same manner as in the $t\textrm{-}0$ model. This derivation differs from the original derivation of Ref.~\cite{Krajnik2022}, but clarifies the distinction between the classical automaton and the $t\textrm{-}0$ model. Specifically, in the classical automaton, the Gaussian nature of $\mathbb{P}_{C}^{\rm CA}[ \Delta N_{R},t]$ arises from $two~propagating~modes$ (see Sec.~\ref{sec:sup_CA} of SM~\cite{SM}), whereas in the $t\textrm{-}0$ model that of $\mathbb{P}_{C}[ \Delta N_{R},t]$ arises from the $infinitely~many~propagating~modes$. The latter fact can be understood well by applying BMFT~\cite{BMFT1,BMFT2,Friedrich2025,Kethepalli2025,Andrew2026} to the $t\textrm{-}0$ model, which is detailed in Sec.~\ref{sec:sup_BMFT} of SM \cite{SM}. From this comparison, we find that the $t\textrm{-}0$ model shares a similar property with the classical automaton (see Eqs.~\eqref{eq:exactPS} and \eqref{eq:exactPS_CA}) but their microscopic propagating modes differ. Despite these microscopic differences, the integrated spin currents of both systems obey the same M-Wright function.

Regarding (ii), it is important to investigate experimental finite-size and finite-time effects on the emergence of the M-Wright function. We numerically study the effects on $\mathbb{P}_{S}[\Delta S_R^z, t] $ with a setup that is realizable in a cold atom experiment. The system size used here is $2N=280$ and we compute the probability $\mathbb{P}_{S}[\Delta S_R^z, t] $ until $t=120$. These system size and timescale are accessible in the recent cold atom experiment of Ref.~\cite{Scherg2021}. The numerical method is the same as that used in Fig.~\ref{fig3}. Figure~\ref{fig4} showcases the time evolution of $\mathbb{P}_{S}[\Delta S_R^z, t] $ for the $t\textrm{-}0$ model. We find that the convergence to $\mathbb{P}_{S}^{\rm typ} [\mathcal{J}_{S}] = \mathbb{P}_{\rm MW}[\mathcal{J}_{S},1/\sqrt{\pi}]$ is fast around the edge, expecting that the non-Gaussian nature of the M-Wright function may be accessible experimentally because the interaction strength can be tuned via Feshbach resonance~\cite{Cheng2010} and integrated currents have already been observed in the cold atom experiment of a spin chain by using quantum gas microscopes~\cite{Wei2022}. 

Regarding (iii), we consider how the emergence of the M-Wright function is affected when the interaction of the Fermi-Hubbard model is finite. To discuss the finite-interaction effect, it is helpful to consider the $t\textrm{-}J$ Hamiltonian, which is given by $\hat{H}_{t\textrm{-}J} \coloneqq \hat{H} + J \hat{P} \hat{H}' \hat{P}$ with the perturbed Hamiltonian $\hat{H}'$ and $J=4/U$. Here, $U$ is the interaction strength of the Fermi-Hubbard model.  Then, the description of the $t\textrm{-}0$ model works well for the time regime $ t \ll U$, and thus we expect that the M-Wright function transiently appears in $ 1 \ll t \ll U$ when $U$ is sufficiently large. 
The detailed discussion is given in Sec.~\ref{sec:discussion} of Ref.~\cite{SM}.

{\it Conclusion and future prospects.--}
We have theoretically studied the probability $\mathbb{P}_{S} [ \Delta S_R^z, t]$ for the integrated spin current in the one-dimensional $t\textrm{-}0$ model. We have carried out the exact microscopic calculation, demonstrating that the scaled probability $t^{1/4}\mathbb{P}_{S} [ t^{1/4} \mathcal{J}_S, t]$ for the initial state of Eq.~\eqref{eq:initial} converges to the M-Wright function $\mathbb{P}_{\rm MW}[\mathcal{J}_S,\sigma]$ in the long-time limit. Furthermore, we have derived $\mathbb{P}_{\rm MW}[\mathcal{J}_S,\sigma]$ for the initial grand-canonical state using GHD and BMFT. On the basis of our results, we have discussed the connection to the previous study~\cite{Krajnik2022} on the classical automaton and the experimental feasibility of observing our theoretical prediction.   

An important direction for future research is to clarify the extent to which anomalous current fluctuations characterized by the M-Wright function are universal in quantum many-body systems. Our exact result suggests that the spin-charge separation, which renders the spin configuration static in time, is essential for the emergence of the M-Wright function. Motivated by this fact, it is highly intriguing to test whether the anomalous current fluctuation persists even in nonintegrable settings by introducing noisy spin-independent potentials into the $t\textrm{-}0$ model. Another natural extension is to consider the SU($N$) $t\textrm{-}0$ model and investigate the robustness of the M-Wright function by systematically varying the internal spin degrees of freedom. These theoretical studies are promising, as they may admit exact solutions by extending our current analytic approach.

Beyond the $t\textrm{-}0$ model, an exact treatment of the XXZ spin chain is of particular interest. While Ref.~\cite{Yoshimura2026} derived the M-Wright function from a hydrodynamic perspective, an exact microscopic analysis based on the Bethe ansatz remains technically challenging. Such exact results would provide invaluable insights into universal aspects of quantum many-body dynamics, including the intriguing KPZ physics of spin chains~\cite{Ljubotina2019,Dipankar2019,Krajnik2020_CS1,Michele2020,Bingtian2022,Dipankar2023,Moca2023,Moca2025,Takeuchi2025}.

\begin{acknowledgments}
The authors are grateful to Ippei Danshita, Masaya Kunimi, and Rina Tazai for helpful discussions.
The work of KF has been supported by JSPS KAKENHI Grant No. JP23K13029. 
The work of TI has been supported by JST SPRING, Japan Grant Number JPMJSP2180.
The work of TS has been supported by JSPS KAKENHI Grants No. JP21H04432, No. JP22H01143, and No. JP23K22414.
TY acknowledges financial support from the Royal Society through the University Research Fellowship (\text{URF\textbackslash R1\textbackslash 251593})
\\

{\it Note added.--}
Anomalous spin current fluctuations in the easy-axis XXZ chain were also explored recently in Ref.~\cite{Yoshimura2026} (appearing in the same arXiv posting as this Letter) where the authors obtained the typical spin current distribution using hydrodynamics. 
After Ref.~\cite{Yoshimura2026} and the present work appeared on arXiv, A. Urilyon et al. reported a numerical study of the XXZ model using a phenomenological hard-rod model, in which the anomalous fluctuations were investigated~\cite{Urilyon2026}.
\end{acknowledgments}

\bibliography{reference}

\begin{thebibliography}{79}
\bibitem [{rev()}]{revise1}%
  \BibitemOpen
\bibfield  {journal} {  }\href@noop {} {\bibinfo  {journal} {We comment on the
  slow convergence around the origin for $\mathbb{P}_{S}[ \Delta S^z_{R},t]$.
  The point of deriving the limiting probability of Eq.~\eqref{eq:Ps_limit} is
  to apply the Stirling's formula to the combination $_{\Delta N_R}C_{n}$ in
  Eq.~\eqref{eq:exactPS}. For $\Delta S_R^z > 0$, we can safely use this
  formula. However, for $\Delta S_R^z=0$, we need careful calculations and then
  have additional small terms that do not appear in the case with $\Delta S_R^z
  > 0$ for the asymptotic analysis. This small terms lead to the slow
  convergence. The detailed mathematical explanation is given in Sec. VI of
  SM~\cite{SM}}\ }\BibitemShut {NoStop}%
\end{thebibliography}
\clearpage

\setcounter{equation}{0}
\setcounter{figure}{0}
\setcounter{section}{0}
\setcounter{table}{0}
\renewcommand{\theequation}{E-\arabic{equation}}
\renewcommand{\thefigure}{E-\arabic{figure}}
\renewcommand{\thetable}{E-\arabic{table}}
\section*{End matter}
\vspace{-2mm}

We shall derive the M-Wright function for the integrated spin current of the $t\textrm{-}0$ model with the initial grand-canonical state of Eq.~\eqref{eq:initial2} by using GHD~\cite{Olalla2016,Bertini2016,Doyon2020_Rev,Alba2021,Bouchoule2022,Essler2023} and BMFT~\cite{BMFT1,BMFT2,Friedrich2025,Kethepalli2025,Andrew2026}. Note that O. Gamayun et al. have applied GHD to the $t\textrm{-}0$ model for the calculation of correlation functions~\cite{Gamayun2023}, and thus some results shown here overlap with the previous work.
\vspace{-2mm}

\subsection{GHD for the $t\textrm{-}0$ model}
First, we introduce the Bethe equation for the $t\textrm{-}0$ model.
We assume that the numbers of the lattice points and the fermions are $L$ and $N_p$, respectively, and the boundary condition is periodic.
Under this setup with $M$ down-spin fermions, the Bethe equation of the $t\textrm{-}0$ model~\cite{Izergin1998,Abarenkova2001,Essler2005} is given by 
\begin{align} 
e^{{\rm i}  \lambda^{\rm (c)}_a L} &= e^{ {\rm i} \Theta }~~~~~~~~~~(a \in \{1,2,...,N_p \}), \label{eq:E_b1}\\
e^{{\rm i}  \lambda^{\rm (s)}_b N_p} &= (-1)^{M+1}~~~(b \in \{1,2,...,M \}) \label{eq:E_b2}
\end{align}
with $\Theta \coloneqq \sum_{b=1}^M \lambda^{\rm (s)}_b$. 
Here, $\lambda^{\rm (c)}_a$ and $\lambda^{\rm (s)}_b$ are the wavenumbers for the charge and spin of the nested Bethe ansatz.  

Second, we take the thermodynamic limit for Eqs.~\eqref{eq:E_b1} and \eqref{eq:E_b2}, deriving the thermodynamic Bethe ansatz equation. 
We define the particle densities for the charge and spin as $\rho^{\rm (\alpha)} (\lambda^{\rm (\alpha)}_c ) \coloneqq \lim_{\rm th} L^{-1}( \lambda^{\rm (\alpha)}_{c+1} - \lambda^{\rm (\alpha)}_{c} )^{-1}$ for $\alpha \in \{ {\rm c, s} \}$. We also denote the corresponding total densities as $\rho^{\rm ( \alpha) }_t (\lambda)$. 
With these notations, the resulting thermodynamic Bethe ansatz equation becomes
\begin{align} 
\begin{pmatrix}
\rho^{\rm (c)}_t (\lambda)  \\
\rho^{\rm (s)}_t (\lambda)  \\
\end{pmatrix}
= \dfrac{1}{2 \pi}
\begin{pmatrix}
1  \\
0  \\
\end{pmatrix}
+ \int_{-\pi}^{\pi} \dfrac{d \lambda}{2\pi}
\begin{pmatrix}
0  \\
\rho^{\rm (c)} (\lambda)  \\
\end{pmatrix}.
\label{eq:E_b3}
\end{align}
When we minimize the free energy with Eq.~\eqref{eq:E_b3}, the occupation numbers $n^{\rm (\alpha)} (\lambda) \coloneqq \rho^{\rm (\alpha)} (\lambda)/\rho^{\rm (\alpha)}_t (\lambda) $ for $\alpha \in \{ {\rm c, s} \}$ are given by
\begin{align} 
n^{\rm (c)} (\lambda) = \dfrac{2}{e^{-\beta(2\cos(\lambda)+\mu)}+2}, \quad n^{\rm (s)} (\lambda) = \dfrac{1}{2}.  \label{eq:E_nc}
\end{align}

Third, we derive the effective velocities for the charge and spin.  
The derivatives of the bare energies $e^{\rm (\alpha)}(\lambda)$ and momentum $p^{\rm (\alpha)}(\lambda)$ for $\alpha \in \{ \rm c, s \}$ are $e^{\rm (c)}(\lambda)' = 2 \sin(\lambda)$, $e^{\rm (s)}(\lambda)' = 0$, $p^{\rm (c)}(\lambda)' = 1$, and $p^{\rm (s)}(\lambda)' = 0$.
Then, the corresponding dressed energies and momentum are defined by
\begin{align} 
\left( e^{\rm (c)}(\lambda)' \right)^{\rm dr} &\coloneqq e^{\rm (c)}(\lambda)' ,\quad \left( p^{\rm (c)}(\lambda)' \right)^{\rm dr} \coloneqq p^{\rm (c)}(\lambda)' ,  \\
\left( e^{\rm (s)}(\lambda)' \right)^{\rm dr} &\coloneqq e^{\rm (s)}(\lambda)' + \int_{-\pi}^{\pi} \dfrac{d\lambda}{2 \pi} n^{\rm (c)} (\lambda) \left( e^{\rm (c)}(\lambda)' \right)^{\rm dr}, \\
\left( p^{\rm (s)}(\lambda)' \right)^{\rm dr} &\coloneqq p^{\rm (s)}(\lambda)' + \int_{-\pi}^{\pi} \dfrac{d\lambda}{2 \pi} n^{\rm (c)} (\lambda) \left( p^{\rm (c)}(\lambda)' \right)^{\rm dr}.
\end{align}
We use the above to derive the effective velocities $v_{\rm eff}^{\rm (c)}(\lambda)$ and  $v_{\rm eff}^{\rm (s)}(\lambda)$ for the charge and spin: 
\begin{align} 
v_{\rm eff}^{\rm (c)}(\lambda) 
&\coloneqq \dfrac{\left( e^{\rm (c)}(\lambda)' \right)^{\rm dr}}{\left( p^{\rm (c)}(\lambda)' \right)^{\rm dr}} = 2 \sin (\lambda), \label{eq:EM_veff_c} \\
v_{\rm eff}^{\rm (s)}(\lambda)
&\coloneqq \dfrac{\left( e^{\rm (s)}(\lambda)' \right)^{\rm dr}}{\left( p^{\rm (s)}(\lambda)' \right)^{\rm dr}} = \dfrac{ \int_{-\pi}^{\pi} d \lambda  ~2 \sin(\lambda) \rho^{\rm (c)} (\lambda) }{ \int_{-\pi}^{\pi} d \lambda~  \rho^{\rm (c)} (\lambda)  } \label{eq:EM_veff_s}.
\end{align}
Note that the same expressions for the effective velocities can also be derived using a Gaudin matrix~\cite{Borsi2020,Pozsgay2020,Borsi2021} as described in Sec.~\ref{sec:veff} of SM~\cite{SM}.

Finally, we derive the equations of motion for the densities $\rho^{\rm (c)} (\lambda)$ and $\rho^{\rm (s)} (\lambda)$.
For this purpose, we allow these densities to depend on space $x$ and time $t$, denoting them as $\rho^{\rm (c)} (\lambda,x,t)$ and $\rho^{\rm (s)} (\lambda,x,t)$.  
Following the framework of GHD~\cite{Olalla2016,Bertini2016,Doyon2020_Rev,Alba2021,Bouchoule2022,Essler2023}, we can derive the equations of motion:
\begin{align} 
&\dfrac{\partial}{\partial t} \rho^{\rm (c)} (\lambda,x,t) + 2 \sin \left(\lambda \right) \dfrac{\partial}{\partial x} \left( \rho^{\rm (c)} (\lambda,x,t) \right) = 0, \label{eq:EM_eom1} \\
&\dfrac{ \partial }{ \partial t } \rho^{\rm (s)} (\lambda,x,t) + \dfrac{\partial}{\partial x} \left( v(x,t) \rho^{\rm (s)} (\lambda,x,t) \right) =0 \label{eq:EM_eom2}
\end{align}
with $v(x,t) \coloneqq  \int d \lambda 2 \sin(\lambda) \rho^{\rm (c)} (\lambda,x,t) /  \int d \lambda \rho^{\rm (c)} (\lambda,x,t) $.

\subsection{Hydrodynamic derivation of the M-Wright function}
We here derive the M-Wright function for the integrated spin current via Eqs.~\eqref{eq:EM_eom1} and \eqref{eq:EM_eom2}.   
Then, we need the equation of motion for the spin in the $z$-direction, $S^z(x,t) \coloneqq \int d\lambda \rho^{\rm (c)} (\lambda,x,t) - 2 \int d\lambda \rho^{\rm (s)} (\lambda,x,t)$. 
Using Eqs.~\eqref{eq:EM_eom1} and \eqref{eq:EM_eom2}, we can readily derive
\begin{align} 
\dfrac{ \partial }{ \partial t } S^z(x,t) + \dfrac{\partial}{\partial x} \left( v(x,t) S^z(x,t) \right) =0 \label{eq:EM_eom3}.
\end{align}
Hence, we can investigate the time evolution of $S^z(x,t)$ by solving Eqs.~\eqref{eq:EM_eom1} and \eqref{eq:EM_eom3} simultaneously.
In the following, we calculate the probability for the typical fluctuations of the integrated spin current by using the hydrodynamic method developed in Refs.~\cite{Yoshimura2025-1,Yoshimura2025-2}.  

First, we introduce the scaled charge density and spin defined by
\begin{align} 
\tilde{\rho}^{\rm (c)}(\lambda,x,t) \coloneqq \rho^{\rm (c)}(\lambda, \sqrt{\tau} x, \tau t), \quad \tilde{S}^z(x,t) \coloneqq S^z(\sqrt{\tau} x, \tau t) \label{eq:EM_scaled_density}.
\end{align}
As to the initial state, we assume
\begin{align} 
\tilde{\rho}^{\rm (c)}(\lambda,x,0) &= {\rho}^{\rm (c)}(\lambda) + \tau^{-1/4} \delta \tilde{\rho}^{\rm (c)}(\lambda,x,0), \\
\tilde{S}^z(x,0) &= \tau^{-1/4} \delta \tilde{S}^z(x,0), 
\end{align}
where $\delta \tilde{S}^z(x,t)$ and $\delta \tilde{\rho}^{\rm (c)}(\lambda,x,t)$ represent the fluctuations from the initial grand-canonical state of Eq.~\eqref{eq:initial2}, which is characterized by the equilibrium density $\rho^{\rm (c)}(\lambda) = n^{\rm (c)}(\lambda)/(2\pi)$ and the zero spin. Furthermore, we assume that the initial fluctuations of $\delta \tilde{\rho}^{\rm (c)}(\lambda,x,0)$ and $\delta \tilde{S}^z(x,0)$ obey Gaussian functions with the following correlations:
\begin{align} 
\langle \delta \tilde{S}^z(x,0) \delta \tilde{S}^z(y,0)  \rangle &= C_{1} \delta(x-y), \label{eq:EM_Crr_SS} \\
\langle \delta \tilde{\rho}^{\rm (c)}(\lambda,x,0) \delta \tilde{\rho}^{\rm (c)}(\nu,y,0)  \rangle &= C_2(\lambda) \delta(x-y) \delta(\lambda-\nu), \label{eq:EM_Crr_CC}\\
\langle \delta \tilde{\rho}^{\rm (c)}(\lambda,x,0) \delta \tilde{S}^z(y,0)   \rangle &= 0 \label{eq:EM_Crr_CS}.
\end{align}
Here, the bracket $\langle \bullet\rangle$ means an average over the Gaussian fluctuations and the functions $C_{1}$ and $C_2(\lambda)$ are given by
\begin{align} 
C_1 &\coloneqq \int_{-\pi}^{\pi} d \lambda \rho^{\rm (c)} (\lambda) ,\label{eq:EM_C1} \\
C_2(\lambda) &\coloneqq \dfrac{1}{2\pi} n^{\rm (c)}(\lambda) (1-n^{\rm (c)}(\lambda) ) \label{eq:EM_C2}, 
\end{align}
which are derived in Sec.~\ref{sec:EM_initial} of SM~\cite{SM}.

Second, we derive the equations of motion for the fluctuations and then solve them via the characteristics. 
From Eqs.~\eqref{eq:EM_eom1} and \eqref{eq:EM_eom3}, one can derive 
\begin{align} 
&\dfrac{\partial}{\partial t} \delta \tilde{\rho}^{\rm (c)}(\lambda,x,t) + 2 \sqrt{\tau} \sin (\lambda)~\dfrac{\partial}{\partial x} \delta \tilde{\rho}^{\rm (c)}(\lambda,x,t) = 0, \label{eq:EM_eom4}\\
&\dfrac{\partial}{\partial t} \delta \tilde{S}^{z}(x,t) + \sqrt{\tau} \dfrac{\partial}{\partial x} \left( \tilde{v}(x,t) \delta \tilde{S}^{z}(x,t) \right) = 0 \label{eq:EM_eom5}
\end{align}
with $\tilde{v}(x,t) \coloneqq \int d \lambda 2 \sin(\lambda) \tilde{\rho}^{\rm (c)} (\lambda,x,t) /  \int d \lambda \tilde{\rho}^{\rm (c)} (\lambda,x,t) $.
Then, the solution of Eq.~\eqref{eq:EM_eom4} is given by
\begin{align} 
\delta \tilde{\rho}^{\rm (c)}(\lambda, x,t) = \delta \tilde{\rho}^{\rm (c)}(\lambda, x - 2 \sqrt{\tau} t \sin(\lambda),0).
\end{align}
To solve Eq.~\eqref{eq:EM_eom5}, we introduce a new variable $ b(x,t) \coloneqq \delta \tilde{S}^{z}(x,t) / \int_{-\pi}^{\pi} d \lambda \tilde{\rho}^{\rm (c)}(\lambda,x,t)$.
As a result, Eq.~\eqref{eq:EM_eom5} becomes
\begin{align} 
\dfrac{\partial}{\partial t} b(x,t) + \sqrt{\tau} \tilde{v}(x,t) \dfrac{\partial}{\partial x} b(x,t) = 0. \label{eq:EM_eom6}
\end{align}
Since the above is the simple convective equation, we can obtain the solution as $b(X_{x,t},t) = b(x,0)$, where $X_{x,t}$ is the solution of the following differential equation:
\begin{align} 
\dfrac{d}{dt} X_{x,t} = \sqrt{\tau} \tilde{v}(X_{x,t},t). 
\end{align}
Solving this equation up to the order of $\tau^{1/4}$, we obtain
\begin{align} 
X_{x,t} &\sim x + X_{t}
\end{align}
with 
\begin{align} 
X_{t} \coloneqq \tau^{1/4}  \dfrac{\int_{0}^{t} dt'\int_{-\pi}^{\pi} d\lambda  ~ 2\sin(\lambda) \delta \tilde{\rho}^{\rm (c)}(\lambda,-2 \sqrt{\tau} t' \sin(\lambda),0 )} { \int_{-\pi}^{\pi} d\lambda \rho^{\rm (c)}(\lambda)}.
\label{eq:EM_Xt}
\end{align} 
We put all the above together, and then the solution of Eq.~\eqref{eq:EM_eom5} becomes
\begin{align} 
\delta \tilde{S}^z(x,t) \sim \delta \tilde{S}^z(x - X_t,0),
\label{eq:EM_sol1}
\end{align}
where $\int_{-\pi}^{\pi} d \lambda \tilde{\rho}^{\rm (c)}(\lambda,x,t)$ is approximated to be the constant value at the equilibrium state because we consider the terms up to the order of $\tau^{1/4}$. 

Finally, we calculate the generating function for $J(\tau)$. 
The integrated spin current $J(\tau)$ can be expressed as
\begin{align} 
J(\tau) \sim \tau^{1/4} \int_0^{X_1} dx \delta \tilde{S}^z(-x,0), 
\end{align}
which can be derived by following Refs.~\cite{Yoshimura2025-1,Yoshimura2025-2}.
From the definition of the generating function, we have
\begin{widetext}
\begin{align} 
\langle e^{ z J(\tau)} \rangle \sim  \dfrac{1}{Z_{\rm hyd}}  \int\mathcal{D} \delta\tilde{S}^z(x,0)  \mathcal{D} \delta\tilde{\rho}^{\rm (c)}(\lambda,x,0) \exp\left[ - \int_{-\infty}^{\infty} dx \left( \dfrac{\delta\tilde{S}^z(x,0) ^2}{2C_{1}} + \int_{-\pi}^{\pi} d \lambda \dfrac{ \delta\tilde{\rho}^{\rm (c)}(\lambda,x,0)^2}{2C_2(\lambda)}  \right) +  z \tau^{1/4} \int_0^{X_1} dx \delta \tilde{S}^z(-x,0) \right] \label{eq:EM_PI}
\end{align}
\end{widetext}
with the normalization constant $Z_{\rm hyd}$. This functional integral is analytically tractable in the same way as in Refs.~\cite{Yoshimura2025-1,Yoshimura2025-2}, and we eventually derive
\begin{align} 
\langle e^{ z J(\tau)} \rangle \sim e^{d(\beta,\mu)^2 z^4 \tau/8} \left[1 + {\rm erf}\left( \dfrac{d(\beta,\mu) z^2 \sqrt{\tau}}{2^{3/2}} \right)\right] \label{eq:EM_Gene}
\end{align}
with $d(\beta,\mu) = \sqrt{ 2 [ 1/(2+e^{-\beta(2+\mu)}) - 1/(2+e^{\beta(2-\mu)}) ]/(\pi \beta) }$ and the error function ${\rm erf}(\bullet)$. 
After a transformation from this generating function to the corresponding probability, we obtain the M-Wright function of Eq.~\eqref{eq:Ps_limit_hydro} of the main text.
The detailed derivation of Eq.~\eqref{eq:EM_Gene} and the M-Wright function is explained in Sec.~\ref{sec:EM_PI} of SM~\cite{SM}.

\clearpage
\widetext

\setcounter{equation}{0}
\setcounter{figure}{0}
\setcounter{section}{0}
\setcounter{table}{0}
\renewcommand{\theequation}{S-\arabic{equation}}
\renewcommand{\thefigure}{S-\arabic{figure}}
\renewcommand{\thetable}{S-\arabic{table}}
\setcounter{page}{1}

\begin{spacing}{1.5}
\begin{center}
{\large \textbf{Supplemental Material for ``Exact Anomalous Current Fluctuations in \\ Quantum Many-Body Dynamics''}}
\vspace{2mm}

Kazuya Fujimoto$^1$, Taiki Ishiyama$^1$, Taiga Kurose$^1$, Takato Yoshimura$^{2,3,4}$, and Tomohiro Sasamoto$^1$
\vspace*{2mm}

$^1$Department of Physics, Institute of Science Tokyo, 2-12-1 Ookayama, Meguro-ku, Tokyo 152-8551, Japan\\
$^2$Department of Mathematics, King's College London \\
$^3$Rudolf Peierls Centre for Theoretical Physics, University of Oxford, 1 Keble Road, Oxford OX1 3NP, U.K.\\
$^4$All Souls College, Oxford OX1 4AL, U.K.

\vspace{2mm}
\end{center}
\end{spacing}

\begin{spacing}{1.2}
\tableofcontents
\end{spacing}
\let\addcontentsline\oldaddcontentsline

\section{Derivation of Eq.~(\ref{eq:Gs})}\label{sec:sup_Gs}

\subsection{Spin-charge separation}
The $t\textrm{-}0$ model is known to be Bethe solvable in Refs.~\cite{Izergin1998,Abarenkova2001,Essler2005} of the main text, and we can write down the exact Bethe many-body function. However, it is convenient to rely on the interesting property of this model, namely spin-charge separation in Refs.~\cite{Zvonarev2009,Gamayun2023,Gamayun2024} of the main text, for the exact calculation of $\mathbb{P}_{S}[\Delta S_{R}^z ,t]$. This property stems from the fact that the on-site interaction is infinite in the $t\textrm{-}0$ model and thus the double occupation is not allowed. Thanks to this property, the initial spin configuration does not evolve over time. To explain this fact in detail, let us denote the Fock basis as $\ket{\bm{x}, \bm{s}}$ with $\bm{x} \coloneqq (x_{1}, ..., x_{N_p})$ and $\bm{s} \coloneqq (s_{1}, ..., s_{N_p})$. Here $N_p$ is the total particle number, and $x_j \in \Lambda$ and $s_j \in \{1, -1 \}$ represent the occupied site label and the spin, respectively. In the $t\textrm{-}0$ model, there are no double occupations, and thus in the following we consider only the ordered coordinate $x_1 < x_2 < \cdots < x_{N_p}$ without loss of generality. Using these notations, we can show 
\begin{align} 
\bra{\bm{x}', \bm{s}'}  \hat{H} \ket{\bm{x}, \bm{s}} = \delta_{\bm{s}', \bm{s}} \bra{\bm{x}'} \hat{H}_{\rm ff} \ket{\bm{x}}, 
\label{t0_a}
\end{align}
where we introduce the Hamiltonian $\hat{H}_{\rm ff}$, defined by 
\begin{align} 
\hat{H}_{\rm ff} \coloneqq - \sum_{j=-N+1}^{N-1} \left( \hat{a}_{j+1}^{\dagger} \hat{a}_{j} + \hat{a}_{j}^{\dagger} \hat{a}_{j+1}  \right).
\end{align}
Here, $\hat{a}_j$ is the annihilation operator of a spinless fermion at site $j \in \Lambda$ and $\ket{\bm{x}}$ is a basis of the spinless fermions.
For the following calculations, we define the notation $ \langle \bm{s}' | \bm{s} \rangle \coloneq \delta_{\bm{s}', \bm{s}} $. Then, the unitary time-evolution operator $\hat{U}(t) = e^{ - {\rm i} \hat{H} t}$ of the $t\textrm{-}0$ model is expressed as
\begin{align} 
\bra{\bm{x}', \bm{s}'}  \hat{U}(t) \ket{\bm{x}, \bm{s}} = \langle \bm{s}' | \bm{s} \rangle \bra{\bm{x}'} e^{ - {\rm i} \hat{H}_{\rm ff} t}  \ket{\bm{x}}. 
\label{eq:S_Unitary}
\end{align}
This matrix element clearly shows that the spin configuration does not change in time.

\subsection{Explicit expression of $G_S(\lambda,t)$}
We calculate the generating function $G_S(\lambda,t)$ exactly. To make this calculation simple, we specify the basis of the spin-1/2 fermions using the notation $\ket{ \bm{x}^{L,l}, \bm{s}^{L,l}; \bm{x}^{R,r}, \bm{s}^{R,r}  }$. Here $\bm{x}^{L, l} \coloneqq (x^{L, l}_l, x^{L, l}_{l-1}, ..., x^{L, l}_1 )$ and $\bm{x}^{R, r} \coloneqq (x^{R, r}_1, x^{R, r}_{2}, ..., x^{R, r}_r )$ are site labels occupied by the fermions, and we require the constraint $ -N+1 \leq   x^{L, l}_l < x^{L, l}_{l-1} < \cdots < x^{L, l}_1 \leq 0 < x^{R,  r}_1 < x^{R, r}_2 < \cdots < x^{R, r}_{r} \leq N$. One can see that the nonnegative integers $l$ and $r$ represent the numbers of fermions in the left and right regions, respectively, where the left (right) region means a set $\{-N+1, ..., 0\}~(\{1,..., N\})$. $\bm{s}^{L,l} \coloneqq (s^{L,l}_l, s^{L,l}_{l-1}, ..., s^{L,l}_1 )$, $\bm{s}^{R, r} \coloneqq (s^{R, r}_1, s^{R, r}_{2}, ..., s^{R, r}_r )$ are the spin values corresponding to fermions occupying the sites of $\bm{x}^{L,l}$ and $\bm{x}^{R,r}$, respectively. Under this notation, the completeness relation is given by 
\begin{align} 
\sum_{l=0}^{N} \sum_{r=0}^{N} \sum_{\bm{s}^{L,l}} \sum_{\bm{s}^{R,r}} \sideset{}{^{'}} \sum_{ \substack{\bm{x}^{L,l} \\ \bm{x}^{R,r}} }  \ket{ \bm{x}^{L,l}, \bm{s}^{L,l}; \bm{x}^{R,r}, \bm{s}^{R,r}  } \bra{ \bm{x}^{L,l}, \bm{s}^{L,l}; \bm{x}^{R,r}, \bm{s}^{R,r}  } = 1, 
\label{eq:S_CR}
\end{align}
where the summation $\displaystyle \sideset{}{^{'}}\sum_{ {\bm{x}^{L,l}, \bm{x}^{R,r}} }$ is taken under the constraint $ -N+1 \leq   x^{L, l}_l < x^{L, l}_{l-1} < \cdots < x^{L, l}_1 \leq 0 < x^{R,  r}_1 < x^{R, r}_2 < \cdots < x^{R, r}_{r} \leq N$.

We first expand the initial density matrix of Eq.~\eqref{eq:initial} of the main text using this completeness relation. The resulting expression becomes
\begin{align} 
\hat{\rho}(0) = \frac{1}{4^N} \sum_{l=0}^{N} \sum_{r=0}^{N}  \frac{1}{2^{l+r}} \sum_{\bm{s}^{L,l}} \sum_{\bm{s}^{R,r}} \sideset{}{^{'}}\sum_{ \substack{\bm{x}^{L,l} \\ \bm{x}^{R,r}} }  \ket{ \bm{x}^{L,l}, \bm{s}^{L,l}; \bm{x}^{R,r}, \bm{s}^{R,r}  } \bra{ \bm{x}^{L,l}, \bm{s}^{L,l}; \bm{x}^{R,r}, \bm{s}^{R,r}  }.
\label{eq:S_rho0}
\end{align}
This can be derived by directly applying Eq.~\eqref{eq:initial} of the main text to Eq.~\eqref{eq:S_CR}.

Next, we expand the generating function $G_{S}(\lambda,t) $ via Eqs.~\eqref{eq:S_Unitary} and \eqref{eq:S_rho0}.
The definition of $G_{S}(\lambda,t)$ given in the main text and Eq.~\eqref{eq:S_rho0} leads to 
\begin{align} 
G_{S}(\lambda,t) =  \dfrac{1}{4^N}\sum_{l=0}^{N} \sum_{r=0}^{N} \left(\dfrac{1}{2}\right)^{l+r} \sum_{ {\bm s}^{{L},l}} \sum_{ {\bm s}^{{R},r} }
\sideset{}{^{'}}\sum_{ \substack{\bm{x}^{L,l} \\ \bm{x}^{R,r}} } e^{-\lambda \sum_{j=1}^{r} s^{ {R},r}_j }
\bra{ {\bm x}^{{L},l},{\bm s}^{{L},l} ; {\bm x}^{{R},r},{\bm s}^{{R},r} } 
\left[ \hat{U}(t)^{\dagger} e^{\lambda \hat{S}^z_R }   \hat{U}(t)   \right]
\ket{ {\bm x}^{{L},l}, {\bm s}^{{L},l}; {\bm x}^{{R},r},{\bm s}^{{R},r} }.
\label{eq:S_GS1}
\end{align}
To calculate the right-hand side, we use the spectral decomposition of $e^{\lambda \hat{S}^z_R }$ given by 
\begin{align} 
e^{\lambda \hat{S}^z_R } 
= \sum_{N_{R}=0}^{N} \sum_{N_{L}=0}^{N} \sum_{\bm{q}^{L,N_{L}}} \sum_{\bm{q}^{R,N_{R}}} \sideset{}{^{'}}\sum_{ \substack{\bm{y}^{L,N_{L}}\\ \bm{y}^{R,N_{R}}} }  e^{\lambda \sum_{j=1}^{N_R} q^{R,N_R}_j }
\ket{ \bm{y}^{L,N_{L}},\bm{q}^{L,N_{L}}; \bm{y}^{R,N_{R}},\bm{q}^{R,N_{R}} } \bra{ \bm{y}^{L,N_{L}}, \bm{q}^{L,N_{L}}; \bm{y}^{R,N_{R}},\bm{q}^{R,N_{R}} }.
\end{align} 
Thus, the part in the summation of Eq.~\eqref{eq:S_GS1} becomes
\begin{align} 
&\bra{ {\bm x}^{{L},l},{\bm s}^{{L},l} ; {\bm x}^{{R},r},{\bm s}^{{R},r} } 
\left[ \hat{U}(t)^{\dagger} e^{\lambda \hat{S}^z_R }   \hat{U}(t)   \right]
\ket{ {\bm x}^{{L},l}, {\bm s}^{{L},l}; {\bm x}^{{R},r},{\bm s}^{{R},r} } \\
&= \sum_{N_{R}=0}^{N} \sum_{N_{L}=0}^{N} \sum_{\bm{q}^{L,N_{L}}} \sum_{\bm{q}^{R,N_{R}}} \sideset{}{^{'}}\sum_{ \substack{\bm{y}^{L,N_{L}}\\ \bm{y}^{R,N_{R}}} }  e^{\lambda \sum_{j=1}^{N_R} q^{R,N_R}_j }
\left| \bra{ \bm{y}^{L,N_{L}},\bm{q}^{L,N_{L}}; \bm{y}^{R,N_{R}},\bm{q}^{R,N_{R}} } \hat{U}(t) \ket{ {\bm x}^{{L},l}, {\bm s}^{{L},l}; {\bm x}^{{R},r}, {\bm s}^{{R},r} } \right|^2. 
\label{eq:S_GS2}
\end{align}
Using Eq.~\eqref{eq:S_Unitary}, we can show 
\begin{align} 
\bra{ \bm{y}^{L,N_{L}}, \bm{q}^{L,N_{L}}; \bm{y}^{R,N_{R}}, \bm{q}^{R,N_{R}} } \hat{U}(t) \ket{ {\bm x}^{{L},l}, {\bm s}^{{L},l}; {\bm x}^{{R},r}, {\bm s}^{{R},r} }   =  \langle  \bm{q}^{L,N_{L}};  \bm{q}^{R,N_{R}}  | {\bm s}^{{L},l}; {\bm s}^{{R},r} \rangle \bra{ \bm{y}^{L,N_{L}}; \bm{y}^{R,N_{R}} } e^{-{\rm i}\hat{H}_{\rm ff}t} \ket{ {\bm x}^{{L},l}; {\bm x}^{{R},r} }, 
\label{eq:S_GS3}
\end{align}
where $\bm{x}$ and $\bm{s}$ in Eq.~\eqref{eq:S_Unitary} correspond to $ ({\bm x}^{{L},l}; {\bm x}^{{R},r} )$ and $( {\bm s}^{{L},l}; {\bm s}^{{R},r} )$, respectively.
Here, we note that $|\bra{ \bm{y}^{L,N_{L}}; \bm{y}^{R,N_{R}} } e^{-{\rm i}\hat{H}_{\rm ff} t} \ket{ {\bm x}^{{L},l}; {\bm x}^{{R},r} }|^2$ is the probability that the spinless free fermions move from the initial position $({\bm x}^{{L},l}; {\bm x}^{{R},r})$ to the final position $(\bm{y}^{L,N_{L}}; \bm{y}^{R,N_{R}})$ at time $t$. Thus, we introduce 
\begin{align} 
\mathbb{P}_C[\bm{y}^{L,N_{L}}; \bm{y}^{R,N_{R}}, {\bm x}^{{L},l}; {\bm x}^{{R},r}, t] \coloneqq \left| \bra{ \bm{y}^{L,N_{L}}; \bm{y}^{R,N_{R}} } e^{-{\rm i}\hat{H}_{\rm ff}t} \ket{ {\bm x}^{{L},l}; {\bm x}^{{R},r} } \right|^2.
\label{eq:S_GS4}
\end{align}
Under the setup, we can eventually derive 
\begin{align} 
&\bra{ {\bm x}^{{L},l},{\bm s}^{{L},l} ; {\bm x}^{{R},r},{\bm s}^{{R},r} } 
\left[ \hat{U}(t)^{\dagger} e^{\lambda \hat{S}^z_R }   \hat{U}(t)   \right]
\ket{ {\bm x}^{{L},l}, {\bm s}^{{L},l}; {\bm x}^{{R},r},{\bm s}^{{R},r} } \\
&= \sum_{N_{R}=0}^{N} \sum_{N_{L}=0}^{N} \sum_{\bm{q}^{L,N_{L}}} \sum_{\bm{q}^{R,N_{R}}} \sideset{}{^{'}}\sum_{ \substack{\bm{y}^{L,N_{L}}\\ \bm{y}^{R,N_{R}}} }  e^{\lambda \sum_{j=1}^{N_R} q^{R,N_R}_j }
\langle  \bm{q}^{L,N_{L}};  \bm{q}^{R,N_{R}}  | {\bm s}^{{L},l}; {\bm s}^{{R},r} \rangle \mathbb{P}_C[\bm{y}^{L,N_{L}}; \bm{y}^{R,N_{R}}, {\bm x}^{{L},l}; {\bm x}^{{R},r}, t].
\label{eq:S_GS5}
\end{align}
Next, in Eq.~\eqref{eq:S_GS5}, we use the fact that $\langle  \bm{q}^{L,N_{L}};  \bm{q}^{R,N_{R}}  | {\bm s}^{{L},l}; {\bm s}^{{R},r} \rangle$ becomes nonzero only when the condition $N_L + N_R = l + r$. Then, we can take the summation over $N_{L}$, obtaining
\begin{align} 
&\bra{ {\bm x}^{{L},l},{\bm s}^{{L},l} ; {\bm x}^{{R},r},{\bm s}^{{R},r} } 
\left[ \hat{U}(t)^{\dagger} e^{\lambda \hat{S}^z_R }   \hat{U}(t)   \right]
\ket{ {\bm x}^{{L},l}, {\bm s}^{{L},l}; {\bm x}^{{R},r},{\bm s}^{{R},r} } \\
&= \sum_{N_{R}=0}^{N} \sum_{\bm{q}^{L, l+r -N_{R}} } \sum_{\bm{q}^{R,N_{R}}} \sideset{}{^{'}}\sum_{ \substack{\bm{y}^{L, l+r -N_{R} }\\ \bm{y}^{R,N_{R}}} }  e^{\lambda \sum_{j=1}^{N_R} q^{R,N_R}_j }
\langle  \bm{q}^{L, l+r -N_{R} };  \bm{q}^{R,N_{R}}  | {\bm s}^{{L},l}; {\bm s}^{{R},r} \rangle \mathbb{P}_C[\bm{y}^{L, l+r -N_{R} }; \bm{y}^{R,N_{R}}, {\bm x}^{{L},l}; {\bm x}^{{R},r}, t] \\
&= \sum_{N_{R}=0}^{N} \sum_{\bm{q}^{L, l+r -N_{R}} } \sum_{\bm{q}^{R,N_{R}}}  e^{\lambda \sum_{j=1}^{N_R} q^{R,N_R}_j }
\langle  \bm{q}^{L, l+r -N_{R} };  \bm{q}^{R,N_{R}}  | {\bm s}^{{L},l}; {\bm s}^{{R},r} \rangle \mathbb{P}_C[N_R, t, {\bm x}^{{L},l}; {\bm x}^{{R},r}], 
\label{eq:S_GS6}
\end{align}
where we define
\begin{align} 
\mathbb{P}_C[N_R, t, {\bm x}^{{L},l}; {\bm x}^{{R},r}]  \coloneqq \sideset{}{^{'}}\sum_{ \substack{\bm{y}^{L, l+r -N_{R} }\\ \bm{y}^{R,N_{R}}} } \mathbb{P}_C[\bm{y}^{L, l+r -N_{R} }; \bm{y}^{R,N_{R}}, {\bm x}^{{L},l}; {\bm x}^{{R},r}, t]. 
\label{eq:S_GS7}
\end{align}
From the expression of Eq.~\eqref{eq:S_GS7}, one can see that $\mathbb{P}_C[N_R, t, {\bm x}^{{L},l}; {\bm x}^{{R},r}]$ is the probability that the system has $N_R$ fermions in the right region ($\{1,2,...,N \}$) at time $t$ when the initial state is $ \ket{{\bm x}^{{L},l}; {\bm x}^{{R},r}} $.
We substitute Eq.~\eqref{eq:S_GS6} into Eq.~\eqref{eq:S_GS1}, getting
\begin{align} 
G_{S}(\lambda,t) = &\dfrac{1}{4^N}\sum_{l=0}^{N} \sum_{r=0}^{N} \left(\dfrac{1}{2}\right)^{l+r} \sideset{}{^{'}}\sum_{ \substack{\bm{x}^{L,l}\\ \bm{x}^{R,r}}} 
 \sum_{N_{R}=0}^{N} \mathbb{P}_C[N_{R},t, {\bm x}^{{L},l}; {\bm x}^{{R},r}] \nonumber \\
 &\hspace{22mm} \times \sum_{ {\bm s}^{{L},l}} \sum_{ {\bm s}^{{R},r} } \sum_{\bm{q}^{L, l+r-N_R }} \sum_{\bm{q}^{R,N_{R}}} e^{ \lambda \sum_{j=1}^{{N}_{R}}  q_j^{R,N_R} - \lambda \sum_{j=1}^{r} s^{ {R},r}_j} \braket{  \bm{q}^{L, l+r-N_R};  \bm{q}^{R,N_{R}} | {\bm s}^{{L},l}; {\bm s}^{{R},r} }.
 \label{eq:S_GS8}
\end{align}
We next proceed with the calculation for the summation of Eq.~\eqref{eq:S_GS8} over the spin variables $\bm{q}^{L, l+r-N_R}$ and $\bm{q}^{R,N_{R}}$ by considering the two cases, namely $N_R \geq  r$ and $N_R < r$. 
Taking these cases into account, we can obtain
\begin{align} 
 e^{ \lambda \sum_{j=1}^{{N}_{R}}  q_j^{R,N_R} - \lambda \sum_{j=1}^{r} s^{ {R},r}_j} \braket{  \bm{q}^{L, l+r-N_R};  \bm{q}^{R,N_{R}} | {\bm s}^{{L},l}; {\bm s}^{{R},r} } =
  \begin{dcases}
    \displaystyle \delta_{\bm{q}, \bm{s}} e^{ \lambda \sum_{j=1}^{N_R - r} s^{L,l}_{j}}         & \displaystyle N_R \geq r \\
    \displaystyle \delta_{\bm{q}, \bm{s}} e^{ -\lambda \sum_{j=1} ^{r -N_R} s^{R,r}_{j} }         & \displaystyle N_R < r. 
  \end{dcases}
 \label{eq:S_GS9}
\end{align}
Using Eq.~\eqref{eq:S_GS9} in Eq.~\eqref{eq:S_GS8}, we can calculate the summations over the spin variables $\bm{q}^{L, l+r-N_R }$ and $\bm{q}^{R,N_{R}}$, and then the generating function becomes
\begin{align} 
G_{S}(\lambda,t) &=  \dfrac{1}{4^N}\sum_{l=0}^{N} \sum_{r=0}^{N} \left(\dfrac{1}{2}\right)^{l+r} \sum_{ {\bm s}^{{L},l}} \sum_{ {\bm s}^{{R},r} } 
\sideset{}{^{'}}\sum_{ \substack{\bm{x}^{L,l}\\ \bm{x}^{R,r}}}
\sum_{  N_{R} \geq r} \mathbb{P}_C[N_{R},t, {\bm x}^{{L},l};{\bm x}^{{R},r}] e^{ \lambda \sum_{j=1}^{N_{R}-r} s^{{L},l}_j } \nonumber \\
&+ \dfrac{1}{4^N}\sum_{l=0}^{N} \sum_{r=0}^{N} \left(\dfrac{1}{2}\right)^{l+r} \sum_{ {\bm s}^{{L},l}} \sum_{ {\bm s}^{{R},r} } 
\sideset{}{^{'}}\sum_{ \substack{\bm{x}^{L,l}\\ \bm{x}^{R,r}}}
\sum_{  N_{R} < r} \mathbb{P}_C[N_{R},t, {\bm x}^{{L},l}; {\bm x}^{{R},r}]  e^{ -\lambda \sum_{j=1}^{r-N_{R}} s^{{R},l}_j } \\
&=  \dfrac{1}{4^N}\sum_{l=0}^{N} \sum_{r=0}^{N} \left(\dfrac{1}{2}\right)^{l+r} \sum_{ {\bm s}^{{L},l}} \sum_{ {\bm s}^{{R},r} }
\sideset{}{^{'}}\sum_{\substack{\bm{x}^{L,l}\\ \bm{x}^{R,r}}}
\sum_{ \ \Delta N_{R} \geq 0} \mathbb{P}_C[r + \Delta N_{R},t, {\bm x}^{{L},l}; {\bm x}^{{R},r}] e^{ \lambda \sum_{j=1}^{ \Delta N_{R}} s^{{L},l}_j } \nonumber \\
&+ \dfrac{1}{4^N}\sum_{l=0}^{N} \sum_{r=0}^{N} \left(\dfrac{1}{2}\right)^{l+r} \sum_{ {\bm s}^{{L},l}} \sum_{ {\bm s}^{{R},r} }
\sideset{}{^{'}}\sum_{\substack{\bm{x}^{L,l}\\ \bm{x}^{R,r}}}
\sum_{  \Delta N_{R} < 0} \mathbb{P}_C[ r + \Delta N_{R},t, {\bm x}^{{L},l}; {\bm x}^{{R},r}] e^{ -\lambda \sum_{j=1}^{ -\Delta N_{R}} s^{{R},l}_j }   \label{eq:S_GS10_1}\\
&=  \dfrac{1}{4^N}\sum_{l=0}^{N} \sum_{r=0}^{N} \left(\dfrac{1}{2}\right)^{l+r} \sum_{ {\bm s}^{{L},l}} \sum_{ {\bm s}^{{R},r} }
\sideset{}{^{'}}\sum_{\substack{\bm{x}^{L,l}\\ \bm{x}^{R,r}}}
  \sum^{l}_{\Delta N_{R} = 0} \mathbb{P}_C[r + \Delta N_{R},t, {\bm x}^{{L},l}; {\bm x}^{{R},r}] e^{ \lambda \sum_{j=1}^{ \Delta N_{R}} s^{{L},l}_j } \nonumber \\
&+ \dfrac{1}{4^N}\sum_{l=0}^{N} \sum_{r=0}^{N} \left(\dfrac{1}{2}\right)^{l+r} \sum_{ {\bm s}^{{L},l}} \sum_{ {\bm s}^{{R},r} } 
\sideset{}{^{'}}\sum_{\substack{\bm{x}^{L,l}\\ \bm{x}^{R,r}}}
 \sum^{-1}_{ \Delta N_{R} = -r}  \mathbb{P}_C[ r + \Delta N_{R},t, {\bm x}^{{L},l}; {\bm x}^{{R},r}] e^{ -\lambda \sum_{j=1}^{ -\Delta N_{R}} s^{{R},l}_j }, 
  \label{eq:S_GS10}
\end{align}
where we introduce $\Delta N_R \coloneqq N_R - r$ in Eq.~\eqref{eq:S_GS10_1} and use the fact that the initial particle number in the right(left) region is $r(l)$ and thus we have the inequality $ -r \leq \Delta N_R \leq l $ in Eq.~\eqref{eq:S_GS10}.
For the further calculation of Eq.~\eqref{eq:S_GS10}, we note the following relations:
\begin{align} 
\sum_{l=0}^{N} \sum_{r=0}^{N}  \sum^{l}_{\Delta N_{R} = 0}     =  \sum^{N}_{\Delta N_{R} = 0}    \sum_{l=\Delta N_{R}}^{N} \sum_{r=0}^{N} ,\quad\quad  \sum_{l=0}^{N} \sum_{r=0}^{N}  \sum^{-1}_{ \Delta N_{R} = -r} =   \sum^{-1}_{ \Delta N_{R} = -N} \sum_{l=0}^{N} \sum_{r=|\Delta N_{R}|}^{N}.  
  \label{eq:S_GS11}
\end{align}
As a result, we derive 
\begin{align} 
G_{S}(\lambda,t) &=  \dfrac{1}{4^N} \sum^{N}_{\Delta N_{R} = 0}    \sum_{l=\Delta N_{R}}^{N} \sum_{r=0}^{N} \left(\dfrac{1}{2}\right)^{l+r}
\sum_{ {\bm s}^{{L},l}} \sum_{ {\bm s}^{{R},r} } 
\sideset{}{^{'}}\sum_{\substack{\bm{x}^{L,l}\\ \bm{x}^{R,r}}}
\mathbb{P}_C[r + \Delta N_{R},t, {\bm x}^{{L},l}; {\bm x}^{{R},r}] e^{ \lambda \sum_{j=1}^{ \Delta N_{R}} s^{{L},l}_j } \nonumber \\
&+ \dfrac{1}{4^N} \sum^{-1}_{ \Delta N_{R} = -N} \sum_{l=0}^{N} \sum_{r=|\Delta N_{R}|}^{N} \left(\dfrac{1}{2}\right)^{l+r}
\sum_{ {\bm s}^{{L},l}} \sum_{ {\bm s}^{{R},r} } 
\sideset{}{^{'}}\sum_{\substack{\bm{x}^{L,l}\\ \bm{x}^{R,r}}} \mathbb{P}_C[ r + \Delta N_{R},t, {\bm x}^{{L},l}; {\bm x}^{{R},r}] e^{ -\lambda \sum_{j=1}^{ -\Delta N_{R}} s^{{R},l}_j } \\
 &=  \dfrac{1}{4^N} \sum^{N}_{\Delta N_{R} = 0} \sum_{l=\Delta N_{R}}^{N} \sum_{r=0}^{N} \left(\dfrac{1}{2}\right)^{l+r} 
\sideset{}{^{'}}\sum_{\substack{\bm{x}^{L,l}\\ \bm{x}^{R,r}}}\mathbb{P}_C[r + \Delta N_{R},t, {\bm x}^{{L},l}; {\bm x}^{{R},r}] \left( 2 \cosh(\lambda) \right)^{\Delta N_{R}} 2^{l-\Delta N_{R} + r} \nonumber \\
&+ \dfrac{1}{4^N} \sum^{-1}_{ \Delta N_{R} = -N} \sum_{l=0}^{N} \sum_{r=|\Delta N_{R}|}^{N} \left(\dfrac{1}{2}\right)^{l+r}
\sideset{}{^{'}}\sum_{\substack{\bm{x}^{L,l}\\ \bm{x}^{R,r}}} \mathbb{P}_C[ r + \Delta N_{R},t, {\bm x}^{{L},l}; {\bm x}^{{R},r}] \left( 2 \cosh(\lambda) \right)^{ |\Delta N_{R}|} 2^{r-|\Delta N_{R}| + l} \\
 &=  \dfrac{1}{4^N} \sum^{N}_{\Delta N_{R} = 0} \left( \cosh(\lambda) \right)^{\Delta N_{R}}    \sum_{l=\Delta N_{R}}^{N} \sum_{r=0}^{N} 
\sideset{}{^{'}}\sum_{\substack{\bm{x}^{L,l}\\ \bm{x}^{R,r}}}\mathbb{P}_C[r + \Delta N_{R},t, {\bm x}^{{L},l}; {\bm x}^{{R},r}]  \nonumber \\
&+ \dfrac{1}{4^N} \sum^{-1}_{ \Delta N_{R} = -N} \left( \cosh(\lambda) \right)^{ |\Delta N_{R}|} \sum_{l=0}^{N} \sum_{r=|\Delta N_{R}|}^{N}
\sideset{}{^{'}}\sum_{\substack{\bm{x}^{L,l}\\ \bm{x}^{R,r}}} \mathbb{P}_C[ r + \Delta N_{R},t, {\bm x}^{{L},l}; {\bm x}^{{R},r}]   \\
 &=  \sum^{N}_{\Delta N_{R} = 0} \left( \cosh(\lambda) \right)^{\Delta N_{R}}  \mathbb{P}_{C}[ \Delta N_{R},t] + \sum^{-1}_{ \Delta N_{R} = -N} \left( \cosh(\lambda) \right)^{ |\Delta N_{R}|}  \mathbb{P}_{C}[  \Delta N_{R},t] \\
 &= \mathbb{P}_{C}[ 0, t] + 2 \sum^{N}_{\Delta N_{R} = 1} \left( \cosh(\lambda) \right)^{\Delta N_{R}}  \mathbb{P}_{C}[ \Delta N_{R},t]. 
   \label{eq:S_GS12}
\end{align}
Here, we use
\begin{align} 
\mathbb{P}_{C}[ \Delta N_{R},t]  &= \dfrac{1}{4^N} \sum_{l=0}^{N} \sum_{r=0}^{N} \sideset{}{^{'}}\sum_{\substack{\bm{x}^{L,l}\\ \bm{x}^{R,r}}} \mathbb{P}_C[ r + \Delta N_{R},t, {\bm x}^{{L},l}; {\bm x}^{{R},r}] \label{eq:S_GS13_1} \\
&=  
\begin{dcases}
    \dfrac{1}{4^N} \sum_{l=\Delta N_R}^{N} \sum_{r=0}^{N} \sideset{}{^{'}}\sum_{\substack{\bm{x}^{L,l}\\ \bm{x}^{R,r}}} \mathbb{P}_C[ r + \Delta N_{R},t, {\bm x}^{{L},l}; {\bm x}^{{R},r}]         & \displaystyle (\Delta N_R \geq 0) \\
    \dfrac{1}{4^N} \sum_{l=0}^{N} \sum_{r=|\Delta N_R|}^{N} \sideset{}{^{'}}\sum_{\substack{\bm{x}^{L,l}\\ \bm{x}^{R,r}}} \mathbb{P}_C[ r + \Delta N_{R},t, {\bm x}^{{L},l}; {\bm x}^{{R},r}]         & \displaystyle (\Delta N_R < 0), 
  \end{dcases}
  \label{eq:S_GS13}
\end{align} 
which is the probability that $\Delta N_R$ fermions are transported into the right region at time $t$ compared to the initial state.
In Eq.~\eqref{eq:S_GS12}, we use the relation $\mathbb{P}_{C}[ \Delta N_{R},t] = \mathbb{P}_{C}[ -\Delta N_{R},t]$. This completes the proof for Eq.~\eqref{eq:Gs} of the main text. Note that the integral formula of $\mathbb{P}_{C}[ \Delta N_{R},t]$ given just below Eq.~\eqref{eq:Gs} of the main text is derived in Sec.~\ref{sec:sup_Pc}.

\section{Derivation of Eq.~(\ref{eq:exactPS})}\label{sec:sup_Ps1}
We derive Eq.~\eqref{eq:exactPS} of the main text using Eq.~\eqref{eq:S_GS12}.
As explained in the main text, the probability $\mathbb{P}_{S}[ \Delta S^z_{R},t]$ can be calculated by implementing the following complex integral:
\begin{align} 
\mathbb{P}_{S}[ \Delta S^z_{R},t] =   \int_{C_r} \dfrac{dz }{2 \pi {\rm i}}~\bar{G}_{S}(z,t) z^{-\Delta S^z_{R}-1}, 
\end{align}
where we define $\bar{G}_{S}(z,t) \coloneqq {G}_{S}(\lambda,t)$ with $z = e^{\lambda}$ and $C_r$ is a circle with a radius $r$ enclosing the origin in the complex plane. When substituting Eq.~\eqref{eq:S_GS12} into the integral, we get
\begin{align} 
\mathbb{P}_{S}[ \Delta S^z_{R},t] &= \dfrac{1}{2 \pi {\rm i}}  \int_{C_r} dz ~z^{-\Delta S^z_{R}-1}  \left( \mathbb{P}_{C}[ 0, t] + 2 \sum^{N}_{\Delta N_{R} = 1}  \dfrac{1}{2^{\Delta N_R}} \left( z + 1/z \right)^{\Delta N_{R}}  \mathbb{P}_{C}[ \Delta N_{R},t] \right) \\
&= \delta_{\Delta S^z_{R}, 0} \mathbb{P}_{C}[ 0, t] + \dfrac{1}{\pi {\rm i}} \sum^{N}_{\Delta N_{R} = 1} \dfrac{1}{2^{\Delta N_R}} \mathbb{P}_{C}[ \Delta N_{R},t]   \sum_{n=0}^{\Delta N_R} ~_{\Delta N_R} C_n  \int_{C_r} dz ~ z^{\Delta N_R - 2n -\Delta S^z_{R}-1} \\
&= \delta_{\Delta S^z_{R}, 0} \mathbb{P}_{C}[ 0, t] +  2 \sum^{N}_{\Delta N_{R} = 1}  \mathbb{P}_{C}[ \Delta N_{R},t]   \sum_{n=0}^{\Delta N_R} \dfrac{ ~_{\Delta N_R} C_n  }{2^{\Delta N_R }}  \delta_{\Delta S^z_{R}, \Delta N_R - 2n } \label{eq:S_PS1}.
\end{align}
This completes the proof for Eq.~\eqref{eq:exactPS} of the main text.

\section{Asymptotic analysis of $\mathbb{P}_{C}[ \Delta N_{R},t]$}\label{sec:sup_Pc}
We shall derive the asymptotic expression of $\mathbb{P}_{C}[ \Delta N_{R},t]$ for $t \gg 1$. 
First, we show that $\mathbb{P}_{C}[ \Delta N_{R},t]$ can be calculated by the unitary dynamics with the Hamiltonian $\hat{H}_{\rm ff}$ starting from the infinite temperature state. 
Second, we demonstrate that the computation of $\mathbb{P}_{C}[ \Delta N_{R},t]$ can be reduced to the unitary dynamics starting from the simple domain-wall initial state. 
Finally, we implement the asymptotic analysis for the expression obtained in the second part. 

\subsection{setup}
We consider the dynamics of the spinless fermions in the $t\textrm{-}0$ model.  
The dynamics is generated by the Hamiltonian $\hat{H}_{\rm ff} $ for the spinless free fermions.  
We denote the density matrix at time $t$ by $\hat{\varpi}(t)$ and we assume that the initial state is the infinite temperature state given by 
\begin{align} 
\hat{\varpi}(0) 
&= \prod_{j=-N+1}^N \left(  \dfrac{1}{2} \hat{a}_{j}^{\dagger} \hat{a}_{j}  + \dfrac{1}{2} (1- \hat{a}_{j}^{\dagger} \hat{a}_{j}) \right) \\
&= \dfrac{1}{4^N} \sum_{l=0}^N \sum_{r=0}^N \sideset{}{^{'}}\sum_{\substack{\bm{x}^{L,l}\\ \bm{x}^{R,r}}} \ket{\bm{x}^{L,l}; \bm{x}^{R,r}} \bra{\bm{x}^{L,l}; \bm{x}^{R,r}}
\end{align}
where we use the completeness relation $\displaystyle \sum_{l=0}^N \sum_{r=0}^N \sideset{}{^{'}}\sum_{\substack{\bm{x}^{L,l}\\ \bm{x}^{R,r}}} \ket{\bm{x}^{L,l}; \bm{x}^{R,r}} \bra{\bm{x}^{L,l}; \bm{x}^{R,r}} = 1$ to derive the last expression.
This expression means that a site is occupied by a fermion with a probability of $1/2$, and thus this state is referred to as the infinite temperature state.

Next, we define the number operator $\hat{N}_{R} \coloneqq \sum_{j =1}^N \hat{a}_j^{\dagger} \hat{a}_j$ in the right region $\{1,2,...,N\}$.
Then, the probability for the integrated charge current $\mathbb{P}_{\rm ff}[ \Delta N_{R},t]$ is defined by
\begin{align} 
\mathbb{P}_{\rm ff}[ \Delta N_R,t] \coloneqq \sum_{N_R = -N}^{N} {\rm Tr} \left[  \hat{P}_{S}(N_R + \Delta N_R) e^{ -{\rm i}t \hat{H}_{\rm ff} t} \hat{P}_{C}(N_R) \hat{\varpi}(0) \hat{P}_{C}(N_R) e^{{\rm i}t \hat{H}_{\rm ff} t } \right], 
\label{eq:S_def_PS}
\end{align}
where $\hat{P}_C(N_R)$ is a projector for the eigenspace with an eigenvalue $N_R$ of $\hat{N}_R$. 
Then, the corresponding generating function $G_{\rm ff}(\lambda,t)$ is defined by
\begin{align} 
G_{\rm ff}(\lambda,t) &\coloneqq \sum_{N_R=-N}^{N} \mathbb{P}_{\rm ff}[ \Delta N_{R},t] e^{\lambda \Delta N_R} \\
&= \sum_{N_R=-N}^{N} {\rm Tr} \left( e^{\lambda \left( \hat{N}_{R} - N_R \right)} e^{ - {\rm i}\hat{H}_{\rm ff}t}  \hat{P}_{C}(N_R) \hat{\varpi}(0) \hat{P}_{C}(N_R) e^{ {\rm i}\hat{H}_{\rm ff}t}  \right).
\label{eq:S_C_Gff} 
\end{align}
We introduce a complex variable $z = e^{\lambda}$ and then get
\begin{align} 
 \mathbb{P}_{\rm ff}[ \Delta N_{R},t] =  \dfrac{1}{2 \pi {\rm i}} \int_{C_r} dz~  \bar{G}_{\rm ff}(z,t) z^{-\Delta N_{R}-1},
\label{eq:S_PXX}
\end{align}
where we define $\bar{G}_{\rm ff}(z,t) \coloneqq G_{\rm ff}(\lambda,t)$. 
In what follows, we exactly compute the generating function of Eq.~\eqref{eq:S_C_Gff} and subsequently derive the asymptotic expression of $ \mathbb{P}_{\rm ff}[ \Delta N_{R},t]$ for $t \gg 1$.

\subsection{Proof of $\mathbb{P}_{\rm ff}[ \Delta N_{R},t] = \mathbb{P}_{C}[ \Delta N_{R},t]$}
We shall show that $\mathbb{P}_{\rm ff}[ \Delta N_{R},t]$ is identical to $\mathbb{P}_{C}[ \Delta N_{R},t]$ of Eq.~\eqref{eq:S_GS13}. This relation makes the asymptotic analysis of $\mathbb{P}_{C}[ \Delta N_{R},t]$ simple, as shown later.

To prove the relation $\mathbb{P}_{\rm ff}[ \Delta N_{R},t] = \mathbb{P}_{C}[ \Delta N_{R},t]$, we first calculate the generating function $G_{\rm ff}(\lambda,t) $.
The straightforward calculation leads to 
\begin{align} 
G_{\rm ff}(\lambda,t) &= \sum_{N_R=-N}^{N} {\rm Tr} \left( e^{\lambda \left( \hat{N}_{R} - N_R \right)} e^{ - {\rm i}\hat{H}_{\rm ff}t}  \hat{P}_{  C}(N_R) \hat{\varpi}(0) \hat{P}_{  C}(N_R) e^{ {\rm i}\hat{H}_{\rm ff}t}  \right)  \\
&= \sum_{N_R=-N}^{N} {\rm Tr} \left( e^{\lambda \left( \hat{N}_{R} - N_R \right)} e^{ - {\rm i}\hat{H}_{\rm ff}t}  \hat{P}_{  C}(N_R) \left[ \dfrac{1}{4^N} \sum_{l=0}^N \sum_{r=0}^N \sideset{}{^{'}}\sum_{\substack{\bm{x}^{L,l}\\ \bm{x}^{R,r}}} \ket{\bm{x}^{L,l}, \bm{x}^{R,r}} \bra{\bm{x}^{L,l}, \bm{x}^{R,r}} \right]\hat{P}_{  C}(N_R) e^{  {\rm i}\hat{H}_{\rm ff}t}  \right) \\
&= \dfrac{1}{4^N} \sum_{l=0}^N \sum_{r=0}^N \sideset{}{^{'}}\sum_{\substack{\bm{x}^{L,l}\\ \bm{x}^{R,r}}}  \sum_{N_R} \bra{\bm{x}^{L,l}, \bm{x}^{R,r}} \hat{P}_{  C}(N_R) e^{ {\rm i}\hat{H}_{\rm ff}t}   e^{\lambda \left( \hat{N}_{R} - N_R \right)} e^{ - {\rm i}\hat{H}_{\rm ff}t}  \hat{P}_{C  }(N_R)  \ket{\bm{x}^{L,l}, \bm{x}^{R,r}}  \\ 
&= \dfrac{1}{4^N} \sum_{l=0}^N \sum_{r=0}^N e^{-\lambda  r} \sideset{}{^{'}}\sum_{\substack{\bm{x}^{L,l}\\ \bm{x}^{R,r}}}  \bra{\bm{x}^{L,l}, \bm{x}^{R,r}} e^{  {\rm i}\hat{H}_{\rm ff}t}   e^{\lambda  \hat{N}_{R} } e^{ - {\rm i}\hat{H}_{\rm ff}t}   \ket{\bm{x}^{L,l}, \bm{x}^{R,r}}, 
\label{eq:S_PC_totyu1}
\end{align}
where we use $\hat{P}_{C  }(N_R)  \ket{\bm{x}^{L,l}, \bm{x}^{R,r}} =  \delta_{N_R,r} \ket{\bm{x}^{L,l}, \bm{x}^{R,r}}$ to derive the last line. 
Next, we use the completeness relation $ \displaystyle\sum_{N_L=0}^N \sum_{N_R=0}^N \sideset{}{^{'}}\sum_{\substack{\bm{y}^{L,N_L}\\ \bm{y}^{R,N_R}}} \ket{\bm{y}^{L,N_L}, \bm{y}^{R,N_R}} \bra{\bm{y}^{L,N_L}, \bm{y}^{R,N_R}} = 1 $, obtaining
\begin{align} 
e^{\lambda  \hat{N}_{R} } e^{ - {\rm i}\hat{H}_{\rm ff}t}   \ket{\bm{x}^{L,l}, \bm{x}^{R,r}} &= e^{\lambda  \hat{N}_{R} } \left( \sum_{N_L=0}^N \sum_{N_R=0}^N \sideset{}{^{'}}\sum_{\substack{\bm{y}^{L,N_L}\\ \bm{y}^{R,N_R}}} \ket{\bm{y}^{L,N_L}, \bm{y}^{R,N_R}} \bra{\bm{y}^{L,N_L}, \bm{y}^{R,N_R}} \right)  e^{ - {\rm i}\hat{H}_{\rm ff}t}   \ket{\bm{x}^{L,l}, \bm{x}^{R,r}} \\
&= \sum_{N_L=0}^N \sum_{N_R=0}^N \sideset{}{^{'}}\sum_{\substack{\bm{y}^{L,N_L}\\ \bm{y}^{R,N_R}}} e^{\lambda  \hat{N}_{R} }  \ket{\bm{y}^{L,N_L}, \bm{y}^{R,N_R}} \bra{\bm{y}^{L,N_L}, \bm{y}^{R,N_R}}   e^{ - {\rm i}\hat{H}_{\rm ff}t}   \ket{\bm{x}^{L,l}, \bm{x}^{R,r}} \\
&=  \sum_{N_R=0}^N \sideset{}{^{'}}\sum_{\substack{\bm{y}^{L,l+r-N_R}\\ \bm{y}^{R,N_R}}} e^{\lambda  \hat{N}_{R} }  \ket{\bm{y}^{L,l+r-N_R}, \bm{y}^{R,N_R}} \bra{\bm{y}^{L,l+r-N_R}, \bm{y}^{R,N_R}}   e^{ - {\rm i}\hat{H}_{\rm ff}t}   \ket{\bm{x}^{L,l}, \bm{x}^{R,r}} \\
&=  \sum_{N_R=0}^N \sideset{}{^{'}}\sum_{\substack{\bm{y}^{L,l+r-N_R}\\ \bm{y}^{R,N_R}}} e^{\lambda  N_R }  \ket{\bm{y}^{L,l+r-N_R}, \bm{y}^{R,N_R}} \bra{\bm{y}^{L,l+r-N_R}, \bm{y}^{R,N_R}}   e^{ - {\rm i}\hat{H}_{\rm ff}t}   \ket{\bm{x}^{L,l}, \bm{x}^{R,r}}.
\label{eq:S_PC_totyu2}
\end{align}
Here, to get the third line, we use $\bra{\bm{y}^{L,N_L}, \bm{y}^{R,N_R}}   e^{ - {\rm i}\hat{H}_{\rm ff}t}   \ket{\bm{x}^{L,l}, \bm{x}^{R,r}} \propto \delta_{N_L+N_R,l+r}$, which can be derived by the particle number conservation for $\hat{H}_{\rm ff}$. 
Plugging Eq.~\eqref{eq:S_PC_totyu2} into Eq.~\eqref{eq:S_PC_totyu1}, we obtain
\begin{align} 
G_{\rm ff}(\lambda,t) &= \dfrac{1}{4^N} \sum_{l=0}^N \sum_{r=0}^N e^{-\lambda  r} \sideset{}{^{'}}\sum_{\substack{\bm{x}^{L,l}\\ \bm{x}^{R,r}}}  \bra{\bm{x}^{L,l}, \bm{x}^{R,r}} e^{  {\rm i}\hat{H}_{\rm ff}t}   \left( \sum_{N_R=0}^N \sideset{}{^{'}}\sum_{\substack{\bm{y}^{L,l+r-N_R}\\ \bm{y}^{R,N_R}}} e^{\lambda  N_R }  \ket{\bm{y}^{L,l+r-N_R}, \bm{y}^{R,N_R}} \bra{\bm{y}^{L,l+r-N_R}, \bm{y}^{R,N_R}}   e^{ - {\rm i}\hat{H}_{\rm ff}t}   \ket{\bm{x}^{L,l}, \bm{x}^{R,r}}  \right) \\
&= \dfrac{1}{4^N} \sum_{N_R=0}^N \sum_{l=0}^N \sum_{r=0}^N  \sideset{}{^{'}}\sum_{\substack{\bm{x}^{L,l}\\ \bm{x}^{R,r}}}  \sideset{}{^{'}}\sum_{\substack{\bm{y}^{L,l+r-N_R}\\ \bm{y}^{R,N_R}}} e^{\lambda (N_R -  r)}  |\bra{\bm{y}^{L,l+r-N_R}, \bm{y}^{R,N_R}}   e^{ - {\rm i}\hat{H}_{\rm ff}t}   \ket{\bm{x}^{L,l}, \bm{x}^{R,r}}|^2.
\label{eq:S_C_Gff_totyu}
\end{align}

Finally, we substitute Eq.~\eqref{eq:S_C_Gff_totyu} into Eq.~\eqref{eq:S_PXX}, getting
\begin{align} 
\mathbb{P}_{\rm ff}[ \Delta N_{R},t] &= \dfrac{1}{4^N} \sum_{N_R=0}^N \sum_{l=0}^N \sum_{r=0}^N  \sideset{}{^{'}}\sum_{\substack{\bm{x}^{L,l}\\ \bm{x}^{R,r}}}  \sideset{}{^{'}}\sum_{\substack{\bm{y}^{L,l+r-N_R}\\ \bm{y}^{R,N_R}}}  \delta_{\Delta {N_R}, N_R -  r}  |\bra{\bm{y}^{L,l+r-N_R}, \bm{y}^{R,N_R}}   e^{ - {\rm i}\hat{H}_{\rm ff}t}   \ket{\bm{x}^{L,l}, \bm{x}^{R,r}}|^2 \\ 
&= \dfrac{1}{4^N}  \sum_{l=0}^N \sum_{r=0}^N  \sideset{}{^{'}}\sum_{\substack{\bm{x}^{L,l}\\ \bm{x}^{R,r}}}  \sideset{}{^{'}}\sum_{\substack{\bm{y}^{L,l-\Delta {N_R}}\\ \bm{y}^{R, \Delta {N_R} +  r}}}   |\bra{\bm{y}^{L,l-\Delta {N_R}}, \bm{y}^{R, \Delta {N_R} +  r}}   e^{ - {\rm i}\hat{H}_{\rm ff}t}   \ket{\bm{x}^{L,l}, \bm{x}^{R,r}}|^2 \\
&=\mathbb{P}_{C}[ \Delta N_{R},t], 
\end{align}
where we use Eq.~\eqref{eq:S_GS13_1}. This equality means that it is sufficient to consider the generating function $G_{\rm ff}(\lambda,t)$ of Eq.~\eqref{eq:S_C_Gff} when we study the asymptotic behavior of $\mathbb{P}_{C}[ \Delta N_{R},t]$.

\subsection{Exact expansion formula of $G_{\rm ff}(\lambda,t)$ }
We describe how to obtain the exact expansion formula of $G_{\rm ff}(\lambda,t)$ that is convenient for its asymptotic analysis with $t \gg 1$.

First, we derive a determinantal expression for Eq.~\eqref{eq:S_C_Gff}. 
By Eq.~\eqref{eq:S_C_Gff}, we obtain
\begin{align} 
G_{\rm ff}(\lambda,t) 
&= {\rm Tr} \left[ e^{  {\rm i}\hat{H}_{\rm ff}t} e^{ \lambda \hat{N}_{R} } e^{ - {\rm i}\hat{H}_{\rm ff}t}   e^{ -\lambda \hat{N}_{R}}  \sum_{ N_{R} } \hat{P}_C(N_{R}) \frac{\hat{I}}{4^{N}} \hat{P}_C(N_{R})     \right]  \\
&= \dfrac{(1+e^{-\lambda})^N }{2^{N}} {\rm Tr} \left[ e^{  {\rm i}\hat{H}_{\rm ff}t} e^{ \lambda \hat{N}_{R} } e^{ - {\rm i}\hat{H}_{\rm ff}t}  \hat{\varpi}_{0}      \right], \label{eq:S_Gff_1}
\end{align}
where we use $\hat{\varpi}(0) = \hat{I}/4^N$ and $\sum_{ N_{R} } \hat{P}(N_{R}) =1$ and define $ \hat{\varpi}_{0} \coloneqq (1+e^{-\lambda})^{-N} 2^{-N} e^{ -\lambda \hat{N}_{R}}  $.
Here we note that $\hat{\varpi}_{0}$ satisfies ${\rm Tr} [ \hat{\varpi}_{0} ] = 1$ because we have
\begin{align} 
{\rm Tr} [ e^{-\lambda \hat{N}_R} ] 
&= {\rm Tr} \left[ \prod_{j=1}^N e^{-\lambda \hat{a}_j^{\dagger}\hat{a}_j } \right] \\
&= \left( \prod_{j=-N+1}^0 {\rm Tr}_j [1] \right) \left( \prod_{j=1}^N {\rm Tr}_j\left[ e^{-\lambda \hat{a}_j^{\dagger}\hat{a}_j} \right] \right) \\
&= 2^N (1 + e^{-\lambda})^N, 
\end{align}
where ${\rm Tr}_j [ \bullet]$ is the trace operation at site $j$. Thus, we can apply the Wick theorem to Eq.~\eqref{eq:S_Gff_1}, obtaining 
\begin{align} 
G_{\rm ff}(\lambda,t) = \dfrac{(1+e^{-\lambda})^N }{2^{N}} \det\left( \delta_{m,n} + (e^{\lambda}-1) C_{m,n}(t)\right)_{m,n \in \{1,2,...,N\}}, 
\label{eq:C_Gff_det}
\end{align}
where the function $D_{m,n}(t)$ is given by 
\begin{align} 
C_{m,n}(t) \coloneqq  {\rm Tr} \left( \hat{a}^{\dagger}_m(t) \hat{a}_n(t) \hat{\varpi}_{0}  \right) 
\label{eq:S_Cdef}
\end{align}
with $\hat{a}_n(t) \coloneqq e^{  {\rm i}\hat{H}_{\rm ff}t}  \hat{a}_n e^{ -{\rm i}\hat{H}_{\rm ff}t}$. 

Second, we calculate the concrete expression of $C_{m,n}(t)$. 
Since the Hamiltonian $\hat{H}_{\rm ff}$ is given as the quadratic form of the spinless fermions, the annihilation operator becomes
\begin{align} 
\hat{a}_n(t) = \sum_{q=-N+1}^{N} \hat{a}_q U_{q,n}(t) \label{eq:S_C_a}
\end{align}
where $U_{q,n}(t)$ is the unitary matrix determined by the Hamiltonian. 
Thus, we substitute Eq.~\eqref{eq:S_C_a} into Eq.~\eqref{eq:S_Cdef}, getting 
\begin{align} 
C_{m,n}(t) 
&= \sum_{p=-N+1}^{N} \sum_{q=-N+1}^{N} U_{p,m}(t)^* U_{q,n}(t) {\rm Tr} \left( \hat{a}_p^{\dagger} \hat{a}_q \hat{\varpi}_{0} \right) \\
&= \dfrac{1}{e^{\lambda}+1} \sum_{p=1}^{N}  U_{p,m}(t)^* U_{p,n}(t) + \dfrac{1}{2} \sum_{p=-N+1}^{0}  U_{p,m}(t)^* U_{p,n}(t) \\
&= \dfrac{1}{e^{\lambda}+1} \delta_{m,n} - \dfrac{1}{e^{\lambda}+1} \sum_{p=-N+1}^{0}  U_{p,m}(t)^* U_{p,n}(t) + \dfrac{1}{2} \sum_{p=-N+1}^{0}  U_{p,m}(t)^* U_{p,n}(t) \\
&= \dfrac{1}{e^{\lambda}+1} \delta_{m,n} + \dfrac{e^{\lambda}-1}{2e^{\lambda}+2} \sum_{p=-N+1}^{0}  U_{p,m}(t)^* U_{p,n}(t) \\
&= \dfrac{1}{e^{\lambda}+1} \delta_{m,n} + \dfrac{e^{\lambda}-1}{2e^{\lambda}+2} D_{m,n}(t) \label{eq:S_C_totyu}
\end{align}
with $D_{m,n}(t) = \sum_{p=-N+1}^{0}  U_{p,m}(t)^* U_{p,n}(t)$. Here we use ${\rm Tr} \left( \hat{a}_p^{\dagger} \hat{a}_q \hat{\varpi}_{0} \right) = \delta_{p,q}/(e^{\lambda}+1)$ for $p > 0$ and ${\rm Tr} \left( \hat{a}_p^{\dagger} \hat{a}_q \hat{\varpi}_{0} \right) = \delta_{p,q}/2$ for $p \leq 0$.
We put Eq.~\eqref{eq:S_C_totyu} into Eq.~\eqref{eq:C_Gff_det}, getting 
\begin{align} 
G_{\rm ff}(\lambda,t) 
&= \dfrac{(1+e^{-\lambda})^N }{2^{N}} \det\left( \delta_{m,n} +  \dfrac{e^{\lambda}-1}{e^{\lambda}+1} \delta_{m,n} +  \dfrac{ (e^{\lambda}-1) (e^{\lambda}-1)}{2e^{\lambda}+2} D_{m,n}(t)  \right)_{m,n \in \{1,2,...,N\}} \\
&= \dfrac{(1+e^{-\lambda})^N }{2^{N}} \det\left( \dfrac{2}{1+e^{-\lambda}} \delta_{m,n} +  \dfrac{ (e^{\lambda/2}-e^{-\lambda/2})^2}{2+2e^{-\lambda}} D_{m,n}(t)  \right)_{m,n \in \{1,2,...,N\}} \\
&= \det\left(  \delta_{m,n} +  \dfrac{ (e^{\lambda/2}-e^{-\lambda/2})^2}{4} D_{m,n}(t)  \right)_{m,n \in \{1,2,...,N\}} \\
&= \det\left(  \delta_{m,n} +  \sinh\left(\dfrac{\lambda}{2}\right)^2  D_{m,n}(t) \right)_{m,n \in \{1,2,...,N\}}
\end{align}
Here, we comment on the expression for $D_{m,n}(t)$. 
Using Eq.~\eqref{eq:S_C_a}, we derive 
\begin{align} 
D_{m,n}(t) 
&= \sum_{p=-N+1}^{0}  U_{p,m}(t)^* U_{p,n}(t) \\
&= \sum_{p=-N+1}^{0} \sum_{q=-N+1}^{0}  U_{p,m}(t)^* U_{q,n}(t) \delta_{p,q} \\
&= \sum_{p=-N+1}^{N} \sum_{q=-N+1}^{N}  U_{p,m}(t)^* U_{q,n}(t)  {\rm Tr}\left( \hat{a}^{\dagger}_p \hat{a}_q \hat{\varpi}_{\rm DW} \right)\\
&=  {\rm Tr}\left( \hat{a}^{\dagger}_m(t) \hat{a}_n(t) \hat{\varpi}_{\rm DW} \right), 
\end{align}
where we define the domain-wall initial state as
\begin{align} 
\hat{\varpi}_{\rm DW} \coloneqq \left( \prod_{n=-N+1}^0 \hat{a}^{\dagger}_n \right) \ket{0} \bra{0} \left( \prod_{n=-N+1}^0 \hat{a}_n \right) 
\end{align}
with the vacuum $\ket{0}$ for the spinless free fermions. Hence, we can calculate the time dependence of $D_{m,n}(t)$ by solving the following differential equation:
\begin{align} 
{\rm i} \dfrac{d}{dt} D_{m,n}(t) = D_{m+1,n}(t) + D_{m-1,n}(t) - D_{m,n+1}(t) - D_{m,n-1}(t),  \label{eq:S_D}
\end{align}
where the initial condition is $D_{m,n}(0) = \delta_{m,n} \chi[ m \in \{-N+1,..., 0 \}]$ with the indicator function $\chi[\bullet]$. The boundary condition is open, namely $D_{N+1,n}(t) = 0$, $D_{m,N+1}(t) = 0$, $D_{-N,n}(t) = 0$, and $D_{m,-N}(t) = 0$. Here, we emphasize that the calculation of $G_{\rm ff}(\lambda,t)$ can be reduced to the full counting statistics for the free-fermion dynamics starting from the simple domain-wall initial state. Thanks to this simplification, we can perform the asymptotic analysis of $\mathbb{P}_{C}[ \Delta N_{R},t]$, as shown in the following subsection.

Third, we take the thermodynamic limit ($N \rightarrow \infty$).
Then, we obtain
\begin{align} 
D_{m,n}(t) = {\rm i}^{n-m}\sum_{l=0}^{\infty} J_{m+l}(2t) J_{n+l}(2t). 
\label{eq:S_D_Bessel}
\end{align}
This can be derived as follows. 
We define a function $\bar{D}(k_1,k_2,t)$ as
\begin{align} 
\bar{D}(k_1,k_2,t) = \sum_{m \in \mathbb{Z}} \sum_{n \in \mathbb{Z}} D_{m,n}(t) e^{{\rm i}(k_1 m - k_2 n) }.
\end{align}
Then, we can expand $D_{m,n}(t)$ as
\begin{align} 
D_{m,n}(t) = \dfrac{1}{4 \pi^2}\int_{0}^{2\pi} dk_1 \int_{0}^{2\pi} dk_2 \bar{D}(k_1,k_2,t) e^{{\rm i}(-k_1 m + k_2 n) }.
\end{align}
After putting it into Eq.~\eqref{eq:S_D}, we get $\bar{D}(k_1,k_2,t) = \bar{D}(k_1,k_2,0) e^{ -{\rm i} 2t \left(\cos(k_1) - \cos(k_2) \right) }$.
Here, the initial condition is given by
\begin{align} 
\bar{D}(k_1,k_2,0) 
&= \sum_{m \in \mathbb{Z}} \sum_{n \in \mathbb{Z}} D_{m,n}(0) e^{{\rm i}(k_1 m - k_2 n) } \\
&= \sum_{m = -\infty}^0 e^{{\rm i}(k_1- k_2 )m }.
\label{eq:S_D0}
\end{align}
Combining all the results above, we obtain
\begin{align} 
D_{m,n}(t) 
&= \dfrac{1}{4 \pi^2} \sum_{l = -\infty}^0 \int_{0}^{2\pi} dk_1 \int_{0}^{2\pi} dk_2   e^{{\rm i}(-k_1 (m-l) - 2t\cos(k_1)  ) } e^{{\rm i}( k_2 (n-l) + 2t \cos(k_2)   ) }\\
&= {\rm i}^{n-m} \sum_{l = 0}^{\infty}  J_{n+l}(2t) J_{m+l}(2t), 
\end{align}
where we use the integral formula for the Bessel function $J_n(x)$ of the first kind:
\begin{align} 
J_n(x) 
&= \dfrac{1}{2\pi {\rm i}^n} \int_0^{2\pi} d\theta \exp\left( {\rm i} x \cos \theta + {\rm i} n \theta  \right) \\
&= \dfrac{ {\rm i}^n}{2\pi} \int_0^{2\pi} d\theta \exp\left( -{\rm i} x \cos \theta - {\rm i} n \theta  \right).
\end{align}
This completes the derivation of Eq.~\eqref{eq:S_D_Bessel}.

Finally, we consider the logarithm of the generating function, deriving the following expansion: 
\begin{align} 
\log G_{\rm ff}(\lambda,t) = -\sum_{a=1}^{\infty} \dfrac{(-1)^a}{a} \omega^a {\rm Tr} D(t)^a 
\label{eq:S_Ea1}
\end{align}
with $\omega \coloneqq {\rm sinh}^2(\lambda/2)$.
According to Appendix A in Ref.~\cite{Moriya2019} of the main text, we obtain
\begin{align} 
{\rm Tr} D(t)^a = \dfrac{1}{4^a}   \left[\prod_{k=1}^{2a} \int_{C_r} \dfrac{dz_k }{2\pi {\rm i} z_k} \int_0^{2t} dt_k\right] ~  \left\{ \prod_{k=1}^{2a}
{\rm exp}\left[ \dfrac{\sqrt{t_k t_{k+1}}}{2} \left( z_k - \dfrac{1}{z_k}\right)\right] \right\}. 
\label{eq:S_Ea2}
\end{align}
This expression is convenient for implementing the asymptotic analysis with $t \gg 1$ because the functions in the complex integral, except for the exponential term, have simple poles only at the origin. In the next subsection, we apply the asymptotic analysis to this expression. 

\subsection{Asymptotic analysis of $\mathbb{P}_{C}[ \Delta N_{R},t]$ via a saddle point approximation}
We analytically derive the asymptotic form of  $\mathbb{P}_{C}[ \Delta N_{R},t]$ with $t \gg 1$ by applying the conventional saddle point approximation to Eqs.~\eqref{eq:S_Ea1} and \eqref{eq:S_Ea2}. First, we obtain the asymptotic expression of ${\rm Tr} D(t)^a$. Second, we take the summation of Eq.~\eqref{eq:S_Ea1}, getting the asymptotic expression of $G_{\rm ff}(\lambda,t)$. Finally, we transform the generating function $G_{\rm ff}(\lambda,t)$ to the probability $\mathbb{P}_{C}[ \Delta N_{R},t]$ via Eq.~\eqref{eq:S_PXX} and then obtain the asymptotic expression of $\mathbb{P}_{C}[ \Delta N_{R},t]$.  In what follows, we set $T\coloneqq 2t$ to simplify the expressions derived here. 

\subsubsection{$a=1$ case}
We shall derive the asymptotic form of ${\rm Tr} D(t)$. 
By definition, we get
\begin{align} 
{\rm Tr} D(t) = \dfrac{1}{4} \int_{C_r} \dfrac{dz_1}{2\pi {\rm i}} \int_{C_r} \dfrac{dz_2}{2\pi {\rm i}} \int_0^{T} dt_1 \int_0^{T} dt_2 ~\dfrac{1}{z_1 z_2} 
{\rm exp}\left[ \dfrac{\sqrt{t_1 t_{2}}}{2} \left( f(z_1) + f(z_2) \right)\right] 
\end{align}
with $f(z) \coloneqq z - z^{-1}$.
We use $z_k = e^{ {\rm i} \theta_k }$, getting
\begin{align} 
{\rm Tr} D(t) = \dfrac{T^2}{4(2\pi)^2}  \int_0^{1}  dx_1 \int_0^{1} dx_2 \int_{0}^{2\pi} d\theta_1 \int_{0}^{2\pi} d\theta_2 
e^{  {\rm i} T \sqrt{x_1 x_{2}} \left[ \sin(\theta_1) + \sin(\theta_2) \right] }.
\end{align}
The extremum of $\theta$ is $\theta = (2n+1)\pi/2~( n \in \mathbb{Z})$, and thus it is enough to consider two of them, which lead to the following expansions:
\begin{align} 
&\sin \left( \dfrac{\pi}{2} + \Delta \right) \sim 1 - \dfrac{1}{2} \Delta^2, \\
&\sin \left( \dfrac{3\pi}{2} + \Delta \right) \sim -1 + \dfrac{1}{2} \Delta^2.
\end{align}
Thus, we have the asymptotic form given by 
\begin{align} 
\int_{0}^{2\pi}   d\theta_1 e^{  {\rm i} T \sqrt{x_1 x_{2}}  \sin(\theta_1)  } 
&\sim \int_{-\infty}^{\infty} d \Delta {\rm exp} \left( {   {\rm i}T \sqrt{x_1 x_2} - {\rm i}t \sqrt{x_1 x_2} \Delta^2/2 } \right) + \int_{-\infty}^{\infty} d \Delta  {\rm exp} \left(  -{\rm i}T \sqrt{x_1 x_2} + {\rm i}t \sqrt{x_1 x_2} \Delta^2/2  \right) \\
&= 2 \sqrt{ \dfrac{2\pi}{T \sqrt{x_1 x_2}} } \cos\left( T \sqrt{x_1 x_2} - \dfrac{\pi}{4} \right).
\end{align}
Putting this into ${\rm Tr} D(t) $, we obtain
\begin{align} 
{\rm Tr} D(t)
& \sim \dfrac{T^2}{4(2\pi)^2} \int_0^{1}  dx_1\int_0^{1}  dx_2    \dfrac{8\pi}{T\sqrt{x_1 x_2}}  \cos\left( T \sqrt{x_1 x_2} - \dfrac{\pi}{4} \right)^2 \\
& = \dfrac{2 T}{\pi} \int_0^1 dy_1 \int_0^1 dy_2 \cos\left( T y_1y_2 - \dfrac{\pi}{4} \right)^2 \\
& = \dfrac{2 T}{\pi} \int_0^1 dy_1 \int_0^1 dy_2  \left[ \dfrac{1}{2} + \cos\left( 2T y_1y_2 - \dfrac{\pi}{2} \right) \right] \\
& \sim \frac{T}{\pi}. 
\end{align}

\subsubsection{$a=2$ case}
We shall consider the asymptotic form of ${\rm Tr} D(t)^2$, but in this case, one will find a difference from the case of $a=1$.  
By definition, we get
\begin{align} 
{\rm Tr} D(t)^2 &=\dfrac{T^4}{4^2}   \left[ \prod_{k=1}^{4} \int_0^1 dx_k \int_0^{2\pi} \dfrac{d\theta_k}{2\pi} \right] \left[\prod_{k=1}^{4} {\rm exp}\left( {\rm i} T \sqrt{x_k x_{k+1}} \sin \theta_k \right) \right].
\end{align}
As in the previous asymptotic analysis, we have
\begin{align} 
\int_{0}^{2\pi}   \dfrac{d\theta_k}{2\pi} e^{  {\rm i} T \sqrt{x_k x_{k+1}}  \sin(\theta_k)  }  \sim \sqrt{ \dfrac{2 }{T \pi \sqrt{x_{k} x_{k+1}}} } \cos\left( T \sqrt{x_{k} x_{k+1}} - \dfrac{\pi}{4} \right).
\end{align}
Then, combining them, we obtain
\begin{align} 
{\rm Tr} D(t)^2 
&\sim \dfrac{T^4}{4^2}   \left[ \prod_{k=1}^{4} \int_0^1 dx_k \right]  \left[ \prod_{k=1}^{4} \sqrt{ \dfrac{2}{ \pi T\sqrt{x_{k} x_{k+1}}} } \cos\left( T \sqrt{x_{k} x_{k+1}} - \dfrac{\pi}{4} \right) \right] \\
&= \left( \dfrac{2T}{\pi} \right)^2  \left[ \prod_{k=1}^{4} \int_0^1 dy_k \right] \left[ \prod_{k=1}^{4} \cos \left( T y_k y_{k+1} - \dfrac{\pi}{4} \right) \right],
\end{align}
where we introduce $y_k = \sqrt{x_k}$. 
We can proceed with the further calculations as
\begin{align} 
{\rm Tr} D(t)^2 &\sim \left( \dfrac{2T}{\pi} \right)^2 \int_0^1 dy_1 \int_0^1 dy_2 \int_0^1 dy_3 \int_0^1 dy_4  \cos \left( T y_1 y_2 - \dfrac{\pi}{4} \right) \cos \left( T y_2 y_3 - \dfrac{\pi}{4} \right) \nonumber\\
&\hspace{59.4mm}\times \cos \left( T y_3 y_4 - \dfrac{\pi}{4} \right) \cos \left( T y_4 y_1 - \dfrac{\pi}{4} \right) \\
&= \left( \dfrac{2T}{\pi} \right)^2 \int_0^1 dy_1 \int_0^1 dy_2 \int_0^1 dy_3 \int_0^1 dy_4 \dfrac{1}{4} \left[ \cos\left( T(y_1+y_3)y_2 - \dfrac{\pi}{2} \right) + \cos\left( T(y_1-y_3)y_2 \right) \right] \nonumber \\
&\hspace{62.02mm}\times\left[ \cos\left( T(y_3+y_1)y_4 - \dfrac{\pi}{2} \right) + \cos\left( T(y_3-y_1)y_4 \right) \right]\\
&\sim \dfrac{1}{4} \left( \dfrac{2T}{\pi} \right)^2 \int_0^1 dy_1 \int_0^1 dy_2 \int_0^1 dy_3 \int_0^1 dy_4  \cos\left( T(y_1-y_3)y_2 \right) \cos\left( T(y_3-y_1)y_4 \right), 
\end{align} 
where we neglect the rapidly oscillating terms to get the last line.
We next perform the integrations with respect to $y_2$ and $y_4$, obtaining
\begin{align} 
{\rm Tr} D(t)^2 &\sim \dfrac{1}{4} \left( \dfrac{2T}{\pi} \right)^2 \int_0^1 dy_1 \int_0^1 dy_3 \left[ \dfrac{ \sin(T(y_1-y_3)) }{ T(y_1-y_3) } \right]^2 \\
&= \dfrac{1}{4T^2} \left( \dfrac{2T}{\pi} \right)^2 \int_0^T dx \int_0^T dy  \left[ \dfrac{ \sin(x-y) }{ x-y } \right]^2 \\
&\sim \dfrac{1}{4T^2} \left( \dfrac{2T}{\pi} \right)^2 \int_0^T dR \int_{-T}^{T} dr  \left( \dfrac{ \sin r }{ r } \right)^2\\
&= \dfrac{T}{\pi}, 
\end{align}
where we neglect the boundary contribution in the third line.

\subsubsection{$a \in \mathbb{N}_{>0}$ case}
We generalize the above result to the general case. 
The calculation is straightforward, and the resulting expression is as follows:
\begin{align} 
{\rm Tr} D(t)^a \sim \left( \dfrac{2T}{\pi} \right)^a  \left[ \prod_{k=1}^{2a} \int_0^1 dy_k \right] \left[ \prod_{k=1}^{2a} \cos \left( T y_k y_{k+1} - \dfrac{\pi}{4} \right) \right].
\end{align}
The multiple integral can be evaluated as follows:
\begin{align} 
I_{2m}(T) &\coloneqq  \left[ \prod_{k=1}^{2a} \int_0^1 dy_k \right] \left[ \prod_{k=1}^{2a}\cos \left( T y_k y_{k+1} - \dfrac{\pi}{4} \right) \right]\\
&= \left[ \prod_{k=1}^{2a} \int_{0}^1 dy_k \right] \left\{ \prod_{k=1}^{a} \left[ \dfrac{1}{2} \cos\left( T(y_{2k-1}+y_{2k+1})y_{2k} - \dfrac{\pi}{2} \right) +  \dfrac{1}{2}\cos\left( T(y_{2k-1}-y_{2k+1})y_{2k}  \right) \right]  \right\} \\
&\sim  \left[ \prod_{k=1}^{2a} \int_{0}^1 dy_k \right] \left[ \prod_{k=1}^{a}  \dfrac{\cos\left( T(y_{2k-1}-y_{2k+1})y_{2k}  \right)}{2}    \right], 
\end{align} 
where we neglect the rapidly oscillating terms in the last line.  
We proceed with this calculation by performing the integrations with respect to $y_{2k}~(k\in \{1,2,...,a\})$, obtaining
\begin{align} 
I_{2m}(T) &\sim \left[ \prod_{k=1}^{a} \int_{0}^1 dy_{2k-1} \right] \left[ \prod_{k=1}^{a}   \dfrac{\sin\left( T(y_{2k-1}-y_{2k+1}) \right) }{2 T( y_{2k-1}-y_{2k+1} )}   \right] \\
&= \dfrac{1}{2^aT^a}  \left[ \prod_{k=1}^{a} \int_{0}^T dz_{k} \right] \left[ \prod_{k=1}^{a}   \dfrac{\sin\left( z_{k}-z_{k+1} \right) }{ z_{k}-z_{k+1} }   \right]\\
&\sim \dfrac{1}{2^aT^a}\int_{0}^T  d\alpha  \left( \int_{-\infty}^{\infty} d\beta \dfrac{\sin \beta}{\beta} \right)^{a-1}\\
&= \dfrac{\pi^{a-1}}{2^a T^{a-1}}. 
\end{align}
Here, we introduce $z_k =  T y_{2k-1}$ in the second line and neglect the boundary contribution to the integral in the third line. 
Hence, we finally get 
\begin{align} 
{\rm Tr} D(t)^a 
\sim \dfrac{T}{\pi}.
\end{align}

\subsubsection{ $G_{\rm ff}(\lambda,t)$ for $t \gg 1$}
We use all the asymptotic results derived above and then obtain 
\begin{align} 
\log G_{\rm ff}(\lambda,t) &= -\sum_{a=1}^{\infty} \dfrac{(-1)^a}{a} \omega^a {\rm Tr} D(t)^a \\
&\sim  -\sum_{a=1}^{\infty} \dfrac{(-1)^a}{a} \omega^a \times \dfrac{T}{\pi} \\
&= \dfrac{T}{\pi} \log \left( \dfrac{1}{2} + \dfrac{1}{4} e^{\lambda} + \dfrac{1}{4} e^{-\lambda} \right)\\
&= \dfrac{2t}{\pi} \log \left( \dfrac{1}{2} + \dfrac{1}{4} e^{\lambda} + \dfrac{1}{4} e^{-\lambda} \right). \label{eq:logG_asy}
\end{align}
Note that the essentially same generating function was reported in Ref.~\cite{Schonhammer2007} of the main text, where scattering matrix techniques were used for the derivation.

\subsubsection{Asymptotic analysis of $\mathbb{P}_{C}[ \Delta N_R,t] $}\label{subsec:sup_Pc}
We derive the asymptotic expression of $\mathbb{P}_{C}[ \Delta N_R,t] $ using Eq.~\eqref{eq:logG_asy}.
We use the previous results and the definition of the cumulant generating function and then obtain
\begin{align} 
\mathbb{P}_{C}[ x , t] &= \mathbb{P}_{\rm ff}[ x , t] \\ 
&= \dfrac{1}{2 \pi {\rm i}} \int_{C_r} dz \dfrac{1}{z^{x+1}} G_{\rm ff}(\lambda,t) ~~~~(z = e^{\lambda})\\
&= \dfrac{1}{2 \pi {\rm i}} \int_{C_r} dz \dfrac{1}{z^{x+1}} \exp\left[ \log G_{\rm ff}(\lambda,t) \right] \\
&= \dfrac{1}{2 \pi {\rm i}} \int_{C_r} dz \dfrac{1}{z} \exp\left[ -x \log z + \sum_{n=2}^{\infty} \dfrac{\kappa_n(t)}{n!} (\log z)^n\right], 
\end{align}
where we denote the $n$th cumulant at time $t$ by $\kappa_n(t)$ and use the property of the cumulant generating function $\log G_{\rm ff}(\lambda,t)$.
Next, we change the integral variable via $z=e^{{\rm i} \theta / \sqrt{t}}~(\theta \in \mathbb{R})$, obtaining
\begin{align} 
\mathbb{P}_{C}[ x , t] = \dfrac{1}{2 \pi \sqrt{t}}  \int_{-\pi \sqrt{t}}^{\pi \sqrt{t}} d\theta  \exp\left[ - {\rm i} \dfrac{x\theta}{ \sqrt{t} }  + \sum_{n=2}^{\infty} \dfrac{\kappa_n(t)}{n!} \dfrac{{\rm i}^n \theta^n}{t^{n/2}} \right].
\label{eq:sup:PC_126}
\end{align}
For $t \gg 1$, we have $\kappa_1(t) \sim 0$ and $\kappa_n(t) \sim c_n t$ with a constant $c_n$ for an integer $ n \geq 2$ from Eq.~\eqref{eq:logG_asy} and thus derive
\begin{align} 
\mathbb{P}_{C}[ \sqrt{t} \mathcal{J}_{C}, t] &\sim \dfrac{1}{2 \pi \sqrt{t}}  \int_{-\infty}^{\infty} d\theta  \exp\left[ - {\rm i} \mathcal{J}_{C} \theta  - \dfrac{1}{2 \pi} \theta^2 \right] \\
&=\dfrac{1}{\sqrt{2t}} \exp\left(  -\dfrac{\pi}{2} \mathcal{J}_{C}^2\right),
\label{eq:sup:PC_asy}
\end{align}
where we use $\kappa_2(t) \sim c_2 t = t/\pi$ for $t \gg 1$.

\section{Asymptotic analysis of $\mathbb{P}_{S}[ \Delta S^z_{R},t]$}\label{sec:sup_Ps2}
We shall investigate the asymptotic behavior of $\mathbb{P}_{S}[ \Delta S^z_{R},t]$ using Eqs.~\eqref{eq:S_PS1} and \eqref{eq:sup:PC_asy}. 
The exact expression of $\mathbb{P}_{S}[ \Delta S^z_{R},t]$ under the thermodynamic limit ($N \rightarrow \infty$) is given by
\begin{align} 
\mathbb{P}_{S}[ \Delta S^z_{R},t] = \delta_{\Delta S^z_{R}, 0} \mathbb{P}_{C}[ 0, t] + 2 \sum^{\infty}_{\Delta N_{R} = 1}  \mathbb{P}_{C}[ \Delta N_{R},t] \sum_{n=0}^{\Delta N_R}   \dfrac{ ~_{\Delta N_R} C_{n}  }{2^{\Delta N_R }} \delta_{\Delta N_{R} - \Delta S^z_{R}-2n,0}. 
\label{eq:sup:PS_exact_tdl}
\end{align}
One can readily show the relation $\mathbb{P}_{S}[ \Delta S^z_{R},t] = \mathbb{P}_{S}[ -\Delta S^z_{R},t]$ because of $ ~_{\Delta N_R} C_{n} = ~_{\Delta N_R} C_{\Delta N_R-n} $ and $\mathbb{P}_{C}[ \Delta N_{R},t] = \mathbb{P}_{C}[ -\Delta N_{R},t]$, and thus we can focus on the case with $\Delta S^z_R \geq 0$ in the following. 

To proceed with the asymptotic analysis of Eq.~\eqref{eq:sup:PS_exact_tdl}, we note that the typical fluctuation of $\mathbb{P}_{C}[ \Delta N_{R},t]$ for $t \gg 1$ is given by Eq.~\eqref{eq:sup:PC_asy}:
\begin{align} 
\mathbb{P}_{C}[ \sqrt{t} \mathcal{J}_{C}, t] \sim \dfrac{1}{\sqrt{2t}} \exp\left(  -\dfrac{\pi}{2} \mathcal{J}_{C}^2\right).
\end{align}
Then, it is convenient to introduce a small bin $\Delta \mathcal{J}_C = 2/\sqrt{t}$, where the factor $2$ is put since the summation of Eq.~\eqref{eq:sup:PS_exact_tdl} over $\Delta N_R$ is taken only for even or odd numbers.  
As a result, we obtain, for $t \gg 1$, 
\begin{align} 
\mathbb{P}_{S}[\Delta S^z_{R},t] &= \delta_{\Delta S^z_{R}, 0} \mathbb{P}_{C}[ 0, t] + \sqrt{t}  \sum^{\infty}_{\Delta N_{R} = 1} \Delta \mathcal{J}_C ~ \mathbb{P}_{C}[ \Delta N_{R},t]  \dfrac{ ~_{\Delta N_R} C_{(\Delta N_{R} - \Delta S^z_{R})/2}  }{2^{\Delta N_R }} \dfrac{1 + (-1)^{ \Delta N_R - \Delta S_R^z}}{2}\theta( \Delta N_{R} - \Delta S^z_{R} )  \label{eq:sup:PS_exact_integral0} \\
&\sim \dfrac{1}{\sqrt{2}} \int^{\infty}_{\Delta S_R^z t^{-1/2}} d\mathcal{J}_C ~ \exp\left(  -\dfrac{\pi}{2} \mathcal{J}_{C}^2\right) 
\sqrt{  \dfrac{ 2\sqrt{t} \mathcal{J}_C}{ \pi t\mathcal{J}_C^2 - \pi(\Delta S_R^z)^2 }  } \left( 1- \dfrac{ (\Delta S^z_R)^2}{ t \mathcal{J}_C^2  } \right)^{-\sqrt{t} \mathcal{J}_C/2} \left( \dfrac{\sqrt{t}\mathcal{J}_C -  \Delta S_R^z}{ \sqrt{t}\mathcal{J}_C + \Delta S_R^z} \right)^{\Delta S^z_R/2}. 
\label{eq:sup:PS_exact_integral1}
\end{align}
Here, we use the asymptotic expression given by
\begin{align} 
~_N C_n \sim  \sqrt{\dfrac{N}{ 2 \pi n (N-n) }} \left( \dfrac{n}{N} \right)^{-n} \left( 1 - \dfrac{n}{N} \right)^{n-N}
\label{eq:sup:combination}
\end{align}
for $N, n, ~ N-n \gg 1$.

We next introduce the scaled integrated spin current $\mathcal{J}_S \coloneqq \Delta S^z_R t^{-\eta}$ with a real parameter $\eta \in (0,1/2)$.
The expression of Eq.~\eqref{eq:sup:PS_exact_integral1} reads
\begin{align} 
\mathbb{P}_{S}[ t^{\eta}\mathcal{J}_S,t] 
&\sim  \int^{\infty}_{ \mathcal{J}_S t^{\eta -1/2}} d\mathcal{J}_C ~ \exp\left(  -\dfrac{\pi}{2} \mathcal{J}_{C}^2\right) 
\sqrt{  \dfrac{ \mathcal{J}_C}{ \pi t^{1/2}\mathcal{J}_C^2 - \pi t^{2\eta-1/2} \mathcal{J}_S^2 }  } \left( 1 - \dfrac{ t^{2\eta-1}\mathcal{J}_S^2 }{ \mathcal{J}_C^2  } \right)^{-\sqrt{t} \mathcal{J}_C/2} 
\left( \dfrac{\mathcal{J}_C -  t^{\eta-1/2}\mathcal{J}_S}{ \mathcal{J}_C +  t^{\eta-1/2}\mathcal{J}_S} \right)^{t^{\eta}\mathcal{J}_S/2} \\
&=  \int^{\infty}_{ \mathcal{J}_S t^{\eta -1/2}} d\mathcal{J}_C ~ 
\sqrt{  \dfrac{ \mathcal{J}_C}{ \pi t^{1/2}\mathcal{J}_C^2 - \pi t^{2\eta-1/2} \mathcal{J}_S^2 }  } 
\exp\left[  -\dfrac{\pi}{2} \mathcal{J}_{C}^2 + \sqrt{t} g(\mathcal{J}_C)  + t^{\eta} \mathcal{J}_S h(\mathcal{J}_C)\right], 
\label{eq:sup:PS_exact_integral2}
\end{align}
where we define the two functions $g(\mathcal{J}_C)$ and $h(\mathcal{J}_C)$ as  
\begin{align} 
&g(\mathcal{J}_C) \coloneqq \dfrac{\mathcal{J}_C}{2} \log \left(  \dfrac{ \mathcal{J}_C^2 }{ \mathcal{J}_C^2 -  t^{2\eta-1} \mathcal{J}_S^2 }  \right), \\
&h(\mathcal{J}_C) \coloneqq \dfrac{1}{2} \log \left( \dfrac{\mathcal{J}_C -  t^{\eta-1/2}\mathcal{J}_S}{ \mathcal{J}_C +  t^{\eta-1/2}\mathcal{J}_S} \right). 
\end{align}
Because the dominant contribution to the integral of Eq.~\eqref{eq:sup:PS_exact_integral2} comes from values of $\mathcal{J}_{C}$ of order unity, due to the factor $e^{-\pi \mathcal{J}_C^2/2}$, we can expand $g(\mathcal{J}_C)$ and $h(\mathcal{J}_C)$ for $t \gg 1$ as
\begin{align} 
&g(\mathcal{J}_C) \sim   t^{2\eta-1}  \frac{\mathcal{J}_S^2}{2\mathcal{J}_C}, \\ &h(\mathcal{J}_C) \sim   -t^{\eta-1/2} \dfrac{\mathcal{J}_S}{\mathcal{J}_C}, 
\end{align}
where $t^{2\eta-1}$ and $t^{\eta-1/2}$ are small because of the assumption $0<\eta<1/2$.
Putting them together, we obtain
\begin{align} 
\mathbb{P}_{S}[ t^{\eta}\mathcal{J}_S,t] \sim \int^{\infty}_{ \mathcal{J}_S t^{\eta -1/2}} d\mathcal{J}_C ~ 
\sqrt{  \dfrac{ \mathcal{J}_C}{ \pi t^{1/2}\mathcal{J}_C^2 - \pi t^{2\eta-1/2} \mathcal{J}_S^2 }  } 
\exp\left[  -\dfrac{\pi}{2} \mathcal{J}_{C}^2 - t^{2\eta-1/2} \dfrac{\mathcal{J}_S^2}{2\mathcal{J}_C}\right]. 
\label{eq:sup:PS_exact_integral3}
\end{align}

Finally, we substitute $\eta =1/4$ into Eq.~\eqref{eq:sup:PS_exact_integral3} to study the typical fluctuations of the integrated spin current, obtaining
\begin{align} 
\mathbb{P}_{S}[ t^{1/4}\mathcal{J}_S,t]
&\sim  \int^{\infty}_{\mathcal{J}_S t^{ -1/4}} d\mathcal{J}_C ~ 
\sqrt{  \dfrac{ \mathcal{J}_C}{ \pi t^{1/2}\mathcal{J}_C^2 - \pi \mathcal{J}_S^2 }  } 
\exp\left[  -\dfrac{\pi}{2} \mathcal{J}_{C}^2 - \dfrac{\mathcal{J}_S^2}{2\mathcal{J}_C}\right] \\
&\sim t^{-1/4}  \dfrac{1}{\sqrt{\pi}} \int^{\infty}_{0} 
\dfrac{ d\mathcal{J}_C  }{  { \sqrt{\mathcal{J}_C }  } } 
\exp\left[  -\dfrac{\pi}{2} \mathcal{J}_{C}^2 - \dfrac{\mathcal{J}_S^2}{2\mathcal{J}_C}\right]. 
\end{align}
Therefore, we derive the following limiting expression for the scaled probability $t^{1/4}  \mathbb{P}_{S}[ t^{1/4}\mathcal{J}_S,t]$:
\begin{align} 
\lim_{t \rightarrow \infty} t^{1/4}  \mathbb{P}_{S}[ t^{1/4}\mathcal{J}_S,t] = \dfrac{1}{\sqrt{\pi}} \int^{\infty}_{0} 
\dfrac{ d\mathcal{J}_C  }{  { \sqrt{\mathcal{J}_C }  } } 
\exp\left[  -\dfrac{\pi}{2} \mathcal{J}_{C}^2 - \dfrac{\mathcal{J}_S^2}{2\mathcal{J}_C}\right].
\end{align}
This completes the derivation of Eq.~\eqref{eq:Ps_limit} of the main text.

\section{Numerical method for the $t\textrm{-}0$ model}\label{sec:numerics}
In Figs.~\ref{fig3} and~\ref{fig4} of the main text, we numerically investigate the M-Wright function for the $t\textrm{-}0$ model. In this section, we explain the details of our numerical calculations. The key feature of our approach is that we do not numerically solve the many-body Schrödinger equation for the $t\textrm{-}0$ model directly, as described below.

Our numerical calculation is based on the following equation:
\begin{align} 
\mathbb{P}_{S}[ \Delta S^z_{R},t] = \delta_{\Delta S^z_{R}, 0} \mathbb{P}_{C}[ 0, t] + 2 \sum^{N}_{\Delta N_{R} = 1}  \mathbb{P}_{C}[ \Delta N_{R},t]   \dfrac{ ~_{\Delta N_R} C_{(\Delta N_R + \Delta S^z_R)/2}  }{2^{\Delta N_R }} \dfrac{ 1 + (-1)^{\Delta N_R + \Delta S^z_R} }{2} \theta \left( \Delta N_R \geq \Delta S_{R}^z \right). 
\end{align} 
The function $\mathbb{P}_{C}[ \Delta N_{R},t]$ is determined by 
\begin{align} 
 \mathbb{P}_{C}[ \Delta N_{R},t] &=  \dfrac{1}{2 \pi {\rm i}} \oint_{C_r} dz~  \bar{G}_{\rm ff}(z,t) z^{-\Delta N_{R}-1}, \\
 \bar{G}_{\rm ff}(z,t) & = \det\left(  \delta_{m,n} +  \dfrac{z + z^{-1}-2}{4}    D_{m,n}(t) \right)_{m,n \in \{1,2,...,N\}}, \\
 {\rm i} \dfrac{d}{dt} D_{m,n}(t) &= D_{m+1,n}(t) + D_{m-1,n}(t) - D_{m,n+1}(t) - D_{m,n-1}(t)  
\end{align}
with the initial condition $D_{m,n}(0) = \delta_{m,n} \chi[ m \in \{-N+1,..., 0 \}]$ and the open boundary conditions $D_{N+1,n}(t) = 0$, $D_{m,N+1}(t) = 0$, $D_{-N,n}(t) = 0$, and $D_{m,-N}(t) = 0$.

\section{Slow convergence of $\mathbb{P}_{S}[0,t]$}
As shown in Fig.~\ref{fig3} of the main text, our numerical results show the slow convergence of $\mathbb{P}_{S}[\Delta S^z_R,t]$ around $S^z_R = 0$. In this section, we theoretically explain the origin of the slow convergence. In short, it originates from the asymptotic formula for the binomial coefficient $_N C_n$, Eq.~\eqref{eq:sup:combination}. The crucial point is that this formula is valid under the condition $N, n, ~ N-n \gg 1$, but this condition is not well satisfied around $\mathcal{J}_S =0$ unless the time $t$ is very large. We elaborate on this point below.

The starting point for deriving the M-Wright function is Eqs.~\eqref{eq:sup:PS_exact_integral0} and \eqref{eq:sup:PS_exact_integral1}. To obtain Eq.~\eqref{eq:sup:PS_exact_integral1}, we apply Eq.~\eqref{eq:sup:combination} to Eq.~\eqref{eq:sup:PS_exact_integral0}. For a $finite$ value of $\mathcal{J}_S$, $\Delta N_R$ is much larger than unity at sufficiently large times because we have the step function $\theta(\Delta N_{R} - \Delta S^z_{R})$ with $\Delta S_R^z = t^{1/4} \mathcal{J}_S$ in Eq.~\eqref{eq:sup:PS_exact_integral0}. Hence, we can safely use the asymptotic expression of Eq.~\eqref{eq:sup:combination}. 

This situation changes, however,  especially in the case with $\mathcal{J}_S=0$ because the step function is $\theta(\Delta N_{R})$ of Eq.~\eqref{eq:sup:PS_exact_integral0} and this does not guarantee that $\Delta N_{R}$ is large. More precisely, substituting $\Delta S_R^z = 0$ into Eq.~\eqref{eq:sup:PS_exact_integral0}, we obtain
\begin{align} 
\mathbb{P}_{S}[0,t] &= \mathbb{P}_{C}[ 0, t] + \sqrt{t}  \sum^{\infty}_{\Delta N_{R} = 1} \Delta \mathcal{J}_C ~ \mathbb{P}_{C}[ \Delta N_{R},t]  \dfrac{ {~_{\Delta N_R} C_{\Delta N_{R}/2} } }{2^{\Delta N_R }} \dfrac{1 + (-1)^{ \Delta N_R }}{2} \\
&= \mathbb{P}_{C}[ 0, t] + \sqrt{t}  \sum_{ \sqrt{t} \mathcal{J}_C \in \{2,4,...\}} \Delta \mathcal{J}_C ~ \mathbb{P}_{C}[ \sqrt{t} \mathcal{J}_C,t]  \dfrac{ {~_{\sqrt{t} \mathcal{J}_C} C_{\sqrt{t} \mathcal{J}_C/2} } }{2^{\sqrt{t} \mathcal{J}_C}} \\
&= \mathbb{P}_{C}[ 0, t] + \sqrt{t}  \sum_{ \sqrt{t} \mathcal{J}_C \in A_1} \Delta \mathcal{J}_C ~ \mathbb{P}_{C}[ \sqrt{t} \mathcal{J}_C,t]  \dfrac{ {~_{\sqrt{t} \mathcal{J}_C} C_{\sqrt{t} \mathcal{J}_C/2} } }{2^{\sqrt{t} \mathcal{J}_C}} + \sqrt{t}  \sum_{ \sqrt{t} \mathcal{J}_C \in A_2} \Delta \mathcal{J}_C ~ \mathbb{P}_{C}[ \sqrt{t} \mathcal{J}_C,t]  \dfrac{ {~_{\sqrt{t} \mathcal{J}_C} C_{\sqrt{t} \mathcal{J}_C/2} } }{2^{\sqrt{t} \mathcal{J}_C}} \label{eq:r1} \\
&\sim   \mathbb{P}_{C}[ 0, t] + 2  \sum_{ \sqrt{t}\mathcal{J}_C \in A_1} ~ \mathbb{P}_{C}[ \sqrt{t} \mathcal{J}_C,t]  \underbrace{ \dfrac{ {~_{\sqrt{t} \mathcal{J}_C} C_{\sqrt{t} \mathcal{J}_C/2} } }{2^{\sqrt{t} \mathcal{J}_C}} }_{\textrm{Eq.~\eqref{eq:sup:combination} cannot be used here.} } + \dfrac{ 1 }{\sqrt{\pi} t^{1/4} }   \int^{\infty}_{(a+2)/\sqrt{t}} d\mathcal{J}_C ~ \dfrac{ 1 }{ \sqrt\mathcal{J}_C} \exp\left(  -\dfrac{\pi}{2} \mathcal{J}_{C}^2\right).
\label{eq:r2}
\end{align}
Here, we have defined the sets $A_1 \coloneqq \{2,4,.., a\}$ and $A_2 \coloneqq \{a+2,a+4,... \}$ with an even constant $a \gg 1$. In Eq.~\eqref{eq:r1}, there are the two binomial coefficients and here we can apply Eq.~\eqref{eq:sup:combination} only to the second one because $\sqrt{t}\mathcal{J}_C$ is sufficiently larger than unity due to the set $A_2$. By contrast, we cannot use Eq.~\eqref{eq:sup:combination} for the first binomial coefficient since $\sqrt{t}\mathcal{J}_C$ is not necessarily large because of the set $A_1$. When one takes the long-time limit in Eq.~\eqref{eq:r2}, the first and second terms vanish because $\mathbb{P}_{C}[ \sqrt{t} \mathcal{J}_C,t]$ decays as $t^{-1/2}$ for large times under the condition where $\sqrt{t} \mathcal{J}_C$ is smaller than $a$. This behavior follows from Eq.~\eqref{eq:sup:PC_126}. Thus, we eventually derive the M-Wright function from the third term of Eq.~\eqref{eq:r2} in the case with $\mathcal{J}_S=0$. The above analysis shows that there are additional corrections that do not appear in the asymptotic analysis with $\mathcal{J}_S \neq 0$, namely the first and second terms in Eq.~\eqref{eq:r2}, which decay as $t^{-1/4}$ for the rescaled probability $t^{1/4}\mathbb{P}_{S}[0,t]$. Therefore, the presence of these terms leads to the slow convergence around $\mathcal{J}_S = 0$.

\section{Asymptotic analysis of the classical automaton}\label{sec:sup_CA}
We shall review the result reported in Ref.~\cite{Krajnik2022} of the main text, where the M-Wright function for the integrated spin current was derived exactly in a classical automaton. We can reproduce the previous result using an analytical method similar to our derivation of the M-Wright function for the $t\textrm{-}0$ model. This derivation is a bit different from the original one in the previous work and is convenient for discussing the difference between the $t\textrm{-}0$ model and the classical automaton.

\subsection{setup}
We review the setup for the classical automaton in Ref.~\cite{Krajnik2022} of the main text. Suppose that we have a one-dimensional lattice labeled by $\mathbb{Z}$ and assign each site three states, namely a up-spin state ($\uparrow$), a down-spin state ($\downarrow$), a vacant state ($\varnothing$). Then a whole state is given by a sequence of these symbols such as $(\cdots \uparrow\varnothing\downarrow\uparrow\uparrow\varnothing\downarrow\uparrow\varnothing \cdots)$.
Note that the authors in Ref.~\cite{Krajnik2022} of the main text use ``charge" instead of ``spin", but we here use ``spin" for a better comparison with the $t\textrm{-}0$ model. 
The time-evolution rule is constructed by a two-site propagator $\hat{U}_{j, j+1}^{\rm CA}$ which induces only the exchanges $(\uparrow \varnothing) \leftrightarrow (\varnothing \uparrow)$ and $(\downarrow \varnothing) \leftrightarrow (\varnothing \downarrow)$. Then, the time-evolution propagator $\hat{U}^{\rm CA}$ is defined by 
\begin{align} 
\hat{U}^{\rm CA} \coloneqq \prod_{j \in \mathbb{Z}} \hat{U}_{2j, 2j+1}^{\rm CA} \times \prod_{k \in \mathbb{Z}} \hat{U}_{2k-1, 2k}^{\rm CA}.
\end{align}
We can obtain a time-evolved state by repeatedly applying $\hat{U}^{\rm CA}$ to the initial state.

\subsection{Probability $\mathbb{P}_{S}^{\rm CA}[\Delta S_R^z, t]$ for the integrated spin current $\Delta S_R^z$}
We derive the explicit expression of $\mathbb{P}_{S}^{\rm CA}[\Delta S_R^z, t]$ for the integrated spin current $\Delta S_R^z$. 
According to Eq.~(5) with $b=0$ in Ref.~\cite{Krajnik2022} of the main text, the generating function $G_{S}^{\rm CA}(\lambda,t) $ associated with $\mathbb{P}_{S}^{\rm CA}[\Delta S_R^z, t]$ is given by
\begin{align} 
G_{S}^{\rm CA}(\lambda,t) 
&= \sum_{l=0}^t \sum_{r=0}^t ~_{t}C_l ~_{t}C_r \rho^{2t - l -r} \bar{\rho}^{l+r} \left( \cosh(\lambda)\right) ^{ |l-r| } \\ 
&= \sum_{\Delta N_{R}=-t}^t  \left( \sum_{l=0}^t \sum_{r=0}^t ~_{t}C_l ~_{t}C_r \rho^{2t - l -r} \bar{\rho}^{l+r} \delta_{\Delta N_R,  r-l} \right) \left( \cosh(\lambda)\right) ^{ |\Delta N_R| }, 
\end{align}
where $\rho$ and $\bar{\rho}$ are the probabilities for the particles and the vacancies.

To make a better comparison between the $t\textrm{-}0$ model and the classical automaton, we define the following function:
\begin{align} 
\mathbb{P}_{C}^{\rm CA}[ \Delta N_{R},t]  \coloneqq \sum_{l=0}^t \sum_{r=0}^t ~_{t}C_l ~_{t}C_r \rho^{2t - l -r} \bar{\rho}^{l+r} \delta_{\Delta N_R,r-l}.
\end{align}
This quantity can be interpreted as the probability that the transferred particle number to the right region ($j \geq 1$) is $\Delta N_R$ by definition. 
Then, we can express $G_{S}^{\rm CA}(\lambda,t) $ as
\begin{align} 
G_{S}^{\rm CA}(\lambda,t) = \sum_{\Delta N_R=-t}^t \mathbb{P}_{C}^{\rm CA}[ \Delta N_{R},t] \left( \cosh(\lambda)\right) ^{ | \Delta N_R| } \\ 
\end{align}
When the initial state has $\rho=1/2$ and $\bar{\rho}=1/2$, we have the relation $\mathbb{P}_{C}^{\rm CA}[ \Delta N_{R},t] = \mathbb{P}_{C}^{\rm CA}[ -\Delta N_{R},t]$.
Then, the generating function becomes
\begin{align} 
G_{S}^{\rm CA}(\lambda,t) = \mathbb{P}_{C}^{\rm CA}[0,t] + 2 \sum_{\Delta N_R =1 }^t \mathbb{P}_{C}^{\rm CA}[ \Delta N_{R},t] \left( \cosh(\lambda)\right) ^{  \Delta N_R }.
\label{eq:sup:SCAM}
\end{align}
In the remaining part of this section, we focus only on the case with $\rho=1/2$ and $\bar{\rho}=1/2$. 

We transform the generating function of Eq.~\eqref{eq:sup:SCAM} into the probability $\mathbb{P}_{S}^{\rm CA}[\Delta S_R^z, t]$, obtaining
\begin{align} 
\mathbb{P}_{S}^{\rm CA}[\Delta S_R^z, t] 
&= \dfrac{1}{2 \pi {\rm i}} \int_{C_r} dz~ G_{S}^{\rm CA}(\lambda,t) z^{-\Delta S_R^z-1} \\
&= \Delta_{\Delta S_R^z,0} \mathbb{P}_{C}^{\rm CA}[0, t] + 2 \sum_{\Delta N_R=1}^t \mathbb{P}_{C}^{\rm CA}[\Delta N_R, t]  \sum_{n=0}^{\Delta N_R} \dfrac{ ~_{\Delta N_R}C_{n} }{2^{\Delta N_R}} \delta_{\Delta S^z_{R},\Delta N_{R}-2n} \label{eq:sup:PSCA}
\end{align}
with $z=e^{\lambda}$.The expression of Eq.~\eqref{eq:sup:PSCA} is formally the same as Eq.~\eqref{eq:exactPS} of the $t\textrm{-}0$ model in the main text. 
This expression implies that the difference between the $t\textrm{-}0$ model and the classical automaton stems only from the charge probabilities, namely $\mathbb{P}_{C}[\Delta N_R, t]$ and $\mathbb{P}_{C}^{\rm CA}[\Delta N_R, t]$. In the following, we derive the M-Wright function using Eq.~\eqref{eq:sup:PSCA} and subsequently clarify this difference.

\subsection{Asymptotic analysis of $\mathbb{P}_{C}^{\rm CA}[ \Delta N_{R},t]$}
To derive the M-Wright function, it is important to investigate the asymptotic behavior of $\mathbb{P}_{C}^{\rm CA}[ \Delta N_{R},t]$. 
For this purpose, we use the generating function $G_{C}^{\rm CA}(\lambda,t)$ corresponding to $\mathbb{P}_{C}^{\rm CA}[ \Delta N_{R},t]$, given by 
\begin{align} 
G_{C}^{\rm CA}(\lambda,t) &\coloneqq \sum_{\Delta N_R = -t }^t \mathbb{P}_{C}^{\rm CA}[ \Delta N_{R},t]  e^{\lambda \Delta N_R} \\
&= \dfrac{1}{4^t} \sum_{\Delta N_R = -t }^t e^{\lambda \Delta N_R} \left( \sum_{l=0}^t \sum_{r=0}^t ~_{t}C_l ~_{t}C_r  \delta_{\Delta N_R, r-l } \right) \\
&= \dfrac{1}{4^t} \sum_{l=0}^t \sum_{r=0}^t ~_{t}C_l ~_{t}C_r    e^{\lambda \left( r-l \right)}.
\label{eq:sup:NCAM}
\end{align}

First, we consider the long-time dynamics ($t \gg1$) for the generating function $G_{C}^{\rm CA}(\lambda,t)$. Then, we have a small quantity $1/t$ and then introduce discrete variables $x_r = r/t$ and $y_l=l/t$. As a result, the summation of Eq.~\eqref{eq:sup:NCAM} can be transformed to be
\begin{align} 
G_{C}^{\rm CA}(\lambda,t) \sim \dfrac{t}{2 \pi \cdot 4^t}  \int_0^1 dx \int_{0}^1 dy  \dfrac{ e^{-t f(x,y)} }{\sqrt{xy(1-x)(1-y)}}, \label{eq:sup:GCCA}
\end{align}
where $x$ and $y$ are the continuous variables corresponding to $x_r$ and $x_l$, respectively, and $f(x,y)$ is defined by
\begin{align} 
f(x,y) \coloneqq \log \left( x^x y^y (1-x)^{1-x} (1-y)^{1-y} \right) + \lambda ( y - x ).
\end{align}
To derive the above, we use the asymptotic formula given by
\begin{align} 
~_{t}C_l \sim \sqrt{ \dfrac{t}{ 2 \pi l (t-l)} } \left( \dfrac{l}{t} \right)^{-l} \left( 1 - \dfrac{l}{t} \right)^{l-t}
\end{align}
for $t \gg 1$, $l \gg 1$, and $t-l \gg 1$. 

Second, we apply the method of steepest descent to Eq.~\eqref{eq:sup:GCCA}. 
Following the conventional procedure, we look for the local minimum point $(x_0,y_0)$ of the function $f(x,y)$, which dominantly contributes to the integral.
After elementary calculations, we obtain 
\begin{align} 
(x_0,y_0) = \left( \dfrac{1}{e^{-\lambda}+1},  \dfrac{1}{e^{\lambda}+1} \right).
\end{align}
Then, the function $f(x,y)$ can be expanded as
\begin{align} 
f(x,y) \sim f(x_0,y_0) + \dfrac{1}{2} \dfrac{(e^{\lambda}+1)^2}{e^{\lambda}} (x-x_0)^2 + \dfrac{1}{2} \dfrac{(e^{\lambda}+1)^2}{e^{\lambda}} (y-y_0)^2
\end{align}
Hence, we apply the method of steepest descent, getting the asymptotic expression of $G_{C}^{\rm CA}(\lambda,t) $ for $t \gg 1$ as
\begin{align} 
G_{C}^{\rm CA}(\lambda,t)  
&\sim \dfrac{ \exp\left( - 2t \log 2 -t f(x_0,y_0) + \log t\right) }{2 \pi \sqrt{x_0y_0(1-x_0)(1-y_0)}}  \int_{-\infty}^{\infty} dx \int_{-\infty}^{\infty} dy  \exp\left(-\dfrac{t}{2} \dfrac{(e^{\lambda}+1)^2}{e^{\lambda}} x^2 - \dfrac{t}{2} \dfrac{(e^{\lambda}+1)^2}{e^{\lambda}} y^2 \right)  \\
&= \exp\left[ - 2t \log 2 -t f(x_0,y_0) + \log t - \log \left(2 \pi \sqrt{x_0y_0(1-x_0)(1-y_0)}\right) + \log \left( \dfrac{2\pi e^{\lambda}}{t (e^{\lambda}+1)^2}  \right) \right]  \\
&\sim \exp \left[- 2t \log 2 -t  f(x_0,y_0) \right] \\
&= \left( \dfrac{1}{2} + \dfrac{1}{4} e^{\lambda} + \dfrac{1}{4} e^{-\lambda} \right)^t, 
\end{align}
where we leave only the leading term in the third line and the relation $x_0 + y_0$ in the last line. As a result, the generating function becomes
\begin{align} 
\log G_{C}^{\rm CA}(\lambda,t) \sim  t \log \left( \dfrac{1}{2} + \dfrac{1}{4} e^{\lambda} + \dfrac{1}{4} e^{-\lambda} \right)
\label{eq:sup:GCA}
\end{align}
for $t \gg 1$. From this expression, one can find that the first cumulant vanishes and the second one becomes $t/2$. 

Third, we derive the probability $\mathbb{P}_{C}^{\rm CA}[ \Delta N_{R},t]$ for the typical fluctuation using Eq.~\eqref{eq:sup:GCA}.
The detailed calculation is almost the same as that for the derivation of Eq.~\eqref{eq:sup:PC_asy}, so we do not repeat the calculation here. The resulting probability is given by 
\begin{align} 
\mathbb{P}_{C}^{\rm CA}[ \sqrt{t} \mathcal{J}_{C}, t] \sim \dfrac{1}{\sqrt{\pi t}} \exp\left(  -\mathcal{J}_{C}^2\right)
\label{eq:sup:PC_CA_asy}
\end{align}
for $t \gg 1$.

Finally, we comment on the difference between $\mathbb{P}_{C}^{\rm CA}[ \Delta N_{R},t]$ and $\mathbb{P}_{C}[ \Delta N_{R},t]$. The apparent difference is the factor of $\mathcal{J}_{C}^2$ in the exponential.
This reflects the different microscopic dynamics for the classical automaton and the $t\textrm{-}0$ models. The classical automaton has only the two propagating modes, while the $t\textrm{-}0$ model has the infinite propagating modes as the eigenstates of the free fermions are characterized by the continuous wavenumber ranging from $-\pi$ to $\pi$. This results in the difference of the factor, and one can understand this fact clearly by Sec.~\ref{sec:sup_BMFT}, where we derive the probability $\mathbb{P}_{C}[ \Delta N_{R},t]$ under a hydrodynamic assumption by highlighting the role of the infinitely many modes.

\subsection{Asymptotic analysis of $\mathbb{P}_{S}^{\rm CA} [\Delta S_R^z,t]$}
We derive the asymptotic expression of $\mathbb{P}_{S}^{\rm CA} [\Delta S_R^z,t]$. 
The derivation is the same as that for the $t\textrm{-}0$ model. 
The similar calculation with Eqs.~\eqref{eq:sup:PSCA} and \eqref{eq:sup:PC_CA_asy} leads to
\begin{align} 
\lim_{t \rightarrow \infty} t^{1/4}  \mathbb{P}_{S}^{\rm CA}[ t^{1/4}\mathcal{J}_S,t] 
&= \dfrac{\sqrt{2}}{\pi} \int^{\infty}_{0}  
\dfrac{ d\mathcal{J}_C }{  { \sqrt{\mathcal{J}_C }  } } 
\exp\left[  -\mathcal{J}_{C}^2 - \dfrac{\mathcal{J}_S^2}{2\mathcal{J}_C}\right].
\end{align}
This is identical to Eq.~(144) with $\Delta^2=1/4$ in Ref.~\cite{Krajnik2022} of the main text, and hence we can reproduce the previous result by employing the asymptotic analysis used in the $t\textrm{-}0$ model.

\section{Application of ballistic macroscopic fluctuation theory to spinless free fermions}\label{sec:sup_BMFT}
We shall give another derivation of Eq.~\eqref{eq:sup:PC_asy} for the probability $\mathbb{P}_{C}[ \Delta N_R, t]$ in Eq.~\eqref{eq:exactPS} of the main text from the viewpoint of hydrodynamics, namely BMFT in Refs.~\cite{BMFT1,BMFT2,Friedrich2025,Kethepalli2025,Andrew2026} of the main text, although Eq.~\eqref{eq:sup:PC_asy} can be derived in the exact manner. As discussed in the main text, this hydrodynamic derivation clarifies the distinction between the $t\textrm{-}0$ model and the classical automaton for Ref.~\cite{Krajnik2022} of the main text, and one can obtain a physically intuitive understanding of the emergence of the M-Wright function for the integrated spin current.

We now apply BMFT to the $t\textrm{-}0$ model to derive the asymptotic probability $\mathbb{P}_{C}[ \Delta N_R ,t]$. As described in Sec.~\ref{sec:sup_Pc}, the probability is determined by the dynamics of free fermions. Thus, the application of BMFT to it is straightforward because solving the equations of motion for GHD in Refs.~\cite{Olalla2016,Bertini2016,Doyon2020_Rev,Alba2021,Bouchoule2022,Essler2023} of the main text is trivial. After applying BMFT to the free fermions, we obtain the following equations for the generating function $G^{\rm BMFT}_{C}(\lambda,t)$ of $\mathbb{P}_{C}[\Delta N_R,t]$ under the BMFT approximation:
\begin{align} 
& \log G^{\rm BMFT}_{C}(\lambda,t) = \dfrac{t}{2 \pi} \int_{-\pi}^{\pi} dk F^{\rm BMFT}_C(\lambda,k,t), \label{eq:E_logG}\\
& F^{\rm BMFT}_{C}(\lambda,k,t) = \dfrac{1}{t} \int_0^{\lambda} d\lambda' \int_0^t d \tau v(k) n(k,0,\tau), \label{eq:E_F}
\end{align}
where we define $v(k) \coloneqq 2\sin k$ and the occupation number as
\begin{align} 
n(k,x,t) \coloneqq  \dfrac{1}{e^{\beta(k,x,t)}+1} \label{eq:E_ON}
\end{align} 
with the local inverse temperature $\beta(k,x,t)$. 
The inverse temperature is determined by 
\begin{align} 
&\dfrac{\partial}{\partial \tau} \beta(k,x,\tau) + v(k) \dfrac{\partial}{\partial x} \beta(k,x,\tau) = 0, \label{eq:E_BMFT1}\\
&\dfrac{\partial}{\partial \tau} H(k,x,\tau) + v(k) \dfrac{\partial}{\partial x} H(k,x,\tau) = 0,  \label{eq:E_BMFT2}\\
& H(k,x,0) = \lambda \theta(x) - \beta(x,k,0),  \label{eq:E_BMFT3}\\
& H(k,x,t) = \lambda \theta(x), \label{eq:E_BMFT4}
\end{align}
where $H(k,x,t)$ is the auxiliary field for the inverse temperature and $\theta(x)$ is the step function. These equations can be derived mainly using the results of Sec. 5.3 in Ref.~\cite{BMFT2} of the main text. Solving the equations for $\beta(k,x,\tau)$ and $H(k,x,\tau)$, we can calculate the cumulant generating function $\log G^{\rm BMFT}_{C}(\lambda,t)$ by Eqs.~\eqref{eq:E_logG}, \eqref{eq:E_F}, and \eqref{eq:E_ON}. 

First, we solve the equation of motion for $\beta(k,x,\tau)$. 
To get the initial condition $\beta(k,x,0)$, we solve Eq.~\eqref{eq:E_BMFT2} and then obtain
\begin{align} 
 H(k,x,\tau) = f(k,x-v(k)\tau) 
\end{align}
with an unknown function $f(\bullet)$.
This equation with Eq.~\eqref{eq:E_BMFT4} leads to 
\begin{align} 
 H(k,x,\tau) = \lambda \theta(x + v(k)t - v(k)\tau). 
\end{align}
Thus, we can derive the initial condition $\beta(x,k,0)$ via Eq.~\eqref{eq:E_BMFT3} as
\begin{align} 
 \beta(k,x,0) = \lambda \theta(x) - \lambda \theta(x + v(k)t). 
\end{align}
Using this initial condition, we can solve Eq.~\eqref{eq:E_BMFT1}, getting
\begin{align} 
 \beta(k,x,\tau) = \lambda \theta(x-v(k) \tau ) - \lambda \theta(x + v(k)t - v(k)\tau). \label{eq:E_BMFT5}
\end{align}

Second, we derive the expression of $\beta(k,0,\tau)$ for $v(k) \geq 0$ with Eq.~\eqref{eq:E_BMFT5} because the computation of Eq.~\eqref{eq:E_F} via Eq.~\eqref{eq:E_ON} needs the inverse temperature only at the origin ($x=0$).
The detailed expression becomes 
\begin{align} 
\beta(k,0,\tau) =
 \begin{dcases}
     -\lambda  & (0 \leq \tau \leq t) \\
     0              & (t < \tau ) \\
     0              & ( \tau < 0).
  \end{dcases}
\end{align}
Thus, the definition of the occupation number of Eq.~\eqref{eq:E_ON} leads to  
\begin{align} 
n(k,0,\tau) =
 \begin{dcases}
     \dfrac{e^{\lambda}}{1+e^{\lambda}}  & (0 \leq \tau \leq t) \\
     1/2              & (t < \tau ) \\
     1/2              & ( \tau < 0).
  \end{dcases}
\end{align}
Hence, we get, for $v(k) \geq 0$, 
\begin{align} 
F^{\rm BMFT}_{C}(\lambda,k,t) = v(k) \log \left( \dfrac{1}{2} + \dfrac{1}{2} e^{\lambda} \right).
\label{eq:E_FP}
\end{align}

Third, we consider the case with $v(k) < 0$. 
Then, using Eq.~\eqref{eq:E_BMFT5}, we obtain, in the same manner as above,  
\begin{align} 
\beta(k,0,\tau) =
 \begin{dcases}
     \lambda  & (0 \leq \tau \leq t) \\
     0              & (t < \tau ) \\
     0              & ( \tau < 0).
  \end{dcases}
\end{align}
This readily leads to 
\begin{align} 
n(k,0,\tau) =
 \begin{dcases}
     \dfrac{e^{-\lambda}}{1+e^{-\lambda}}  & (0 \leq \tau \leq t) \\
     1/2              & (t < \tau ) \\
     1/2              & ( \tau < 0).
  \end{dcases}
\end{align}
As a result, we get, for $v(k) < 0$, 
\begin{align} 
F^{\rm BMFT}_{C}(\lambda,k,t) = -v(k) \log\left( \dfrac{1}{2} + \dfrac{1}{2}e^{-\lambda} \right).
\label{eq:E_FN}
\end{align}

Finally, we substitute Eqs.~\eqref{eq:E_FP} and \eqref{eq:E_FN} into Eq.~\eqref{eq:E_logG}, obtaining
\begin{align} 
\log G^{\rm BMFT}_{C}(\lambda,t) 
&= \dfrac{t}{2 \pi} \int_{0}^{\pi} dk v(k) \log \left( \dfrac{1}{2} + \dfrac{1}{2} e^{\lambda} \right) \left( \dfrac{1}{2} + \dfrac{1}{2} e^{-\lambda} \right) \\
&= \dfrac{2t}{\pi} \log \left( \dfrac{1}{2} + \dfrac{1}{4}e^{\lambda} + \dfrac{1}{4}e^{-\lambda} \right).
\end{align}
This is identical to the cumulant generating function obtained in the asymptotic analysis based on the exact expression (see Eq.~\eqref{eq:logG_asy}). Note that the essentially same generating function was reported in Ref.~\cite{Schonhammer2007} of the main text, where scattering matrix techniques were used. Therefore, following the same calculations given in Sec.~\ref{subsec:sup_Pc}, we can derive the probability of  Eq.~\eqref{eq:sup:PC_asy} for the typical fluctuations of the integrated charge current.

This hydrodynamic calculation clearly shows that the infinite number of propagating modes $k$ contribute to the cumulant generating function $G^{\rm BMFT}_{C}(\lambda,t)$. This situation is very different from the classical automaton in Ref.~\cite{Krajnik2022} of the main text, where there are only two propagating modes. This point is explained in the main text and Sec.~\ref{sec:sup_CA}. 

\section{Discussion on the Fermi-Hubbard model with finite interactions}\label{sec:discussion}

We now discuss how a finite interaction strength in the Fermi–Hubbard model affects the emergence of the M-Wright function. Before considering this effect, we first describe the transport properties of the $t\textrm{-}0$ model, which is the key ingredient in the later discussion. As shown in the main text, the probability for the typical fluctuations of the integrated spin current obeys the scaling form with the scaled current $\mathcal{J}_S = \Delta S^z_R/t^{1/4}$. Crucially, the $t^{1/4}$ scaling is the consequence of diffusive transport in the $t\textrm{-}0$ model.

In light of this diffusive behavior in the $t\textrm{-}0$ model, we expect the absence of the M-Wright function for the $finite$ interaction case in the $long\textrm{-}time~limit$ because the previous numerical studies [see Fig.~5 in Ref.~\cite{Michele2020} of the main text and Fig.~6 in Ref.~\cite{Moca2023} of the main text] found KPZ-like superdiffusion in the Fermi-Hubbard model with finite interactions and then this behavior means that the description of the $t\textrm{-}0$ model is not expected to remain valid in the $long\textrm{-}time~limit$. Furthermore, the numerical study for classical spin chains in Ref.~\cite{Krajnik2024-2} of the main text reported the absence of the M-Wright function when the spin transport is KPZ-like superdiffusion. However, we emphasize that these facts do not necessarily imply the absence of the M-Wright function in the finite interaction case; the M-Wright function can still appear in the transient dynamics. 

To elaborate on the above in detail, let us consider a system described by the Fermi-Hubbard model with large interactions. The Hamiltonian of the Fermi-Hubbard model is given by
\begin{align} 
\hat{H}_{\rm FH} \coloneqq -\sum_{\sigma \in \{ \uparrow, \downarrow \}} \sum_{j} \left( \hat{f}_{\sigma,j+1}^{\dagger} \hat{f}_{\sigma,j} + {\rm h.c.}  \right) + U \sum_{j} \hat{n}_{\uparrow,j} \hat{n}_{\downarrow,j}, 
\label{FH}
\end{align}
where $\hat{f}_{\sigma,j}$ is the fermion annihilation operator with spin $\sigma$ at site $j$, $U$ is the interaction strength, and $\hat{n}_{\sigma,j} = \hat{f}_{\sigma,j}^{\dagger} \hat{f}_{\sigma,j}$ is the number operator with spin $\sigma$ at site $j$. When $U$ is much larger than unity, one can derive the effective model, namely the $t\textrm{-}J$ model, whose Hamiltonian is given by~\cite{Essler2005}
\begin{align} 
\hat{H}_{t \textrm{-} J} &\coloneqq \underbrace{- \hat{P}\sum_{\sigma \in \{ \uparrow, \downarrow \}} \sum_{j} \left( \hat{f}_{\sigma,j+1}^{\dagger} \hat{f}_{\sigma,j} + {\rm h.c.}  \right) \hat{P}}_{\textrm{$t \textrm{-} 0$  Hamiltonian} } \nonumber \\
&+ \underbrace{\dfrac{4}{U} \hat{P} \sum_{j}\left( \hat{\vec{S}}_j \cdot \hat{\vec{S}}_{j+1} - \dfrac{1}{4} \hat{n}_{j} \hat{n}_{j+1} \right) \hat{P} +  \dfrac{2}{U}\sum_{j,k,l} t_{j,k}t_{k,l} \left( \sum_{\alpha,\beta,\gamma}\hat{f}^{\dagger}_{\beta,j} \sigma^{\alpha}_{\beta,\gamma}\hat{f}_{\gamma,l} \hat{S}_k^{\alpha}  - \dfrac{1}{2} \sum_{\beta}\hat{f}^{\dagger}_{\beta,j} \hat{f}_{\beta,l} \hat{n}_k \right) }_{\textrm{perturbed Hamiltonian}}, 
\label{tJ}
\end{align}
where $\hat{P}$ is the projector for the Hilbert space without double occupancies, $\hat{\vec{S}}_j$ is the spin operator at site $j$, $\hat{n}_j = \hat{n}_{\uparrow,j} + \hat{n}_{\downarrow,j}$ is the number operator at site $j$, $\sigma^{\alpha}_{\beta,\gamma}$ is a matrix element of the Pauli matrix with $\alpha \in \{x,y,z\}$, and we have defined $t_{j,k} \coloneqq -\delta_{j,k+1} - \delta_{j,k-1}$ . In the following, we set the Dirac's constant $\hbar$ to be unity.

Under this setup, we consider the two time scales for the effective Hamiltonian of Eq.~\eqref{tJ}, discussing the time range for the appearance of the M-Wright function. One time scale is $1$ for the particle hopping induced by the $t\textrm{-}0$ Hamiltonian and the other is the timescale $U$ for the perturbed Hamiltonian. Hence, when the time range of our interest is $1 \ll t \ll U$, the dynamics should be well described by the $t\textrm{-}0$ Hamiltonian, and then we expect that the behavior of the M-Wright function associated with the diffusive transport can transiently appear in this time range. On the other hand, when the time range is $U \ll t$, the effect of the perturbed Hamiltonian can become important and then the M-Wright function is expected to disappear. To summarize the main points, we expect the following in the large $U$ case:
\begin{itemize}
    \item {\bf Transient-time range ($1 \ll t \ll U$)}\\
    The dynamics is well governed by the $t\textrm{-}0$ model and the M-Wright function is expected to appear transiently.
    
    \item {\bf Long-time range ($U \ll t $)} \\
    The description of the $t\textrm{-}0$ model does not work well and the M-Wright function is not expected to persist.
\end{itemize}

\section{Details for the hydrodynamic derivation of the M-Wright function}
\subsection{Derivation of the effective velocity via a Gaudin matrix}\label{sec:veff}
We derive the effective velocity of Eqs.~\eqref{eq:EM_veff_c} and \eqref{eq:EM_veff_s} of End Matter using a Gaudin matrix. The method used here is based on Refs.~\cite{Borsi2020,Pozsgay2020,Borsi2021} of the main text.

We consider the setup used in End Matter of the main text. Then, the Bethe equation of the $t\textrm{-}0$ model~\cite{Izergin1998,Abarenkova2001,Essler2005} is given by 
\begin{align} 
e^{{\rm i}  \lambda^{\rm (c)}_a L} &= e^{ {\rm i} \Theta }~~~~~~~~~~(a \in \{1,2,...,N_p \}), \\
e^{{\rm i}  \lambda^{\rm (s)}_b N_p} &= (-1)^{M+1}~~~(b \in \{1,2,...,M \}) \\
\end{align}
with $\Theta = \sum_{b=1}^M \lambda^{\rm (s)}_b$. 
When taking the logarithm of the above, we get 
\begin{align} 
 &\lambda^{\rm (c)}_a L - \sum_{b=1}^M \lambda^{\rm (s)}_b = 2\pi I^{\rm (c)}_a, \\
 &\lambda^{\rm (s)}_b N_p - \pi(M+1) = 2\pi I^{\rm (s)}_b, 
\end{align}
where $ I^{\rm (c)}_a \in \mathbb{Z}$ and $ I^{\rm (s)}_b \in \mathbb{Z}$ are the Bethe quantum numbers.

According to Refs.~\cite{Borsi2020,Pozsgay2020,Borsi2021} of the main text, there is a conjecture for the form factor of current operators in multicomponent systems.
To be specific, the averages of a current operator $\hat{J}_{m}$ corresponding to a local conserved quantity $\hat{Q}_{m}$ is given by 
\begin{align} 
\bra{ \{ \lambda ^{ (\alpha )}_a \} } \hat{J}_{m} \ket{ \{\lambda ^{(\alpha )}_a \} } 
 &=  \left(\bm{e}'\right)^{\rm T} \cdot G^{-1} \cdot \bm{h}_{m} \label{eq:S_formfactor}  \\
 &= \dfrac{1}{L}\sum_{\alpha, a}v_{\rm eff}(\lambda ^{ (\alpha )}_a) h_m^{ (\alpha )} ( \lambda ^{ (\alpha )}_a ), 
\end{align}
where $h_m^{ (\alpha )}$ is the one-particle eigenvalue of $\hat{Q}_m$ and $G^{-1}$ is the inverse of the Gaudin matrix $G_{(\alpha,j),(\beta, k)}~~(\alpha, \beta \in \{\rm c, s\})$ defined by
\begin{align} 
 G_{( {\rm c}, j), ({\rm c}, k)} &\coloneqq 2\pi \dfrac{\partial  I^{\rm (c)}_j }{ \partial \lambda^{\rm (c)}_k }~~~~~(j,k \in \{1,2,...,N_p \}),  \\
 G_{( {\rm c}, j), ({\rm s}, k)} &\coloneqq 2\pi \dfrac{\partial  I^{\rm (c)}_j }{ \partial \lambda^{\rm (s)}_k }~~~~~(j \in \{1,2,...,N_p \}, ~k \in \{1,2,...,M \}), \\
 G_{( {\rm s}, j), ({\rm c}, k)} &\coloneqq 2\pi \dfrac{\partial  I^{\rm (s)}_j }{ \partial \lambda^{\rm (c)}_k }~~~~~(j \in \{1,2,...,M \},~k \in \{1,2,...,N_p \}), \\
 G_{( {\rm s}, j), ({\rm s}, k)} &\coloneqq 2\pi \dfrac{\partial  I^{\rm (s)}_j }{ \partial \lambda^{\rm (s)}_k }~~~~~(j,k \in \{1,2,...,M \}).
\end{align}
Here, we assume that the order of the labels for the Gaudin matrix is defined as $({\rm c},1) < ({\rm c},2) < .... < ({\rm c},L) < ({\rm s},1) < ({\rm s},2) < ... < ({\rm s},N_p).$
The vector $\bm{e}'$ and the effective velocity $\bm{v}_{\rm eff}(\lambda_a^{\rm (\alpha)})$ are defined by 
\begin{align} 
\bm{e}' &\coloneqq ( \underbrace{2 \sin( \lambda^{\rm (c)}_1), ..., 2 \sin( \lambda^{\rm (c)}_{N_p})}_{N_p}, \underbrace{0, ..., 0}_{M})^{\rm T}, \\
{v}_{\rm eff}(\lambda_a^{\rm (\alpha)}) &\coloneqq L \sum_{b, \beta} e' ( \lambda_b^{\rm (\beta)}  ) G^{-1}_{ ({\rm \beta}, b), ({\rm \alpha}, a)}.
\end{align}
The detailed expression of the Gaudin matrix is given by 
\begin{align} 
G = 
\begin{pmatrix}
G_{1} & G_{3} \\
O       & G_{2} \\
\end{pmatrix}
,
\end{align}
where $O$ is the $M \times N_p$ the matrices whose entries are zero and $G_{1}$, $G_{2}$, and $G_{3}$ are defined by
\begin{align} 
&G_1 \coloneqq {\rm diag}( \underbrace{L, L, ..., L}_{N_p} )  \eqqcolon  L E_N, \\
&G_2 \coloneqq {\rm diag}( \underbrace{N_p, N_p, ..., N_p}_M ) \eqqcolon N_p E_M, \\ 
&G_3 \coloneqq 
\begin{pmatrix}
-1 & -1 & \cdots & -1 \\
-1 & -1 & \cdots & -1 \\
\vdots & \vdots & \ddots & \vdots \\
-1 & -1 & \cdots & -1
\end{pmatrix}
.
\end{align}
Here $G_3$ is $N_p \times M$ matrix, and $E_{N_p}$ and $E_M$ are the unit $N_p \times N_p$ and $M \times M$ matrices, respectively.  
Thus, the inverse of $G$ is given by
\begin{align} 
G^{-1} &= 
\begin{pmatrix}
G_{1}^{-1} & - G_{1}^{-1}G_{3}G_{2}^{-1} \\
O       & G_{2}^{-1} \\
\end{pmatrix}
\\
&= 
\begin{pmatrix}
L^{-1} E_{N_p} & - L^{-1}{N_p}^{-1} G_{3} \\
O       & {N_p}^{-1} E_M \\
\end{pmatrix}
,
\end{align}
where $G_1^{-1}$ and $G_2^{-1}$ are the inverses of $G_1$ and $G_2$. 
As a result, we obtain
\begin{align} 
\bm{v}_{\rm eff}(\bm{\lambda}) = \left( \underbrace{2 \sin( \lambda^{\rm (c)}_1), ..., 2 \sin( \lambda^{\rm (c)}_{N_p})}_{N_p}, \underbrace{\dfrac{2}{N_p} \sum_a \sin( \lambda^{\rm (c)}_a), ..., \dfrac{2}{N_p} \sum_a \sin( \lambda^{\rm (c)}_a)}_M \right)^T.
\label{eq:sup:veff}
\end{align}
Finally, we take the thermodynamic limit in Eq.~\eqref{eq:sup:veff}, obtaining 
\begin{align} 
 v_{\rm eff}^{\rm c}(\lambda) &= 2 \sin( \lambda ), \\
 v_{\rm eff}^{\rm s}(\lambda) &= \lim_{\rm th} \dfrac{ \displaystyle \dfrac{2}{L} \sum_a\sin( \lambda_a )}{ \displaystyle \dfrac{N_p}{L}} \\
 &= \dfrac{ \displaystyle \int_{-\pi}^{\pi} d \lambda  ~2 \sin(\lambda) \rho^{\rm (c)} (\lambda) }{ \displaystyle\int_{-\pi}^{\pi} d \lambda~  \rho^{\rm (c)}(\lambda)  }.
\end{align}
These expressions are identical to the effective velocities of Eqs.~\eqref{eq:EM_veff_c} and \eqref{eq:EM_veff_s} in End Matter.

\subsection{Detailed explanation for the initial correlations of Eqs.~\eqref{eq:EM_Crr_SS}, \eqref{eq:EM_Crr_CC}, and \eqref{eq:EM_Crr_CS} }\label{sec:EM_initial}
We consider the initial correlations for the charge and spin of the $t\textrm{-}0$ model. 
The initial state addressed here is the grand-canonical density matrix of Eq.~\eqref{eq:initial2} of the main text, which is given by 
\begin{align} 
\hat{\rho}(0) = \frac{1}{Z} e^{-\beta (\hat{H} - \mu \hat{N} ) },
\end{align}
where $\beta$ and $\mu$ are the inverse nonzero temperature and the chemical potential, and we define $\hat{N} = \hat{P} \sum_{j \in \Lambda} \left( \hat{n}_{\uparrow,j} + \hat{n}_{\downarrow,j} \right) \hat{P} $ and $Z = {\rm Tr}~e^{-\beta (\hat{H} - \mu \hat{N} )}$. 
To simplify notions in the following calculations, we introduce $ \langle \bullet \rangle_{\rm ini} \coloneqq {\rm Tr} \left[  \hat{\rho}(0) \bullet \right] $.

\subsubsection{Initial correlation of Eq.~\eqref{eq:EM_Crr_SS}}
The Hamiltonian $\hat{H}$ does not have any terms with $\hat{s}^z_j$, and thus the spin state is described by the infinite temperature state.
As a result, we derive 
\begin{align} 
\langle  \hat{s}^z_j \rangle_{\rm ini} &= 0, \\
\langle  \hat{s}^z_j \hat{s}^z_k \rangle_{\rm ini} &=  \langle  \hat{n}_j  \rangle_{\rm ini} \delta_{j,k}, 
\end{align}
where we use $\hat{s}^z_j = \hat{n}_{\uparrow,j} - \hat{n}_{\downarrow,j}$ and the constraint $\hat{n}_{\uparrow,j} \hat{n}_{\downarrow,j} = 0$ for no double occupancy. 
This readily leads to
\begin{align} 
\langle  \hat{s}^z_j \hat{s}^z_k \rangle_{\rm ini} - \langle  \hat{s}^z_j \rangle_{\rm ini} \langle  \hat{s}^z_k \rangle_{\rm ini} = \langle  \hat{n}_j  \rangle_{\rm ini} \delta_{j,k}.
\end{align}
Then, we expect 
\begin{align} 
\langle \delta \tilde{S}^z(x,0) \delta \tilde{S}^z(y,0)  \rangle = C_1 \delta(x-y)
\end{align}
with a constant $C_1$. 

We next determine $C_1$ using a microscopic calculation. The hydrodynamic variable $S^z(x,0)$ is the averaged spin density at a position $x$ and thus the operator counterpart is 
\begin{align} 
\hat{S}^z \coloneqq \dfrac{1}{2N}\sum_{j \in \Lambda} \hat{s}^{z}_j.
\end{align}
We can compute the spin fluctuation, getting
\begin{align} 
\langle  (\hat{S}^z )^2 \rangle_{\rm ini} - \langle  \hat{S}^z \rangle_{\rm ini} \langle  \hat{S}^z \rangle_{\rm ini} 
 &= \dfrac{1}{4N^2} \sum_{j \in \Lambda} \langle\hat{n}_j\rangle_{\rm ini}  \\
 &\sim \dfrac{1}{2N} \int_{-\pi}^{\pi} d\lambda \rho^{\rm (c)}(\lambda)~~(N \gg 1). \label{eq:sup:Css1}
\end{align}
Here we note that $2N$ in Eq.~\eqref{eq:sup:Css1} is the system size and can be regarded as the typical spatial scale of hydrodynamics. 
Thus, $2N$ corresponds to $\sqrt{\tau}$ in Eq.~\eqref{eq:EM_scaled_density} and then we get
\begin{align} 
C_1 = \int_{-\pi}^{\pi} d\lambda \rho^{\rm (c)}(\lambda).
\end{align}
This completes the explanation of Eq.~\eqref{eq:EM_Crr_SS}.

\subsubsection{Initial correlation of Eq.~\eqref{eq:EM_Crr_CC}}
We can show that the correlation function $ \langle  \hat{n}_j \hat{n}_k \rangle_{\rm ini} - \langle  \hat{n}_j \rangle_{\rm ini} \langle  \hat{n}_k \rangle_{\rm ini}$ at nonzero temperature decays exponentially as a function of $|j-k|$. Furthermore, the rapidity $\lambda$ in the thermodynamic Bethe ansatz can be interpreted as the wavenumber of the free fermions. 
Thus, in the description of hydrodynamics, we expect 
\begin{align} 
\langle \delta \tilde{\rho}^{\rm (c)}(\lambda,x,0) \delta \tilde{\rho}^{\rm (c)}(\nu,y,0)  \rangle = F_{\rm charge}(\lambda,\nu) \delta(x-y) .
\label{eq:sup:Crr_CC1}
\end{align} 
with a function $F_{\rm charge}(\lambda,\nu)$.

We next determine the form of $F(\lambda,\mu)$ in Eq.~\eqref{eq:sup:Crr_CC1} using the microscopic calculation of the $t\textrm{-}0$ model. 
As commented above, the rapidity $\lambda$ corresponds to the wavenumber of the free fermions, and thus the microscopic object of ${\rho}^{\rm (c)}(\lambda)$ is $\hat{N}(\lambda)$ defined by
\begin{align} 
\hat{N}(\lambda_{\alpha}) \coloneqq \dfrac{1}{2\pi}   \hat{A}^{\dagger}(\lambda_{\alpha}) \hat{A}(\lambda_{\alpha}), 
\end{align} 
where the operator $\hat{A}(\lambda_{\alpha})$ is defined by
\begin{align} 
\hat{A}(\lambda_{\alpha}) \coloneqq \dfrac{1}{\sqrt{2 N}} \sum_{j \in \Lambda} \hat{a}_j e^{ -{\rm i} \lambda_{\alpha} j}
\end{align} 
with $\lambda_{\alpha} = \pi \alpha /N~(\alpha \in \Lambda)$. Using this notation, we can show
\begin{align} 
\dfrac{1}{2N} \sum_{j \in \Lambda}\hat{a}^{\dagger}_j \hat{a}_j = \dfrac{1}{2N} \sum_{\alpha \in \Lambda}\hat{A}^{\dagger}(\lambda_{\alpha}) \hat{A}(\lambda_{\alpha}) 
\xrightarrow[N \rightarrow \infty]{}  \int_{-\pi}^{\pi}  \dfrac{ d \lambda }{2\pi}  \hat{A}^{\dagger}(\lambda) \hat{A}(\lambda), 
\end{align} 
which explains the reason why $\hat{N}(\lambda_{\alpha})$ corresponds to ${\rho}^{\rm (c)}(\lambda)$.
Under this setup, we can diagonalize the Hamiltonian $\hat{H}_{\rm ff}$ under the periodic boundary condition
\begin{align} 
\hat{H}_{\rm ff} = \sum_{\alpha \in \Lambda} ( -2 \cos(\lambda_{\alpha}) ) \hat{A}^{\dagger}(\lambda_{\alpha}) \hat{A}(\lambda_{\alpha}).
\end{align} 
Then, by definition, we obtain 
\begin{align} 
\langle  \hat{N}(\lambda_{\alpha}) \rangle_{\rm ini} &= \dfrac{ \displaystyle {\rm Tr} \left[ \hat{A}^{\dagger}(\lambda_{\alpha}) \hat{A}(\lambda_{\alpha})  \exp\left( {  \sum_{\alpha \in \Lambda} ( 2 \beta \cos(\lambda_{\alpha}) + \beta\mu + \log 2) \hat{A}^{\dagger}(\lambda_{\alpha}) \hat{A}(\lambda_{\alpha})  }   \right)  \right] }
{\displaystyle 2\pi  {\rm Tr} \left[  \exp\left( {  \sum_{\alpha \in \Lambda} ( 2 \beta \cos(\lambda_{\alpha}) + \beta\mu +  \log 2) \hat{A}^{\dagger}(\lambda_{\alpha}) \hat{A}(\lambda_{\alpha})  }   \right)  \right] } \\
& = \dfrac{1}{2\pi} \dfrac{2}{\displaystyle e^{-\beta(2\cos(\lambda_{\alpha} )+\mu)} + 2} \\
& = \rho^{\rm (c)} (\lambda_{\alpha}). 
\end{align} 
We follow a similar calculation with the Wick theorem, obtaining 
\begin{align} 
\langle  \hat{N}(\lambda_{\alpha}) \hat{N}(\nu_{\beta}) \rangle_{\rm ini} - \langle  \hat{N}(\lambda_{\alpha}) \rangle_{\rm ini} \langle \hat{N}(\nu_{\beta}) \rangle_{\rm ini} 
&=  \dfrac{ \delta_{\lambda_{\alpha}, \nu_{\beta}} }{2 \pi}  \dfrac{1}{2 \pi} n^{\rm (c)} (\lambda_{\alpha} ) (1 - n^{\rm (c)} (\nu_{\beta} ) ) \\
&\sim \dfrac{1}{2N} \delta(\lambda - \nu)  \dfrac{1}{2 \pi} n^{\rm (c)} (\lambda ) (1 - n^{\rm (c)} (\nu ) ) ~~(N \gg 1).
\end{align} 
As explained in Eq.~\eqref{eq:sup:Css1}, the system size $2N$ corresponds to the spatial scale $\sqrt{\tau}$ of the hydrodynamics and thus we can get 
\begin{align} 
F_{\rm charge}(\lambda,\nu) = \delta(\lambda-\nu) C_2(\lambda)
\end{align} 
with 
\begin{align} 
C_2(\lambda) = \dfrac{1}{2\pi} n^{\rm (c)}(\lambda) (1 - n^{\rm (c) }(\lambda) ).
\end{align} 
This completes the explanation of Eq.~\eqref{eq:EM_Crr_CC}.

\subsubsection{Initial correlation of Eq.~\eqref{eq:EM_Crr_CS}}
The Hamiltonian of the $t\textrm{-}0$ model and the particle number operator $\hat{n}_j = \hat{n}_{\uparrow,j} + \hat{n}_{\downarrow,j}$ are invariant under the transformation of $z \rightarrow -z$, while the spin operator $\hat{s}^z_j = \hat{n}_{\uparrow,j} - \hat{n}_{\downarrow,j}$ is transformed into $- \hat{s}^z_j$.
Then, the correlation for the charge and spin becomes
\begin{align} 
\langle  \hat{n}_j \hat{s}^z_j \rangle_{\rm ini} - \langle  \hat{n}_j \rangle_{\rm ini} \langle  \hat{s}^z_j \rangle_{\rm ini} = 0.
\end{align} 
Hence, we expect that the initial correlation in the hydrodynamics is given by Eq.~\eqref{eq:EM_Crr_CS} of the main text:
\begin{align} 
\langle \delta \tilde{\rho}^{\rm (c)}(\lambda,x,0) \delta \tilde{S}^z(y,0)   \rangle &= 0.
\end{align}

\subsection{Detailed calculation for the functional integral of Eq.~\eqref{eq:EM_PI}}\label{sec:EM_PI}
Our starting point is the functional integral of Eq.~\eqref{eq:EM_PI} in End Matter. 
First, we rewrite the functional integral as follows: 
\begin{align} 
\langle e^{ z J(\tau)} \rangle \sim \dfrac{1}{Z_{\rm charge} Z_{\rm spin}} \int \mathcal{D} \delta\tilde{S}^z(x,0)  \mathcal{D} \delta\tilde{\rho}^{\rm (c)}(\lambda,x,0) \exp \bigg[ &- \dfrac{1}{2} \int_{-\infty}^{\infty} dx \left( \dfrac{ \delta\tilde{S}^z(x,0) ^2}{C_{1}} + \int_{-\pi}^{\pi} d \lambda \dfrac{\delta\tilde{\rho}^{\rm (c)}(\lambda,x,0)^2}{C_2(\lambda)}  \right) \nonumber \\
&+  z \tau^{1/4} \int_{-\infty}^{\infty} dx \delta \tilde{S}^z(x,0) \chi[ -|X_1| < x < 0]  \bigg]. \label{eq:sup_PI1}
\end{align}
Here, we use the relation $Z_{\rm hyd} = Z_{\rm charge} Z_{\rm spin}$ with 
\begin{align} 
Z_{\rm charge} &\coloneqq \int   \mathcal{D} \delta\tilde{\rho}^{\rm (c)}(\lambda,x,0) \exp \bigg( - \dfrac{1}{2} \int_{-\infty}^{\infty} dx  \int_{-\pi}^{\pi} d \lambda \dfrac{\delta\tilde{\rho}^{\rm (c)}(\lambda,x,0)^2}{C_2(\lambda)}     \bigg), \\
Z_{\rm spin} &\coloneqq  \int \mathcal{D} \delta\tilde{S}^z(x,0)   \exp \bigg( - \dfrac{1}{2} \int_{-\infty}^{\infty} dx  \dfrac{\delta\tilde{S}^z(x,0)^2}{C_{1}}    \bigg).
\end{align}

Second, we implement the functional integral with respect to $\delta\tilde{S}^z$. Then, we introduce the new variable defined by $ A(x) \coloneqq \delta\tilde{S}^z(x,0) - C_1 z \tau^{1/4} \chi[ -|X_1| < x < 0] $ with the indicator function $\chi[\bullet]$.
As a result, we get
\begin{align} 
\langle e^{ z J(\tau)} \rangle \sim \dfrac{1}{Z_{\rm charge}} \int  \mathcal{D} \delta\tilde{\rho}^{\rm (c)}(\lambda,x,0) \exp \bigg( &- \dfrac{1}{2} \int_{-\infty}^{\infty} dx  \int_{-\pi}^{\pi} d \lambda \dfrac{\delta\tilde{\rho}^{\rm (c)}(\lambda,x,0)^2}{C_2(\lambda)}  +  \dfrac{C_1 z^2 \tau^{1/2}}{2} |X_1|  \bigg). \label{eq:sup_PI2}
\end{align}

Third, we put Eq.~\eqref{eq:EM_Xt} with $t=1$ into Eq.~\eqref{eq:sup_PI2}, calculating the functional integral with respect to $\delta\tilde{\rho}^{\rm (c)}$. 
This substitution leads to 
\begin{align} 
\langle e^{ z J(\tau)} \rangle \sim \dfrac{1}{Z_{\rm charge}} \int  \mathcal{D} \delta\tilde{\rho}^{\rm (c)}(\lambda,x,0) &\exp \bigg( - \dfrac{1}{2} \int_{-\infty}^{\infty} dx  \int_{-\pi}^{\pi} d \lambda \dfrac{\delta\tilde{\rho}^{\rm (c)}(\lambda,x,0)^2}{C_2(\lambda)} \bigg) \nonumber \\
\times  &\exp \bigg(  \dfrac{z^2 \tau^{1/4}}{2} \left| \int_{-\pi}^{\pi} d\lambda \int_0^{2 \sqrt{\tau} \sin \lambda} dx \delta\tilde{\rho}^{\rm (c)}(\lambda,-x,0)  \right| \bigg). \label{eq:sup_PI3}
\end{align}
To proceed with the functional integral of Eq.~\eqref{eq:sup_PI3}, we define $Y \coloneqq \int_{-\pi}^{\pi} d\lambda \int_0^{2 \sqrt{\tau} \sin \lambda} dx \delta\tilde{\rho}^{\rm (c)}(\lambda,-x,0)$ and $B \coloneqq z^2 \tau^{1/4}/2$, and then obtain
\begin{align} 
\langle e^{ z J(\tau)} \rangle \sim \left\langle e^ {  B |Y| }  \right\rangle_{\rm charge}, 
 \label{eq:sup_PI4}
\end{align}
where we define 
\begin{align} 
\left\langle \bullet  \right\rangle_{\rm charge} \coloneqq \dfrac{1}{Z_{\rm charge}} \int  \mathcal{D} \delta\tilde{\rho}^{\rm (c)}(\lambda,x,0) \exp \bigg( - \dfrac{1}{2} \int_{-\infty}^{\infty} dx  \int_{-\pi}^{\pi} d \lambda \dfrac{\delta\tilde{\rho}^{\rm (c)}(\lambda,x,0)^2}{C_2(\lambda)} \bigg) \bullet.
\end{align} 
To evaluate Eq.~\eqref{eq:sup_PI4}, we consider $\left\langle e^ {  B Y }  \right\rangle_{\rm charge}$. 
Following almost the same calculation as the derivation of Eq.~\eqref{eq:sup_PI2} and taking into account the sign of $\sin \lambda$ for the upper limit of the integral, we obtain
\begin{align} 
\left\langle e^ {  B Y }  \right\rangle_{\rm charge} &= \exp\left( \dfrac{B^2 \sqrt{\tau}}{2 \pi} \int_{-\pi}^{\pi} d\lambda |\sin \lambda| n^{\rm (c)}(\lambda) (1-n^{\rm (c)}(\lambda))  \right)  = \exp\left( \dfrac{B^2 }{2} ( \tau^{1/4}d(\beta,\mu))^2 \right), 
\end{align}
where we use 
\begin{align} 
d(\beta,\mu) = \sqrt{ \dfrac{2}{\pi \beta} \left( \dfrac{1}{ 2+e^{-\beta(2+\mu)}} - \dfrac{1}{ 2+e^{\beta(2-\mu)}}  \right)  }.
\end{align} 
This result means that the probability of $Y$ obeys the Gaussian distribution with the zero average and the variance $\tau^{1/2}d(\beta,\mu)^2$ because the cumulant generating function $\log \left \langle e^ {  B Y } \right\rangle_{\rm charge}$ is the quadratic form of $B$. Therefore, we get
\begin{align} 
\langle e^{ z J(\tau)} \rangle &\sim \left\langle e^ {  B |Y| }  \right\rangle_{\rm charge} \\
&= \int_{-\infty}^{\infty} dy \dfrac{1}{\sqrt{2\pi \tau^{1/2} d(\beta,\mu)^2}} \exp \left( - \dfrac{y^2}{2 \tau^{1/2} d(\beta,\mu)^2 } + B |y| \right) \\
&= \exp \left( \dfrac{  d(\beta,\mu)^2 z^4 \tau  }{8} \right) \left[ 1 + {\rm erf}\left( \dfrac{  d(\beta,\mu) z^2 \sqrt{\tau} }{2^{3/2}}\right) \right]
\label{eq:sup_GS}
\end{align}
with the error function ${\rm erf}(\bullet)$. This completes the derivation of Eq.~\eqref{eq:EM_Gene} of End Matter.

Finally, we show that Eq.~\eqref{eq:sup_GS} leads to the M-Wright function in Eq.~\eqref{eq:Ps_limit_hydro} of the main text. 
We denote the probability of the integrated spin current under the hydrodynamic assumption by $\mathbb{P}_S^{\rm hyd}[\Delta S^z_R,\tau]$, obtaining
\begin{align} 
\mathbb{P}_S^{\rm hyd}[\Delta S^z_R,\tau] 
&\coloneqq \dfrac{1}{2\pi} \int_{-\infty}^{\infty} dz e^{-{\rm i} \Delta S^z_R z } \langle e^{ {\rm i} z J(\tau)} \rangle  \\
&= \dfrac{1}{2\pi} \int_{-\infty}^{\infty} dz e^{-{\rm i} \Delta S^z_R z } \exp \left( \dfrac{  d(\beta,\mu)^2 z^4 \tau  }{8} \right) \left[ 1 + {\rm erf}\left( \dfrac{ - d(\beta,\mu) z^2 \sqrt{\tau} }{2^{3/2}}\right) \right] \\
&= \dfrac{1}{2\pi} \int_{-\infty}^{\infty} dz e^{-{\rm i} \Delta S^z_R z } \exp \left( \dfrac{  d(\beta,\mu)^2 z^4 \tau  }{8} \right)  {\rm erfc} \left( \dfrac{ d(\beta,\mu) z^2 \sqrt{\tau} }{2^{3/2}} \right)  \label{eq:sup:Ps1}\\
&= \dfrac{1}{4\pi^{3/2}} \int_{-\infty}^{\infty} dz e^{-{\rm i} \Delta S^z_R z } \int_{-\infty}^{\infty} dy \exp \left[   - \dfrac{y^2}{4} - \left( \dfrac{ d(\beta,\mu) z^2 \sqrt{\tau} }{2^{3/2}} \right) |y|    \right] \label{eq:sup:Ps2} \\
&= \sqrt{ \dfrac{\sqrt{2}}{8\pi^2 d(\beta,\mu) \sqrt{\tau}} } \int_{-\infty}^{\infty} dy \dfrac{1}{\sqrt{|y|}} \exp \left[  - \dfrac{y^2}{4} - \dfrac{(\Delta S^z_R)^2}{\sqrt{2} d(\beta,\mu) |y| \tau^{1/2} }    \right], 
\end{align}
where we use the formula $2\sqrt{\pi} e^{x^2} {\rm erfc} (x) = \int_{-\infty}^{\infty} dy \exp \left( -x|y| - y^2/4\right)$ of the error function for Eq.~\eqref{eq:sup:Ps1} with the complementary error function ${\rm erfc}(\bullet)$ and perform the Gaussian integral of Eq.~\eqref{eq:sup:Ps2} with respect to $z$. To investigate the typical fluctuation of $\Delta S_R^z$, we consider $\tau^{1/4} \mathbb{P}_S^{\rm hyd}[\tau^{1/4} \mathcal{J}_S,\tau] $ and then get
\begin{align} 
\tau^{1/4} \mathbb{P}_S^{\rm hyd}[\tau^{1/4} \mathcal{J}_S,\tau]   
&= \sqrt{ \dfrac{\sqrt{2}}{8\pi^2 d(\beta,\mu)} } \int_{-\infty}^{\infty}  \dfrac{dy}{\sqrt{|y|}} \exp \left[  - \dfrac{y^2}{4}  - \dfrac{\mathcal{J}_S^2}{\sqrt{2} d(\beta,\mu) |y|  }  \right] \\
&=  \dfrac{1}{\pi d(\beta,\mu)}  \int_{0}^{\infty}  \dfrac{d \mathcal{J}_C}{\sqrt{\mathcal{J}_C}} \exp \left[  - \dfrac{\mathcal{J}_C^2}{2d(\beta,\mu)^2} - \dfrac{\mathcal{J}_S^2}{ 2  \mathcal{J}_C  }    \right].
\end{align}
This completes the derivation for Eq.~\eqref{eq:Ps_limit_hydro} of the main text.

\end{document}